\documentclass[a4paper,pop,fleqn]{w-art}
\usepackage{amsmath,amsfonts,amssymb,w-thm,times,graphicx,subfigure,listings,tocvsec2,w-greek}

\def\be#1#2\ee{\begin{equation}\label{#1}#2\end{equation}}
\def\bea#1#2\eea{\begin{eqnarray}\label{#1}#2\end{eqnarray}}
\def\note#1{\marginpar{\raggedright\if@twoside\ifodd\c@page\raggedleft\fi\fi\sf\scriptsize Note: #1}}

\newcommand{\bN}{\mathbb{N}}
\newcommand{\bQ}{\mathbb{Q}}

\newcommand{\bZ}{\mathbb{Z}}
\newcommand{\bZZ}{\bZ_2\!\times\!\bZ_2}

\newcommand{\cA}{\mathcal{A}}

\newcommand{\cN}{\mathcal{N}}

\newcommand{\ga}{\alpha}

\newcommand{\del}{\partial}
\newcommand{\gsb}{\bar{\sigma}}
\newcommand{\Hess}{\mathrm{Hess}}
\newcommand{\hodge}{\star}
\newcommand{\hypeq}{\stackrel{!}{=}}
\newcommand{\ov}{\overline}
\newcommand{\pii}{\frac{1}{2\pi i}}

\newcommand{\rep}[1]{\mathbf{#1}}

\newcommand{\N}{\rep{N}}
\newcommand{\1}{\rep{1}}
\newcommand{\2}{\rep{2}}
\newcommand{\3}{\rep{3}}

\newcommand{\fivebar}{\rep{\bar{5}}}
\newcommand{\ten}{\rep{10}}
\newcommand{\fifteen}{\rep{15}}
\newcommand{\Sym}{\rep{Sym}}
\newcommand{\Anti}{\rep{Anti}}

\newcommand{\figwidth}{0.60\linewidth}
\newcommand{\fig}[3]{%
  \begin{figure}[th]%
  \begin{center}%
  \includegraphics[width=\figwidth]{#1}%
  \caption{#3}%
  \label{#2}%
  \end{center}%
  \end{figure}%
}
\newcommand{\figclip}[3]{%
  \begin{figure}[th]%
  \begin{center}%
  \includegraphics[width=\figwidth,trim=0mm 10mm 0mm 10mm,clip]{#1}%
  \caption{#3}%
  \label{#2}%
  \end{center}%
  \end{figure}%
}
\newcommand{\twofig}[4]{%
  \begin{figure}[th]%
  \begin{center}%
  \subfigure[\label{#3_a}]{%
    \includegraphics[width=0.48\linewidth]{#1}%
  }%
  \hfill%
  \subfigure[\label{#3_b}]{%
    \includegraphics[width=0.48\linewidth]{#2}%
  }%
  \caption{#4}%
  \label{#3}%
  \end{center}%
  \end{figure}%
}

\newcommand{\fourfigclip}[6]{%
  \begin{figure}[th]%
  \begin{center}%
  \subfigure[\label{#5_a}]{%
    \includegraphics[width=0.49\linewidth,trim=0mm 10mm 0mm 10mm,clip]{#1}%
  }%
  \subfigure[\label{#5_b}]{%
    \includegraphics[width=0.49\linewidth,trim=0mm 10mm 0mm 10mm,clip]{#2}%
  }\\%
  \subfigure[\label{#5_c}]{%
    \includegraphics[width=0.49\linewidth,trim=0mm 10mm 0mm 10mm,clip]{#3}%
  }%
  \subfigure[\label{#5_d}]{%
    \includegraphics[width=0.49\linewidth,trim=0mm 10mm 0mm 10mm,clip]{#4}%
  }%
  \caption{#6}%
  \label{#5}%
  \end{center}%
  \end{figure}%
}

\providecommand{\href}[2]{#2}
\begin{document}
\pagespan{1}{}
\bigskip

\begin{flushright}
MPP-2006-103\\LMU-ASC 54/06\\hep-th/0608227
\end{flushright}


\title{Gauge sector statistics of intersecting D-brane models}

\author[F. Gmeiner]{Florian Gmeiner}
\address{%
Max-Planck-Institut f\"ur Physik\\%
F\"ohringer Ring 6, 80805 M\"unchen, Germany\\[1.5ex]%
Arnold-Sommerfeld-Center for Theoretical Physics\\%
Department f\"ur Physik, Ludwig-Maximilians-Universit\"at  M\"unchen\\%
Theresienstra{\ss}e 37, 80333 M\"unchen, Germany\\[1.5ex]%
E-Mail: \textsl{flo@mppmu.mpg.de}}

\begin{abstract}
In this article, which is based on the first part of my PhD thesis,
I review the statistics of the open string sector in $T^6/(\bZZ)$ orientifold
compactifications of the type IIA string.
After an introduction to the orientifold setup, I discuss the two different
techniques that have been developed to analyse the gauge sector statistics,
using either a saddle point approximation or a direct computer based method.
The two approaches are explained and compared by means of eight- and
six-dimensional toy models. In the four-dimensional case the results are
presented in detail.
Special emphasis is put on models containing phenomenologically interesting
gauge groups and chiral matter, in particular those containing a standard
model or SU(5) part.
\end{abstract}

\maketitle
\vspace{1cm}
\tableofcontents

%
%
\section{Introduction}
Starting from early observations of Lerche, L\"ust and
Schellekens~\cite{lls86}, it has become clear over the years that
string theory does provide us not only with one consistent low energy
effective theory, but with a multitude of solutions.
This phenomenon has been given the name
``the landscape''~\cite{su03,sc06}
(for a recent essay on the subject see also~\cite{do06}).

It was known from the very first approaches to compactification of
string theory to four dimensions
that there exist many families of solutions due to the so-called moduli.
These scalar fields parametrise the geometric properties of possible
compactification manifolds and their values are generically not fixed. It was
believed for a long time that some stabilisation mechanism for these moduli
would finally lead to only one consistent solution. Even though it is way too
early to completely abandon this idea, recent developments suggest that even
after moduli stabilisation there exists a very large number of consistent
vacuum solutions.
Especially the studies of compactifications with
fluxes (see e.g.~\cite{gr05} and references therein)
clarified the situation. The effective potential induced by these
background fluxes, together with non-perturbative effects, allow to fix
the values of some or even all of the moduli at a supersymmetric minimum.
What is surprising is the number of possible minima, which has been
estimated~\cite{bopo00}\footnote{Note that in this estimate not all effects
from the process of moduli stabilisation have been taken into account.}
to be of the order of $10^{500}$.
So it seems very likely that there exists a very large number of stable
vacua in string theory that give rise to low energy theories which
meet all our criteria on physical observables.

After the initial work of Douglas~\cite{do03}, who pointed out that facing
these huge numbers the search for \emph{the} vacuum is no longer feasible,
recent research has
started to focus on the statistical distributions of string vacua.
This approach relies on the conjecture that, given such a huge number of
possible vacua, our world can be realized in many different ways and only
a statistical analysis might be possible.
Treating physical theories on a statistical basis is a provocative statement
and it has given rise to a sometimes very emotional debate. Basic
criticism is expressed in~\cite{ba04,bdg03}, where the authors emphasise
the point that, as long as we do not have a non-perturbative description of
string theory, such reasoning seems to be premature.
Moreover such an approach immediately rises philosophical questions.
How can we talk seriously about the idea to abandon unambiguous predictions of
reality and replace it with statistical reasoning? One is reminded to
similar questions concerning quantum mechanics, but there is a major
difference to this problem. In the case of quantum mechanics there is a
clean definition of observer and measurement. Most importantly, measurements
can be repeated and therefore we can make sense out of a statistical
statement. In the case of our universe we have just one measurement and
there is no hope to repeat the experiment.

At the moment there are two roads visible that might lead to a solution of
these problems. One of them is based on anthropic arguments~\cite{su03},
which have already been used outside
string theory to explain the observed value of the cosmological
constant~\cite{we87}. Combined with the landscape picture this gives rise to
the idea of a multiverse, where all possible solutions for a string vacuum
are actually realised~\cite{kali02} (for a recent essay on the
cosmological constant problem and the string landscape see also~\cite{po06}).
Anthropic reasoning is not very satisfactory, especially within the framework
of a theory that is believed to be unique. Another possible way to deal with
the landscape might therefore be the assignment of an entropy to
the different vacuum solutions and their interpretation in terms
of a Hartle-Hawking wave function~\cite{ovv05,clp06}. A principle of
extremisation of the entropy could then be used to determine the correct
vacuum.

We do not dwell into philosophical aspects of the landscape
problem in this work, but rather take a very pragmatic point of view,
following Feynman's ``shut up and calculate'' attitude.
In this endeavour a lot of work has been done to analyse the properties and
define a suitable statistical measure in
the closed string sector of the theory~\cite{asdo03,dedo041,ddf04,%
gkt04,dgt04,mina04,coqu04,dedo042,dgkt04,ddg04,dos05,adv05,diva05,dolu05,%
ku06}.
In this work we are focusing on the statistics of the open string
sector~\cite{bghlw04,kuwe04,kuwe05,adk05,va05,gbhlw05,gm05,gmst06,kuwe06,%
dgs06,di06}.
We are not trying to take the most general point of view and analyse a
generic statistical distribution, but focus instead on a very specific
class of models. In this small region of the landscape we are
going to compute almost all possible solutions and give an estimate for
those solutions we were not able to take into account.

There are several interesting questions one can ask, given a large set of
possible models. One of them concerns the frequency distribution of
properties, like the total rank of the gauge group or the occurrence of
certain gauge factors. Another question concerns the correlation of
observables in these models.
This question is particularly interesting, since
a non-trivial correlation of properties could lead to the exclusion of
certain regions of the landscape or give hints where to look for realistic
models.
It should be stressed that in our analysis of realistic four-dimensional
compactifications we are not dealing with an abstract statistical measure,
but with explicit constructions.

\subsection{Outline}\label{intro_outline}
This paper is based on the first part of the author's PhD thesis~\cite{gm06}
and is structured as follows.
In section~\ref{ch_landscape} we
prepare the stage, introducing the special class of type~II orientifold
models that are our objects of interest. Moreover
we explain the two methods we use to analyse these models. On the
one hand, the saddle point approximation and on the other hand a brute force
computer algorithm for explicit calculations. Concerning this algorithm, we
comment on its computational complexity, which touches a more general
issue about computations in the landscape.
In the last section we discuss another fundamental problem of the
statistical analysis, namely the finiteness of vacua. An analytic proof
of finiteness seems to be out of reach, but we give several numerical
arguments that support the conjecture that the total number of solutions
is indeed finite.

In section~\ref{ch_stat} we apply the described methods to type~II orientifold
models. We begin with general questions about the frequency distributions of
properties of the gauge sector in compactifications to
six and four dimensions. After that we select several
subsets of models for a more detailed analysis. We choose those subsets that
could provide us with a phenomenologically interesting low energy gauge group.
This includes first of all the standard model, but in addition constructions
of Pati-Salam, $SU(5)$ and flipped $SU(5)$ models.
In the case of standard model-like constructions we investigate the relations
and frequency distributions of the gauge coupling constants and compare the
results with a recent analysis of Gepner models~\cite{dhs041,dhs042}.
In the last part of this section the question
of correlations of gauge sector observables is explored.

Finally we sum up our results and give an outlook to further directions of
research in section~\ref{ch_conclusions}.
In appendix~\ref{app_models} we summarise some useful formulae for the
different orientifold models.
Appendix~\ref{app_pa} contains details about the implementation of the
computer algorithm, used to construct the models that have been analysed.

%
%
\section{Models and methods}\label{ch_landscape}
As explained in the introduction, our program to classify a subset of the
landscape of string vacua is performed on a very specific set of
models. In this section, we want to set the stage for the analysis, explain
the construction and the constraints on possible solutions. Moreover, we
have to develop the necessary tools of analysis.

In the first part of this section we give a general introduction to the
orientifolds we are planning to analyse. We focus on the consistency
conditions that have to be met by any stable solution. In particular
these are the tadpole conditions for the R-R fields, the supersymmetry
conditions on the three-cycles wrapped by D-branes and orientifold planes
and restrictions coming from the requirement of anomaly cancellation.

In the second part we develop the
tools for a statistical analysis and test them on a very simple
compactification to eight dimensions. There are two methods that we use
for six- and four-dimensional models in the next section, namely
an approximative method and a direct, brute force analysis.
The first method relies on the saddle point approximation, which we
explain in detail and compare it with known results from number theory.
For the second method we describe an algorithm that can be used for a large
scale search performed on several computer clusters. To estimate the amount
of time needed to generate a suitable amount of solutions, we analyse
the computational complexity of this algorithm.

In the last part of this section we investigate the problem of
finiteness of the number of solutions, an issue that is important to judge
the validity of the statistical statements.

\subsection{Orientifold models}\label{ls_orientifolds}
Let us give a brief introduction to the
orientifold models we use in the following to do a statistical
analysis. We will not try to give a complete introduction to the subject,
for readers with interest in more background material, we refer to the
available textbooks~\cite{po981,po982,jo03} and reviews~\cite{pcj96,po96,fo01}
for a general introduction and the recent review~\cite{bcls05} for an account
of orientifold models and their phenomenological aspects.

Our analysis is based on the study of supersymmetric
toroidal type~II orientifold
models with intersecting D-branes~\cite{bdl96,arsh96,bale96,ohto97,sh97}.
These models are, of course, far from being the most general compactifications,
but they have the great advantage of being very well understood.
In particular, the basic constraints for model building, namely the tadpole
cancellation conditions, the supersymmetry and K-theory constraints, are well
known. It is therefore possible to classify almost all possible solutions for
these constructions.

The orientifold models we consider can be described in type~IIB string
theory using space-filling D9-branes with background gauge fields on their
worldvolume. An equivalent description can be given in the T-dual
type~IIA picture, where the D9-branes are replaced by D6-branes, which
intersect at non-trivial angles.
This point of view is geometrically appealing and goes under the name
of intersecting D6-branes. We use this description in the following.

The orientifold projection is given by $\Omega\gsb(-1)^{F_L}$, where
$\Omega: (\sigma,\tau)\to(-\sigma,\tau)$ defines
the world-sheet parity transformation and $\gsb$ is an isometric
anti-holomorphic involution, which we choose to be simply complex conjugation
in local coordinates: $\gsb:z\to\bar{z}$. $F_L$ denotes the left-moving
space-time fermion number.
This projection introduces topological defects in the geometry,
the so-called orientifold O6-planes.
These are non-dynamical objects, localised at the fixed point locus of $\gsb$,
which carry tension and charge under the
R-R seven-form, opposite to those of the D6-branes\footnote{It is
also possible to introduce orientifold planes with different charges, but
we consider only those with negative tension and charge in this work.}.

Both, the O6-planes and D6-branes
wrap three-cycles $\pi\in H_3(M,\bZ)$ in the internal Calabi-Yau manifold
$M$, which, in order to preserve half of the
supersymmetry, have to be special Lagrangian.
Since the charge of the orientifolds is fixed and we are dealing with a
compact manifold, the induced R-R and NS-NS tadpoles have to be cancelled
by a choice of D6-branes. These two conditions, preserving supersymmetry and
cancelling the tadpoles, are the basic model building constraints we have to
take into account.

The homology group $H_3(M,\bZ)$ of three-cycles in the compact manifold
$M$ splits under the action of $\Omega\bar{\sigma}$ into an even
and an odd part, such that the only non-vanishing intersections are
between odd and even cycles. We can therefore choose a symplectic
basis $(\alpha_I,\beta_I)$ and expand $\pi_a$ and $\pi'_a$ as
\be{eq_base}
  \pi_a = \sum_{I=1}^{b_3/2}\left(X_a^I\alpha_I+Y_a^I\beta_I\right),\qquad
  \pi'_a = \sum_{I=1}^{b_3/2}\left(X_a^I\alpha_I-Y_a^I\beta_I\right),
\ee
and $\pi_{O6}$ as
\be{eq_base2}
  \pi_{O6} = \frac{1}{2}\sum_{I=1}^{b_3/2}L^I\alpha_I,
\ee
where $b_3$ is the third Betti-Number of $M$, counting the number of
three-cycles.

\subsubsection{Chiral matter}\label{sb_chmatter}
Chiral matter arises at the intersection of branes wrapping different
three-cycles. Generically we get bifundamental representations
$(\N_a,\overline{\N}_b)$ and $(\N_a,\N_b)$
of $U(N_a)\times U(N_b)$
for two stacks with $N_a$ and $N_b$ branes. The former arise at the
intersection of brane $a$ and brane $b$, the latter at the intersection
of brane $a$ and the orientifold image of brane $b$, denoted by $b'$.
An example is shown in figure~\ref{fig_isbranes}.

\fig{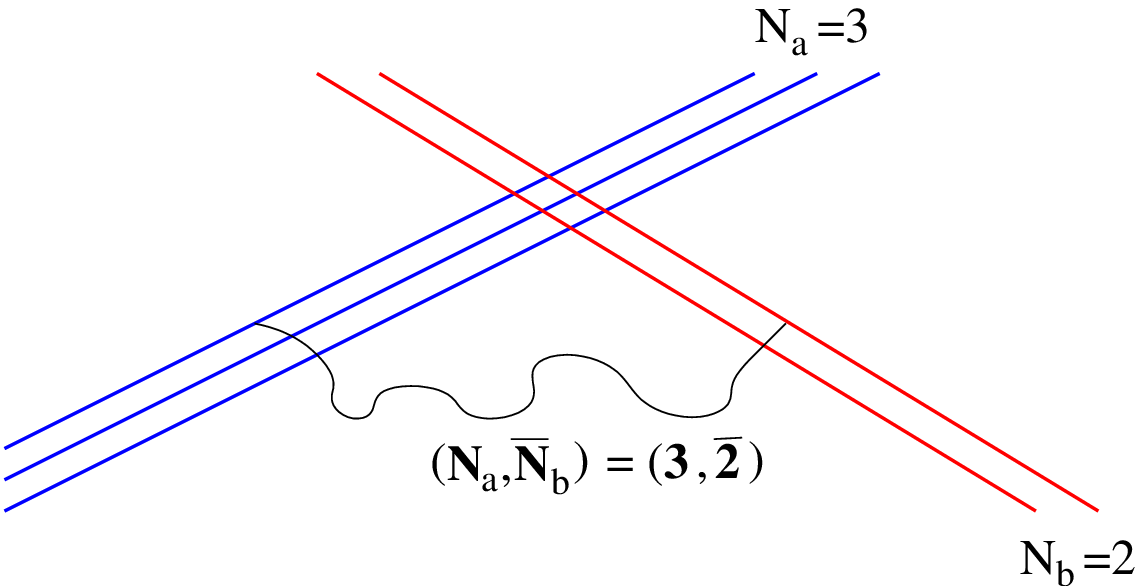}{fig_isbranes}{%
We find chiral matter at the intersection of two stacks of branes. The
representation is given in terms of the number of branes of each stack.%
}

In addition we get matter
transforming in symmetric or antisymmetric representations of the
gauge group for each individual stack.
The multiplicities of these representations are given by the intersection
numbers of the three-cycles,
\be{eq_isn}
I_{ab} := \pi_a\circ \pi_b
        = \sum_{I=1}^{b_3/2}\left(X_a^IY_b^I-X_b^IY_a^I\right).
\ee
The possible representations are summarized in table \ref{tab_reps}, where
$\Sym_a$ and $\Anti_a$ are the symmetric and antisymmetric representations of
$U(N_a)$.
\begin{table}[h]
\begin{center}
\begin{tabular}{|c|c|}\hline
representations & multiplicity \\\hline
$(\N_a,\overline{\N}_b)$ &
  $\pi_a\circ\pi_b=I_{ab}$\\\hline
$(\N_a,\N_b)$ &
  $\pi'_a\circ\pi_b=I_{ab'}$ \\\hline
$\Sym_a$ &
  $\frac{1}{2}\left(\pi_a\circ\pi'_a-\pi_a\circ\pi_{O6}\right)
  =\frac{1}{2}(I_{aa'}-I_{a{\rm O}6})$\\\hline
$\Anti_a$ &
  $\frac{1}{2}\left(\pi_a\circ\pi'_a+\pi_a\circ\pi_{O6}\right)
  =\frac{1}{2}(I_{aa'}+I_{a{\rm O}6})$\\\hline
\end{tabular}%
\caption{Multiplicities of the chiral spectrum.}
\label{tab_reps}
\end{center}
\end{table}

\subsubsection{Tadpole cancellation conditions}\label{sb_tadpoles}
The D6-branes in our models are charged under a R-R seven-form \cite{po95}.
Since the internal manifold is compact, as a simple consequence of the Gauss
law, all R-R charges have to add up to zero.
These so-called tadpole cancellation conditions
can be obtained considering the part of the supergravity
Lagrangian that contains the corresponding contributions.
In particular we do not only get contributions from $k$ stacks of branes,
wrapping cycles $\pi_a$, but in addition terms from
the orientifold mirrors of these branes, wrapping
cycles $\pi'_a$, and the O6-planes.
\be{eq_C7coupling}
  S = -\frac{1}{4\kappa^2}\int\limits_{X\times M}dC_7\wedge\hodge dC_7
      +\mu_6\sum_{a=1}^k N_a\left( \int\limits_{X\times\pi_a}C_7
                           +\int\limits_{X\times\pi'_a}C_7\right)
      -4\mu_6\int\limits_{X\times\pi_{O6}}C_7,
\ee
where the ten dimensional gravitational coupling is given by
$\kappa^2=\frac{1}{2}(2\pi)^7(\alpha')^4$, the R-R charge is denoted by
$\mu_6=(\alpha')^{-\frac{7}{2}}(2\pi)^{-6}$ and $X$ denotes the
uncompactified space-time.

From this we can derive the equations of motion for the R-R field strength
$G_8=dC_7$ to be
\be{eq_G8eom}
  d\hodge G_8 = \kappa^2\mu_6\left(\sum_{a=1}^k N_a
    \left(\delta(\pi_a)+\delta(\pi'_a)\right)-4\delta(\pi_{O6})\right).
\ee
In this equation $\delta(\pi)$ denotes the Poincar\'e dual three form of a
cycle $\pi$. Noticing that the left hand side of~(\ref{eq_G8eom}) is exact,
we can rewrite this as a condition in homology as
\be{eq_tad}
  \sum_{a=1}^k N_a(\pi_a +\pi'_a)=4\pi_{O6}
\ee

We do not have to worry about the NS-NS tadpoles, as long as we are
considering supersymmetric models, since the supersymmetry conditions
together with R-R tadpole cancellation ensure that there are no NS-NS
tadpoles. In the following we consider supersymmetric models only.

\subsubsection{Supersymmetry conditions}\label{sb_susy}
Since we want to analyse supersymmetric models, it is crucial that the
D-branes and O-planes preserve half of the target-space supersymmetry.
It can be shown~\cite{mmms99} that this requirement is equivalent to a
calibration condition on the cycles,
\be{eq_susy}
\Im (\Omega_3)|_{\pi_a}=0,\qquad
\Re (\Omega_3)|_{\pi_a}>0,
\ee
where $\Omega_3$ is the holomorphic 3-form. The second equation
in~(\ref{eq_susy}) excludes anti-branes from the spectrum.

Written in the symplectic basis~(\ref{eq_base}),
these equations read
\be{eq_susy2}
\sum_{I=1}^{b_3/2}Y_a^If_I = 0,\qquad
\sum_{I=1}^{b_3/2}X_a^Iu_I > 0,
\ee
where we defined
\be{eq_susyhelp}\nonumber
f_I := \int_{\beta_I}\Omega_3,\qquad
u_I := \int_{\alpha_I}\Omega_3.
\ee

\subsubsection{Anomalies and K-theory constraints}%
\label{sb_anomalies}\label{sb_ktheory}
If the tadpole cancellation conditions~(\ref{eq_tad}) are satisfied, there
are no cubic anomalies of $SU(N)$ gauge groups in our models.
What we do have to worry about are mixed anomalies, containing abelian
factors. The mixed anomaly for branes stretching between two stacks $a$ and $b$
with $N_a=1$ and $N_b>1$ branes per stack, looks like
\be{eq_mixanom}
  \cA_{U(1)_a-SU(N)_b} \simeq N_a(I_{ab}+I_{ab'})c_2(N_b)
   =  -2N_a\vec{Y_a}\vec{X_b}c_2(N_b),
\ee
where $c_2(N_b)$ denotes the value of the quadratic Casimir operator for the
fundamental representation of $SU(N_b)$.

The cubic anomaly consisting of three abelian factors is
cancelled by the Green-Schwarz mechanism. This makes these $U(1)$s
massive and projects them out of the low energy spectrum.
But in some cases, for example in the case of a standard model-like gauge group
or for flipped $SU(5)$ models, we want to get a massless $U(1)$ factor.
A sufficient condition to get such a massless $U(1)$ in one of our models
is that the anomaly~(\ref{eq_mixanom}) vanishes.

This can be archived, if the $U(1)$, defined in general by a combination of
several $U(1)$ factors as
\be{eq_u1def}
  U(1)=\sum_{a=1}^k x_aU(1)_a,
\ee
fulfills the following relations,
\be{eq_u1rel}
  \sum_{a=1}^kx_aN_a\vec{Y}_a=0.
\ee
Inserting this into~(\ref{eq_mixanom}) shows that $\cA$ vanishes.

Besides these local gauge anomalies, there is also the potential danger of
getting a global gauge anomaly, which would make the whole model inconsistent.
This anomaly arises if a $\bZ_2$-valued K-theory charge is not
conserved~\cite{ur00}. In our case this anomaly can be derived by introducing
$Sp(2)$ probe branes on top of the orientifold planes and compute their
intersection numbers with all branes in the model. This intersection number
has to be even, otherwise we would get an odd number of fermions, transforming
in the fundamental representation of $Sp(2)$~\cite{wi82}.

\subsection{Methods of D-brane statistics}\label{ls_osstats}
To analyse a large class of models in the orientifold setting described in
the last section, we have to develop some tools that allow us to generate
as many solutions to the supersymmetry, tadpole and K-theory conditions
as possible.
It turns out that the most difficult part of this problem can be reduced
to a purely number theoretical question, namely the problem of
counting partitions of natural numbers. This insight allows us to use
an approximative method, the saddle point approximation that we introduce
in section~\ref{ls_saddle} and apply to a simple toy-model in~\ref{ls_saddle2}.
Unfortunately it turns out that this method is not
very well suited to study the most interesting compactifications, namely those
down to four dimensions.
Therefore we have to change the method of analysis in that case
to a more direct one, using a brute force, exact computer analysis.
The algorithm used to do so is described in section~\ref{ls_comp}.

\subsubsection{Introduction to the saddle point approximation}\label{ls_saddle}
As an approximative method to analyse the gauge sector of type~II orientifolds,
the saddle point approximation has been introduced in~\cite{bghlw04}.
In the following we begin with a very simple, eight-dimensional model,
in order to explain the method.

In the most simple case, a compactification to eight dimensions on $T^2$,
the susy conditions reduce to $Y_a=0$ and $X_a>0$ and the tadpole cancellation
conditions are given by
\be{eq_tad8}
  \sum_{a=1}^k N_a\, X_a = 16,
\ee
as shown in appendix~\ref{app_t2}.

The task to count the number of solutions to this equation for an arbitrary
number of stacks $k$ is a combination of a partitioning and factorisation
problem.
Let us take things slowly and start with a pure partitioning problem,
namely to count the unordered solutions of
\be{eq_part}
  \sum_{a=1}^k  N_a  = L.
\ee
This is nothing else but the number of unordered partitions of $L$.
Since we are not interested in an exact solution, but rather an approximative
result, suitable for a statistical analysis and further generalisation to the
more ambitious task of solving the tadpole equation, let us attack this by
means of the saddle point approximation~\cite{an76,wo89}.

As a first step to solve~(\ref{eq_part}), let's consider
\be{eq_part2}
  \sum_{k=1}^\infty  k\,  n_k  = L,
\ee
where we do not have to worry about the ordering problem.
We can rewrite this as
\be{eq_part3}
  \cN(L) = \sum_{all} \delta_{\sum_{k}\,   k  n_k-L ,0}
         = \pii\oint dq \frac{1}{q^{L+1}}\sum_{n_k=0}^\infty
           q^{\sum_{k}\,  k  n_k}
         = \pii\oint dq \frac{1}{q^{L+1}}
           \prod_{k=1}^\infty \left(\frac{1}{1-q^k}\right).
\ee
To evaluate integrals of this type in an asymptotic expansion, the saddle point
method is a commonly used tool. In the following we describe its
application in detail.
The last line of~(\ref{eq_part3}) can be written as
\be{eq_part4}
  \cN(L)=\pii\oint{dq}\,\exp(f(q)),\quad
  \mbox{with}\quad f(q)=-\sum_{k=1}^\infty\log(1- q^k)-(L+1)\,\log q.
\ee

Now we are going to assume that the main contributions to this integral
come from saddle points $q_i$,
determined by $\left.df/dq\right|_{q_i}=0$. In the following we work
with only one saddle point at $q=q_0$, the generalisation to many points is
always straightforward.
Using the decomposition $q=\rho\exp(i\varphi)$ we get
\be{eq_partfinal}
  \cN(L)=\frac{1}{2\pi}\int\limits_{-\pi}^\pi d\varphi\,q\,\exp(f(q)).
\ee
Performing a Taylor expansion in $\varphi$
\be{eq_parttaylor}
  f(\rho_0,\varphi)=f(q_0)+\left.\frac{1}{2}\,\frac{\del^2f}{\del\varphi^2}
                    \right|_{q_0}\,\varphi^2+\ldots,
\ee
we can compute~(\ref{eq_partfinal}) to arbitrary order by inserting the
corresponding terms from~(\ref{eq_parttaylor}).

The leading order term is simply given by
\be{eq_partleading}
  \cN^{(0)}(L) = \exp(f(q_0)),
\ee
and the first correction at next-to-leading order by
\be{eq_nexttoleading}
  \cN^{(2)}_{corr}(L) =  \frac{1}{2\pi}
    \int\limits_{-q_0 \pi}^{q_0 \pi}dx\,\exp\left(-\frac{1}{2}\left.
    \frac{\del^2 f}{\del q^2}\right|_{q_0} x^2\right),
\ee
where we defined $x:=q_0\varphi$ and used that
$(\del^2f/\del\varphi^2)_{q_0}=- q^2\,(\del^2f/\del q^2)_{q_0}$.
For $\del^2f/\del q^2$ large enough we finally obtain the result for the
saddle point approximation including next-to-leading order corrections
\be{eq_partnextfinal}
  \cN^{(2)}(L) = \frac{1}{2\pi}\,\exp(f(q_0))\left(\left.
    \frac{\del^2f}{\del q^2} \right|_{q_0}\right)^{-1/2}.
\ee

The same procedure can also be performed for functions of several variables.
The integral to approximate this situation looks like
\be{eq_partextint}
  \cN(\vec L)=\pii \oint \prod_{I=1}^n {d\vec{q}} \,\exp(f(\vec{q})),
\ee
with $f$ being of the form
\be{eq_partextf}
  f(\vec{q}) = g(\vec{q}) - \sum_{I=1}^N (L_I+1)\, \log q_I.
\ee
We can perform the saddle point approximation around
$\nabla f(\vec q)|_{\vec q_0}=0$ in the same way as above and obtain the
following result at next-to-leading order
\be{eq_partextfinal}
  \cN^{(2)}(\vec L)= (2\pi)^{-n/2}\,\exp(f(\vec{q_0}))
    \left(\left.\det\Hess f(\vec{q})\right|_{\vec{q_0}}\right)^{-1/2}.
\ee

In the simple case discussed so far, contrary to the more complicated cases
we encounter later, an analytic evaluation of the leading
order contribution is possible.
For large $L$ the integrand of~(\ref{eq_part3}) quickly approaches infinity
for $q<1$ and $q\simeq 1$. One expects a sharp minimum close to 1, which
would be the saddle point we are looking for.

Close to $q\simeq 1$ we can write the first term in~(\ref{eq_part4}) as
\be{eq_sad2}
  -\sum_{k=1}^\infty\log(1-q^k) = \sum_{k,m>0} \frac{1}{m} q^{km}
  \simeq \frac{1}{1-q} \sum_{m>0}\frac{1}{m^2}
  = \frac{\pi^2}{6}\,\frac{1}{1-q},
\ee
such that we can approximate $f(q)$ by
\be{eq_sad3}
  f(q) \simeq \frac{\pi^2}{6}\,\frac{1}{1-q} - (L+1)\, \log q.
\ee
For large values of $L$, the minimum of this function is approximately at
$q_{0}\simeq1-\sqrt{\frac{\pi^2}{6L}}$ which leads to
$f(q_{0}) \simeq \pi \sqrt{2L/3}$.
Inserting this into~(\ref{eq_partleading}) gives
a first estimate of the growth of the partitions for large $L$
to be
\be{eq_sad4}
  \cN(L) \simeq \exp\left(\pi \sqrt{2L/3}\right).
\ee
This is precisely the leading term in the Hardy-Ramanujan formula~\cite{hara18}
for the asymptotic growth of the number of partitions
\be{eq_hara}
  \cN(L)^{(HR)} \simeq \frac{1}{4 L \sqrt 3}\, \exp\left(\pi\sqrt{2L/3}\right).
\ee

\fig{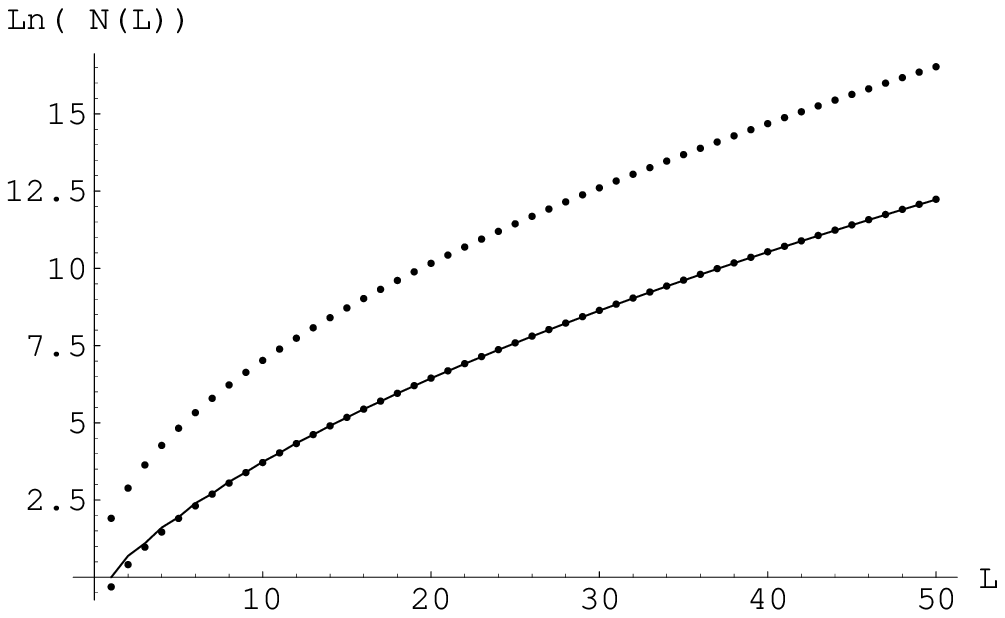}{fig_partcomp}{%
Comparison of the number of partitions obtained by an exact calculation
(solid line) and a saddle point approximation to leading (upper dotted line)
and next-to-leading order (lower dotted line).%
}

In figure~\ref{fig_partcomp} the results of an exact calculation, using the
partition algorithm described in appendix~\ref{app_pa}, and the saddle
point approximation in leading and next-to-leading order are shown.

\subsubsection{A first application of the saddle point approximation}%
\label{ls_saddle2}
After this introduction to the saddle point method let us come back to our
original problem. To solve equation~(\ref{eq_tad8}), we first have to
transfer our approximation method to~(\ref{eq_part}) and then include
the factorisation in the computation. This last step turns out not to be
too difficult, but in order to use the
technique developed above, we have to be a bit careful
about the ordering of solutions.

In the example we presented to introduce the method, we did not have to worry
about the ordering, since it was solved implicitly by the definition of the
partition function. This is not the case for~(\ref{eq_part}), such that
by simply copying from above the result is too large.
We should divide the result by the product of the number of
possibilities to order each partition.
Obtaining this factor precisely is very difficult and since we are only
interested in an approximative result anyway, we should try to estimate
the term.
Such an estimate can be made dividing by $k!$, where $k$ is the total number
of stacks.
This restricts the number of solutions more than necessary,
because the factor is too high for partitions that contain the same element
more than once. Let us nevertheless calculate the result with this rough
estimate and see what comes out.

Repeating the steps from above, we can rewrite~(\ref{eq_part}) to obtain
\bea{eq_partb}\nonumber
  \widetilde{\cN}(L) & \simeq & \pii \oint dq \frac{1}{q^{L+1}}
     \sum_{k=1}^\infty\frac{1}{k!}\,\prod_{i=1}^k\left(
     \sum_{N_i=1}^\infty q^{\sum_a \, N_a} \right)
  = \pii \oint dq \frac{1}{q^{L+1}} \sum_{k=1}^\infty \frac{1}{k!}\,
     \left(\sum_{N=1}^\infty  q^{N}\right)^k \\
  & = & \pii \oint dq \frac{1}{q^{L+1}} \sum_{k=1}^\infty \frac{1}{k!}\,
     \left( \frac{q}{1-q}\right)^k
  = \pii \oint dq \frac{1}{q^{L+1}} \exp\left(\frac{q}{1-q}\right).
\eea
Applying the saddle point approximation as explained above for the
function
\be{eq_partb2}
  \tilde{f}(q)=\frac{q}{1-q}-(L+1)\log q,
\ee
we get for the number of solutions of~(\ref{eq_part}) the estimate
\be{eq_partb3}
  \widetilde{\cN}(L) \simeq \exp(2\sqrt{L}).
\ee
Comparing this result with ~(\ref{eq_sad4}) shows that we get the correct
exponential growth, but the coefficient is too small by a factor
\be{eq_partb4}
  \frac{\log\cN}{\log\widetilde{\cN}}=\frac{\pi}{\sqrt{6}}\simeq 1.28.
\ee

\fig{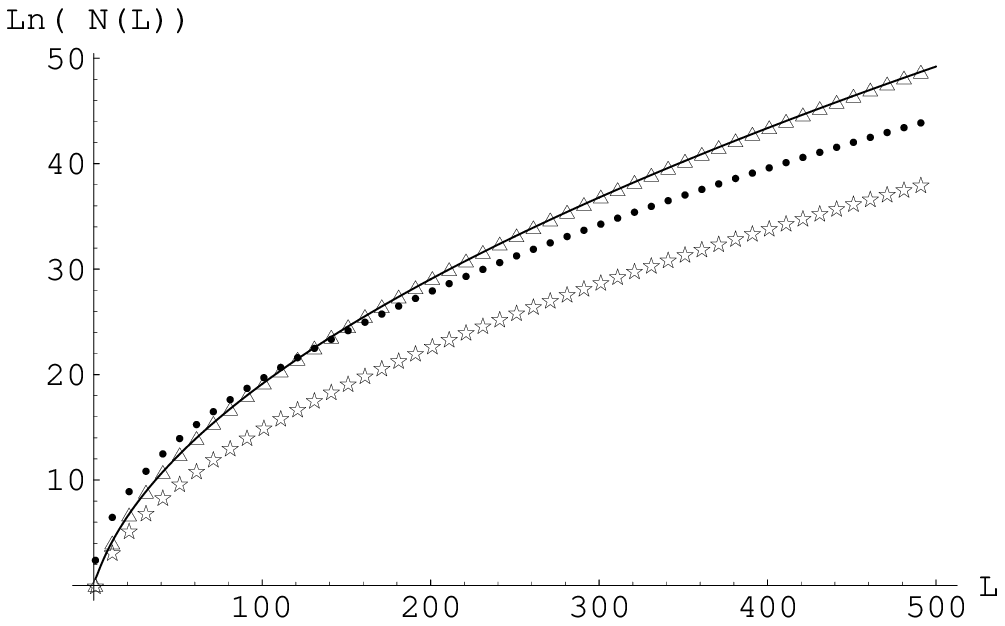}{fig_partcomp2}{%
Comparing the results for the number of partitions of $L$. The solid line is
the exact result, the dotted line is the saddle point approximation to leading
order. The stars and triangles show the next-to-leading order result, without
and including the additional analytic factor 1.28, respectively.%
}

In figure~\ref{fig_partcomp2} we compare the results for the leading and
next-to-leading order results of the computation above with the exact
result. As already expected, the value for the second order approximation
is too small, since our suppression factor $k!$ is too big.
Nevertheless, qualitatively the results are correct.
Since we are not aiming at exact results, but rather
at an approximative method to get an idea of the frequency distributions
of properties of the models under consideration, this is not a big problem.

Let us finally come back to the full tadpole equation~(\ref{eq_tad8}).
It can be treated in the same way as the pure partition problem and
analogous to~(\ref{eq_partb}) we can write
\bea{eq_partc}\nonumber
  \cN(L) & \simeq & \pii\oint dq \frac{1}{q^{L+1}}
     \sum_{k=1}^\infty\frac{1}{k!}\,\prod_{i=1}^k\left(
     \sum_{N_i=1}^\infty \sum_{X_i=1}^L q^{\sum_a \, N_aX_a} \right)\\
  & = & \pii \oint dq \frac{1}{q^{L+1}} \sum_{k=1}^\infty \frac{1}{k!}\,
     \left(\sum_{X=1}^L  \frac{q^X}{1-q^X}\right)^k,
\eea
such that we obtain for $f$
\be{eq_partc2}
  f(q)=\sum_{X=1}^L\frac{q^X}{1-q^X}-(L+1)\log q.
\ee
Close to $q\simeq 1$ we can approximate this to
\be{eq_partc3}
  f(q)\,\simeq\,\frac{1}{1-q}\sum_{X=1}^L\frac{1}{X}-L\log q
    \,\simeq\,\frac{\log L}{1-q}-L\log q.
\ee
The minimum can then be found at $q_0 \simeq 1-\sqrt{\frac{\log L}{L}}$,
which gives for the number of solutions
\be{eq_partc4}
  \cN(L)\simeq\exp(2\sqrt{L\log L}).
\ee
The additional factor of $\log L$ in the scaling behaviour compared
to~(\ref{eq_partb3}) can be explained by a result from number theory.
It is known that the function $\sigma_0(n)$, counting number of divisors
of an integer $n$, has the property
\be{eq_divsc}
  \frac{1}{L}\sum_{n=1}^L\sigma_0(n)\simeq\log L+(2\gamma_E-1),
\ee
where $\gamma_E$ is the Euler-Mascheroni constant.

\fig{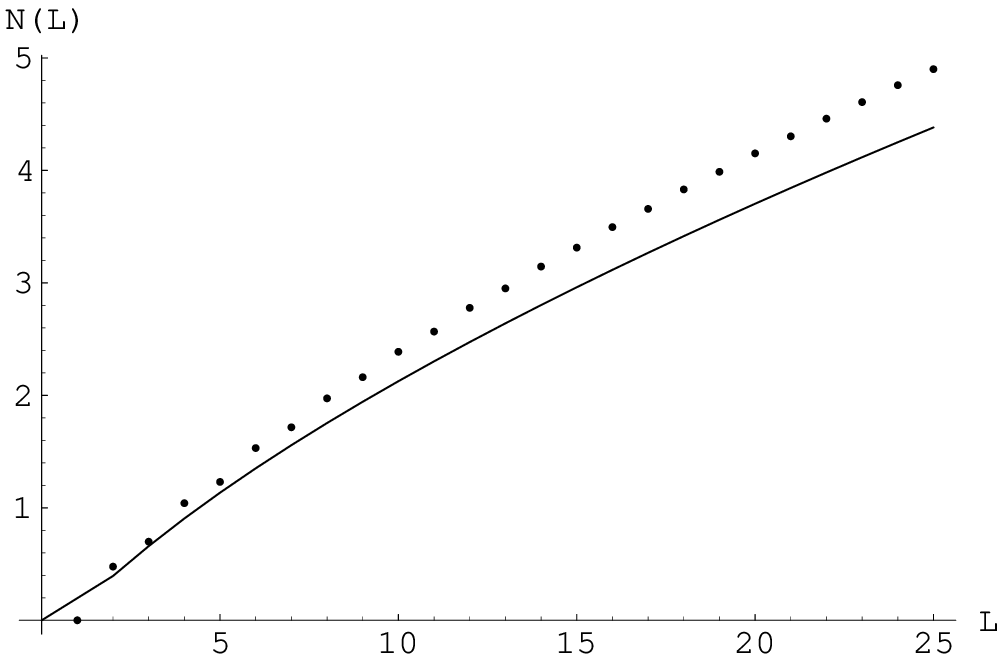}{fig_8dcount}{Logarithmic plot of the number of solutions to
the supersymmetry and tadpole equations for compactifications on $T^2$. The
dotted line shows the exact results, the solid line is the result of a
next-to-leading order saddle point approximation.%
}

Let us compare the result~(\ref{eq_partc4}) with the exact number of solutions,
obtained with a brute force computer analysis. This is shown in
figure~\ref{fig_8dcount}.
As expected from the discussion above, the estimate using the saddle point
approximation is too small, but it has the correct scaling behaviour and
should therefore be suitable to qualitatively analyse the properties of the
solutions.

We can use the saddle point approximation method introduced above to
analyse several properties of the gauge sector of the models. To show how
this works, we present two examples in the simple eight-dimensional
case, before applying these methods in section~\ref{ls_6d} to models on
$T^4/\bZ_2$.

One interesting observable is the probability to find an $SU(M)$ gauge factor
in the total set of models. Using the same reasoning as in the computation
of the number of models this is given by
\bea{eq_sad_um}\nonumber
  P(M,L) &\simeq& \frac{1}{2\pi i\cN(L)}\oint dq\frac{1}{q^{L+1}}
    \sum_{k=1}^\infty\frac{1}{(k-1)!}\left(\sum_{X=1}^L
    \frac{q^X}{1-q^X}\right)^{k-1}\sum_{X=1}^L\sum_{N=1}^\infty q^{NX}
    \delta_{N,M}\\
  & = & \frac{1}{2\pi i\cN(L)}\oint dq \frac{1}{q^{L+1}}
    \exp\left(\sum_{X=1}^L\frac{q^X}{1-q^X}\right)q^M
    \frac{1-q^{ML}}{1-q^M}.
\eea
The saddle point function is therefore given by
\be{eq_sad_umf}
  f(q)=\sum_{X=1}^L\frac{q^X}{1-q^X}+\log\left(q^M\frac{1-q^{ML}}{1-q^M}\right)
    -(L+1)\log q.
\ee
A comparison between exact computer results and the saddle point approximation
to second order is shown in figure~\ref{fig_8dhrkum_a}.

\twofig{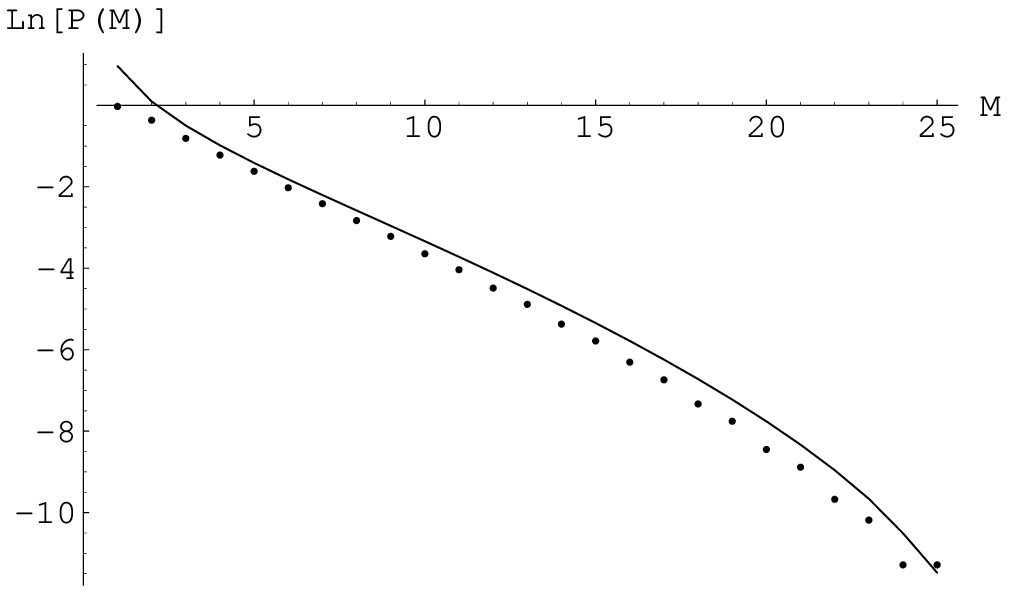}{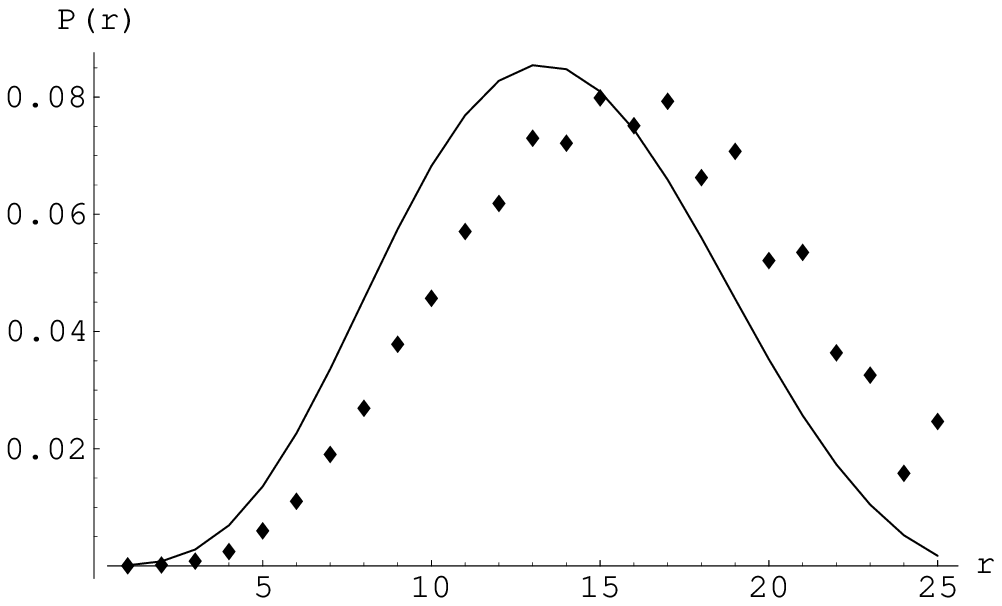}{fig_8dhrkum}{%
Distributions for compactifications on $T^2$. The solid lines are the exact
result, the dotted lines represent the second order saddle point approximation.
(a) Probability to find at least one $SU(M)$ gauge factor. (b) Frequency
distribution of the total rank.%
}

Another observable we are interested in is the distribution of the total rank
of the gauge group in our models. This amounts to including a constraint
\be{eq_sad_rkcon}
  \sum_{a=1}^\infty N_a=r,
\ee
that fixes the total rank to a specific value $r$. This constraint can be
accounted for by adding an additional
delta-function, represented by an additional contour integral to our formula.
We obtain
\bea{eq_sad_rk}\nonumber
  P(r,L) &\simeq& \frac{1}{2\pi i\cN(L)}\oint dq\frac{1}{q^{L+1}}
    \oint dz\frac{1}{z^{r+1}}\sum_{k=1}^\infty\frac{1}{k!}
    \prod_{i=1}^k\left(\sum_{N_i=1}^\infty\sum_{X_i=1}^L q^{\sum_a N_aX_a}
    z^{\sum_a N_a}\right)\\
  & = &  \frac{1}{2\pi i\cN(L)}\oint dq\frac{1}{q^{L+1}}
    \oint dz\frac{1}{z^{r+1}}\exp\left(\sum_{X=1}^L\frac{zq^X}{1-zq^X}\right),
\eea
with saddle point function
\be{eq_sad_rkf}
  f(q,z) =\sum_{X=1}^L\frac{zq^X}{1-zq^X}-(L+1)\log q -(r+1)\log z.
\ee
As we can see in figure~\ref{fig_8dhrkum_b}, where we also show the exact
computer result, we get a Gaussian distribution.

\subsubsection{Exact computations}\label{ls_comp}
Instead of using an approximative method, it is also possible to
directly calculate possible solutions to the constraining equations.
At least for models on $T^2$ or $T^4/\bZ_2$,
this is much more time-consuming than the saddle point approximation,
and, what is even more important, cannot be done completely for models on
$T^6/\bZZ$.
The reason why a complete classification is not possible has to do with
the fact that the problem to find solutions to the supersymmetry and tadpole
equations belongs to the class of NP-complete problems, an issue that we
elaborate on in section~\ref{ls_nosol}.
Despite these difficulties, it turns out to be necessary to use an explicit
calculation for four-dimensional compactifications, the ones we are most
interested in, since the saddle point method does not lead to reliable
results in that case.

In the eight-dimensional case the algorithmic solution to the tadpole
equation
\be{eq_t2_tad_text}
  \sum_a N_a\, X_a = L,
\ee
can be formulated as a two-step algorithm.
First calculate all possible unordered partitions of $L$,
then find all possible factorisations to obtain solutions for $X$ and $N$.
The task of partitioning is solved by the algorithm explained in
appendix~\ref{app_pa}, the factorisation can only be
handled by brute force. In this way we are not able to calculate solutions
up to very high values for $L$, but for our purposes, namely to check the
validity of the saddle point approximation (see section~\ref{ls_saddle2}),
the method is sufficient.

In the case of compactifications to six dimensions we can still use the same
method, although we now have to take care of two additional constraints.
First of all we exclude multiple wrapping, which gives
an additional constraint on the wrapping numbers $X_1, X_2, Y_1$ and $Y_2$,
defined in appendix~\ref{app_t6}. This constraint can be formulated in terms
of the greatest common divisors of the wrapping numbers -- we will come back
to this issue in section~\ref{ls_6d}. Another difference compared to the eight
dimensional case is that we have to take different values for the complex
structure parameters $U_1$ and $U_2$ (see appendix~\ref{app_t6} for a
definition) into account.
As it is shown in section~\ref{ls_nosol}, these are bounded
from above and we have to sum over all possible values, making sure that we
are not double counting solutions with wrapping numbers which allow for
different values of the complex structures.

In~(\ref{eq_t6_xydef}) the wrapping numbers $\vec{X}$ and
$\vec{Y}$ are defined as integer valued quantities
in order to implement the supersymmetry~(\ref{eq_t6_susy}) and
tadpole~(\ref{eq_t6_tad}) conditions in a fast computer algorithm.
From the equations we can derive the following inequalities
\be{eq_ineq}
  0 < \sum_{I=0}^3 X^I\, {U}_I \le \sum_{I=0}^3 L^I\, {U}_I.
\ee

The algorithm to find solutions to these equations and the additional
K-theory constraints~(\ref{eq_t6_ktheory}) consists of four steps.
\begin{enumerate}
\item First we choose a set of complex structure variables $U_I$. This is done
systematically and leads to a loop over all possible values.
Furthermore, we have to check for redundancies, which might exist because
of trivial symmetries under the exchange of two of the three two-tori.

\item In a second step we determine all possible values for the wrapping
numbers $X^I$ and
$Y^I$, using (\ref{eq_ineq}) for the given set of complex structures,
thereby obtaining all possible supersymmetric branes. In this step we also
take care of the multiple wrapping constraint, which can be formulated,
analogously to the six dimensional case, in terms of the greatest
common divisors of the wrapping numbers.

\item In the third and most time-consuming part, we use the tadpole
equations~(\ref{eq_t6_tad}), which after a summation can be written as
\be{eq_tadsum}
  \sum_{a=0}^k S_a=\Lambda \quad\mbox{with}\quad
  S_a:=\sum_{I=0}^3 N_a {U}_I X_a^I\quad\mbox{and}\quad
  \Lambda:=\sum_I L^I {U}_I.
\ee
To solve this equation, we note that all $S_a$ and $\Lambda$ are positive
definite integers, which allows us to use the partition algorithm to obtain
all possible combinations. The algorithm is improved by using only those
values for the elements of the partition which are in the list of values we
computed in the second step.
For a detailed description of the explicit algorithm we
used, see appendix~\ref{app_pa}.
Having obtained the possible $S_a$, we have to factorise them into values
for $N_a$ and $X_a^I$.

\item Since (\ref{eq_tadsum}) is only a necessary but no sufficient condition,
we have to check in the fourth and last step, if the obtained results indeed
satisfy all constraints, especially the individual tadpole cancellation
conditions and the restrictions from K-theory, which up to this
point have not been accounted for at all.
\end{enumerate}

The described algorithm has been implemented in C and was put on several
high-performance computer clusters, using a total CPU-time of about
$4\times 10^5$ hours. The solutions obtained in this way have been saved in a
database for later analysis.

The main problem of the algorithm described in the last section lies in the
fact that its complexity scales exponentially with the complex structure
parameters. Therefore we are not able to compute
up to arbitrarily high values for the $U_I$.
Although we tried our best, it may of course be possible to improve the
algorithm in many ways, but unfortunately the exponential behaviour cannot
be cured unless we might have access to a quantum computer. This
is due to the fact that the number of possib le solutions to the
Diophantine equations
we are considering grows exponentially with $U$.
In fact, this is quite a severe issue since the Diophantine structure of the
tadpole equations encountered here is not at all exceptional, but
very  generic for the topological constraints also
in other types of string  constructions.
The problem seems indeed to appear generically in computations of landscape
statistics, see~\cite{dedo06} for a general account on this issue.

As we outlined in the previous section, the computational effort to generate
the solutions to be analysed in the next section took a significant amount of
time, although we used several high-end computer clusters.
To estimate how many models could be computed in principle, using a computer
grid equipped with contemporary technology
in a reasonable amount of time,
the exponential behaviour of the problem has to be taken into account.
Let us be optimistic and imagine that we would have a total number of
$10^5$ processors at our disposal which are
twice as fast as
the ones we have been using.
Expanding our analysis to cover a range of complex structures which is
twice as large as the one we considered  would,
in a very rough estimate, still take us of the order of 500 years.

Note that in principle there can be a big difference in the estimated computing
time for the two computational problems of finding all string vacua in a
certain class on the one hand, and of looking for configurations with
special properties, that lead to additional constraints, on the other hand.
As we explore in section~\ref{ls_su53stack} the computing time can be
significantly reduced if we restrict ourselves to a maximum number of stacks
in the hidden sector and take only configurations of a specific visible
sector into account (in the example we consider we look for grand unified
models with an $SU(5)$ gauge group). Nevertheless, although a much larger
range of complex structures can be covered,
the scaling of the algorithm remains unchanged. This means in particular that
a cutoff on the $U_I$, even though it might be at higher values, has to be
imposed.

\subsection{Finiteness of solutions}\label{ls_nosol}
It is an important question whether or not the number of solutions is
infinite. Making statistical statements about an infinite set of models is
much more difficult than to deal with a finite sample, because we would have to
rely on properties that reoccur at certain intervals, in order to be able
to make any
valuable statements at all. If instead the number of solutions is finite, and
we can be sure that the solutions we found form a representative sample,
it is possible to draw conclusions by analysing the frequency distributions
of properties without worrying about their pattern of occurrence within the
space of solutions.

In the case of compactifications to eight dimensions, the results are
clearly finite, as can be seen directly from the fact that the variables
$X$ and $N$ have to be positive and $L$ has a fixed value. Note however,
although such an eight-dimensional model is clearly not realistic, that
the complex structures are unconstrained. This means that if we do not
invoke additional methods to fix their values, each solution to the tadpole
equation represents in fact an infinite family of solutions.

\subsubsection{The six-dimensional case}
In the six dimensional case, the finiteness of the number of solutions
is not so obvious, but it can be rigorously proven.
In order to do so, we have to show that possible
values for the complex structure parameters $U_1$ and $U_2$ are bounded from
above. If this were not be the case, we could immediately deduce from
equations~(\ref{eq_t6_susy}) that infinitely many brane configurations would
be possible.

In contrast to the eight-dimensional toy-model that we explored in
section~\ref{ls_saddle2}, in this case, and also for the four-dimensional
compactifications, we do not want to allow branes that wrap the torus
several times. To exclude this, we can derive the following condition
on the wrapping numbers (for details see appendix~\ref{app_t4_mw}),
\be{eq_t4_mw_maintext}
  \gcd(X^1,Y^2)\gcd(X^2,Y^2)=Y^2.
\ee
This condition implies that all $\vec{X}$ and $\vec{Y}$ are non-vanishing.
Additional branes, which wrap the same cycles as the orientifold planes,
are given by $\vec{X}\in\left\{(1,0),(0,1)\right\}$, with $\vec{Y}=\vec{0}$
in both cases.

From~(\ref{eq_t6_susy}) we conclude that all non-trivial solutions have to
obey $U_1/U_2\in\bQ$. Therefore we can restrict ourselves to coprime values
\be{eq_t4_u1u2def}
  (u_1,u_2) \quad \mbox{with} \quad u_i:=\frac{U_i}{\gcd(U_1,U_2)}.
\ee
With these variables we find from the supersymmetry conditions that
$Y^1 = u_2\,\alpha$, for some $\alpha\in\bZ$. Now we can use the
relation~(\ref{eq_t4_XYrel}) to get
\be{eq_t4_finite}
  X^1\,X^2 = u_1\,u_2\,\alpha^2.
\ee

In total we get two classes of possible branes,
those where $X^1$ and $X^2$ are both
positive and those where one of them is $0$. The latter are those where the
branes lie on top of the orientifold planes.

For fixed values of $u_1$ and $u_2$ the tadpole cancellation
conditions~(\ref{eq_t4_tad})
admit only a finite set of solutions. Since all quantities in these
equations are positive, we can furthermore deduce
from~(\ref{eq_t4_finite}) that solutions which contain at least one brane
with $X^1,X^2 > 0$ are only possible if the complex
structures satisfy the bound
\be{eq_t4_csbound}
  u_1u_2 \le L_1 L_2.
\ee
In figure~\ref{fig_t4_uvalues} we show the allowed values for $u_1$ and $u_2$
that satisfy equation~(\ref{eq_t4_csbound}).

\fig{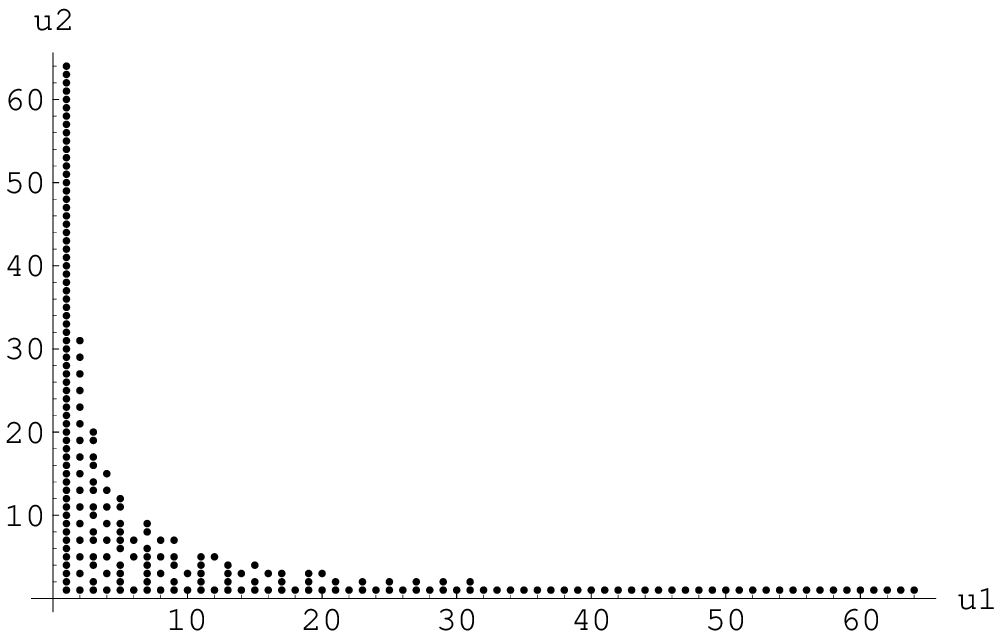}{fig_t4_uvalues}{%
Allowed values for the complex structure parameters $u_1$ and $u_2$ for
compactifications to six dimensions.%
}

In the case that only branes with one of the $X^i$ vanishing are present in
our model, the complex structures are not bounded from above, but since 
there exist only two such branes in the case of coprime wrapping numbers,
all solutions of this type are already contained in the set of solutions
which satisfy~(\ref{eq_t4_csbound}).
Therefore we can conclude that the overall number of solutions to the
constraining equations in the case of compactifications to six dimensions
is finite\footnote{Note however, that in the case
where all branes lie on top of the orientifold planes, we are in an
analogous situation for the eight-dimensional compactifications.
Unless we invoke additional methods of moduli stabilisation, the complex
structure moduli represent flat directions and we get infinite
families of solutions.}.

\subsubsection{Compactifications to four dimensions}\label{ls_fin_4d}
The four dimensional case is very similar to the six-dimensional one discussed
above, but some new phenomena appear. In particular, we see that the
wrapping numbers can have negative values, which is the crucial point that
prevents us from proving the finiteness of solutions.
Although we were not able to obtain an analytic proof,
we present some arguments and numerical results, which provide evidence
and make it very
plausible that the number of solutions is indeed finite\footnote{An analytic
proof of this statement has recently been found~\cite{dota06}.}.

As in the $T^4/\bZ_2$ case, we can derive a condition on the (rescaled)
wrapping numbers
$\vec{X}$ and $\vec{Y}$, defined by~(\ref{eq_t6_xytilderel}),
to exclude multiple wrapping. The derivation is
given in appendix~\ref{app_t6_mw} and the result is\footnote{We have to use
rescaled wrapping numbers, as defined by~(\ref{eq_t6_xytilderel}), to write
the solution in this simple form.}
\be{eq_t6_mw_maintext}
  \prod_{i=1}^3\gcd(\tilde{Y}^0,\tilde{X}^i)=(\tilde{Y}^0)^2.
\ee

From the relations~(\ref{eq_t6_xyrel}), it follows that either one, two or all
four $X^I$ can be non-vanishing. The case with only one of them vanishing is
excluded.
Let us consider the three possibilities in turn and see what we can say
about the number of possible solutions in each case.

\begin{enumerate}
\item In the case that only one of the $X^I\neq 0$, the corresponding brane
lies on top of one of the orientifold planes on all three $T^2$. This situation
is equivalent to the eight-dimensional case and can be included in the
discussion of the next possibility.
\item If two $X^I\neq 0$, we are in the situation discussed for the
compactification to six dimensions. The two $X^I$ have to be positive by means
of the supersymmetry condition and one of the complex structures is fixed at
a rational number. Together with the eight-dimensional branes, the same proof
of finiteness we have given for the $T^4/\bZ_2$-case can be applied.
\item A new situation arises for those branes where all $X^I\neq 0$. Let us discuss this a bit more in detail.
\end{enumerate}
From the relations~(\ref{eq_t6_xyrel}) we deduce that an odd number of them has
to be negative. In the case that three would be negative and one positive --
let us without loss of generality choose $X_0>0$ -- we can write the
supersymmetry condition~(\ref{eq_t6_susy}) as
\be{eq_t6_xn1}
  \sum_{I=0}^3 Y^I \frac{1}{U_I}=\frac{Y^0}{U_0}\left( 1+\sum_{i=1}^3
    \frac{X^0\,U_0}{X^i\,U_i}\right) = 0,
\ee
which implies $X^i\, U_i < -X^0\, U_0\,\,\forall\,i\in\{1,2,3\}$.
This contradicts the second supersymmetry condition,
\be{eq_t6_xn2}
  X^0\, U_0\left( 1+\sum_{i=1}^3 \frac{X^i\, U_i}{X^0\, U_0}\right)>0.
\ee
Therefore, we conclude that the only remaining possibility is to have one
of the $X^I < 0$. Again we choose $X^0$ without loss of generality.
We can now use~(\ref{eq_t6_xn1}) to express $X^0$ in terms of the other three
wrapping numbers as
\be{eq_t6_xn3}
  X^0=-\left(\sum_i \frac{U_0}{U_i\, X^i}\right)^{-1}.
\ee
Furthermore, we can use the inequality~(\ref{eq_ineq}) and derive an upper bound
\be{eq_t6_xn4}
  \sum_{I=0}^3 L_I\, U_I \,\ge\, X^0\ U_0 +\sum_{i=1}^3 X^i\, U_i
    \,>\, X^j\, U_j \,>\, 0, \quad\forall\,j\in\{1,2,3\}.
\ee

As in the six-dimensional case, we can use the argument that the
complex structures are fixed at rational values,
as long as we take a sufficient number of branes.
So we can write them, in analogy
to~(\ref{eq_t4_u1u2def}) as $u_{I,2}/u_{I,1}$. Using this definition, we can
write~(\ref{eq_t6_xn4}) as
\be{eq_t6_xn5}
  1 \le X_i \le \frac{\sum_{P=0}^3   u_{P,2}u_{Q,1}u_{R,1}u_{S,1}L_P}
    {u_{i,2}u_{J,1}u_{K,1}u_{L,1}},
\ee
for $P\ne Q\ne R\ne S\ne P$ and $i\ne J\ne K\ne L\ne i$.

From this we conclude that as long as the complex structures are fixed,
we have only a finite number of possible brane configurations, i.e.
only a finite number of solutions.
This is unfortunately not enough to conclude that we have only a finite number
of solutions in general. We would have to show, as in the six-dimensional case,
that there exists an upper bound on the complex structures. Since we were not
able to find an analytic proof that such a bound exists, we have to rely on
some numerical hints that it is in fact the case. We present some of these
hints in the following.

\fig{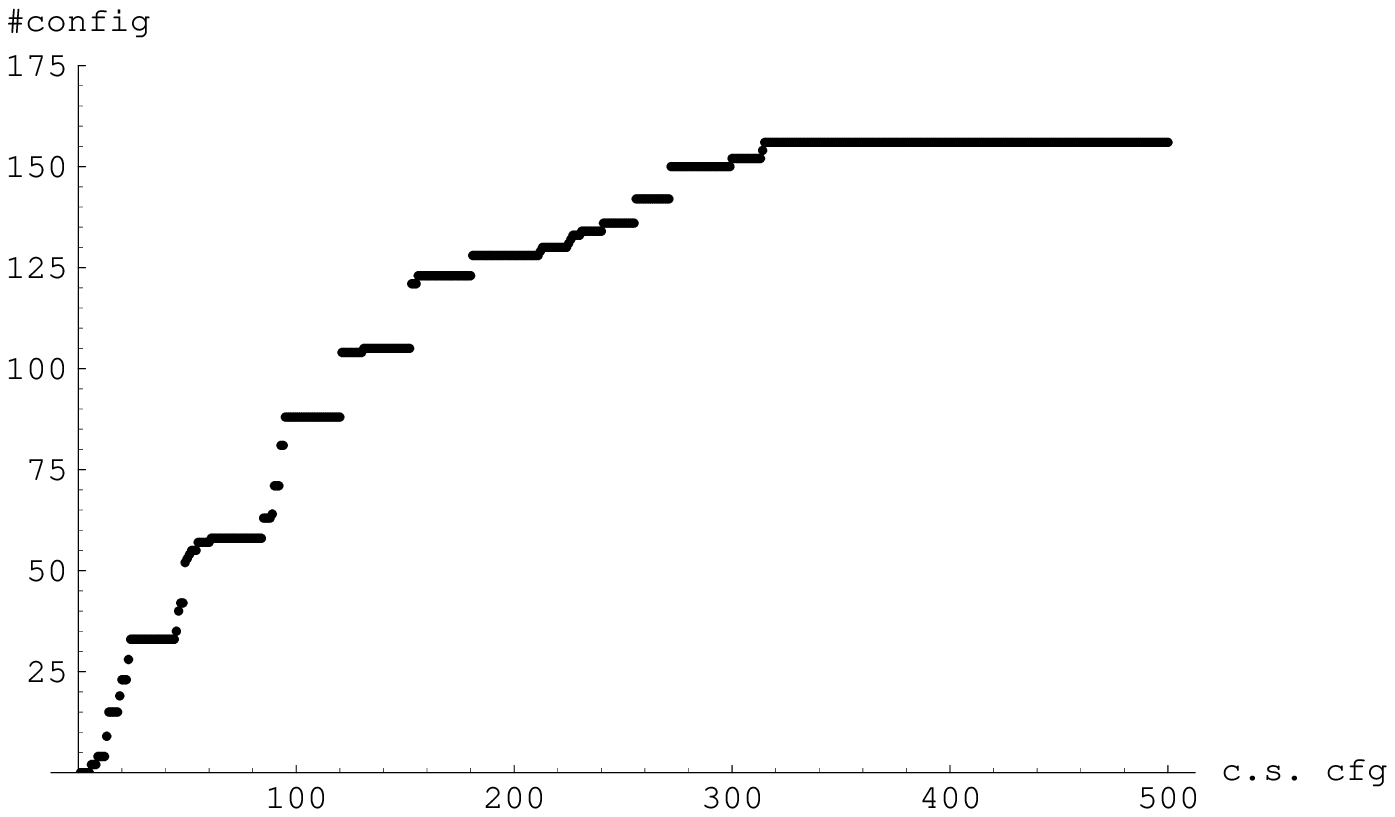}{fig_numsol_a}{%
The number of unique solutions for compactifications on $T^6/\bZZ$,
taking $L_I=2$ $\forall I\in\{0,\ldots,3\}$.
The horizontal axis shows combinations of the $U_I$, ordered by their absolute
value $|U|$. For each of these values we plotted the cumulative set of
solutions obtained up to this point.%
}

Figure~\ref{fig_numsol_a} shows how the total number of mutually
different brane
configurations for $L=2$ increases and saturates,
as we include more and more combinations
of values for the complex structures $U_I$ into the set for which we construct solutions.
For this small value of $L$ our algorithm actually admits pushing the
computations up to
those complex structures  where obviously no additional brane
solutions exist.

\fig{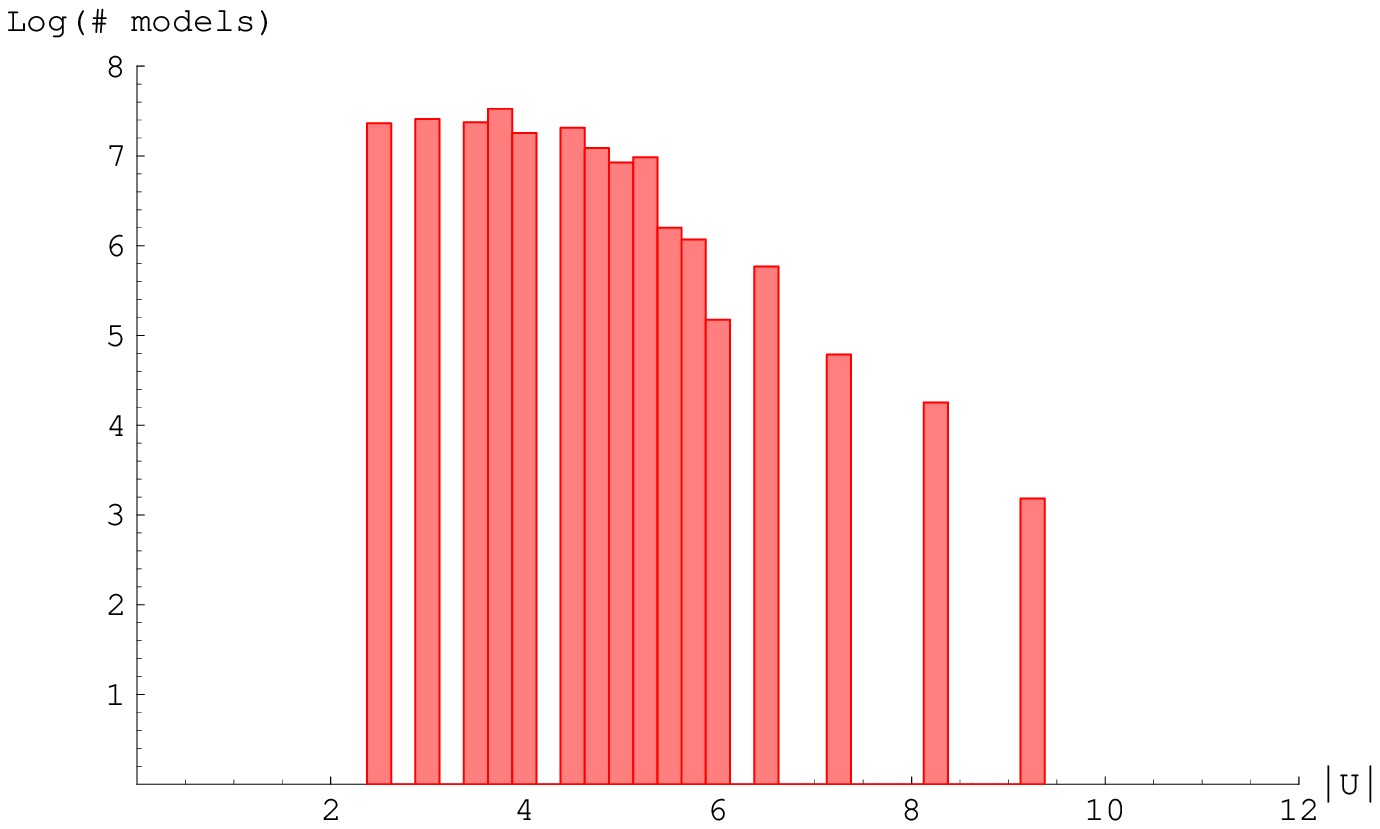}{fig_numsol_b}{%
Logarithmic plot of the absolute number of solutions for compactifications
on $T^6/\bZZ$ using
the physical values $L_I=8$ $\forall I\in\{0,\ldots,3\}$ against the
absolute value $|U|$. The cutoff is set at $|U|=12$. In this plot, as in all
other plots of this paper, we use a decadic logarithm.%
}

For the physically relevant case of $L=8$ the total number of models compared
to the absolute value  $|U|$  of the complex structure variables scales
as displayed
in figure~\ref{fig_numsol_b}.
The plot shows all complex structures we have actually been able to analyse
systematically.
We find that the number of solutions falls
logarithmically for increasing values of $|U|$.
In order to interpret this result, we observe that the complex structure
moduli $U_I$ are only defined up to 
an overall rescaling by the volume modulus of the compact space.
We have chosen all radii and thereby also all $U_I$ to be integer valued,
which means that large $|U|$ correspond to large coprime values of
$R_1^{(i)}$ and $R_2^{(i)}$. This comprises on the 
one hand decompactification limits which have to  be discarded in any case
for phenomenological reasons, 
but on the other hand also tori which are slightly distorted, e.g. almost
square tori with $R_2^{(i)}/R_1^{(i)}=0.99$.

%
%
\section{Statistical analysis of orientifold models}\label{ch_stat}
After preparing the stage in the last section, introducing the models 
and methods of analysis, we are now going to analyse some specific
constructions of phenomenological interest.
At the end of this section we want to arrive at a point where we can make
some meaningful statistical statements about the probability to find
realisations of the standard model or GUT models in the
specific set of models we are considering.

However, it is important to mention, that our results cannot be regarded
to be complete. First of all we neglect the impact of fluxes, which
does not change the distributions completely, but definitely has
some influence.
Secondly, we are considering only very specific geometries.
Since the construction of the orientifolds, especially the choice of the
orbifold group which in our case is always $\bZ_2$, has a strong impact on
the constraining equations, it is very probable that the results
change significantly once we use a different compactification space.
Nevertheless we think that these results are one step towards a
deeper understanding of open string statistics.

In the first part of this section we discuss some general aspects
of compactifications to six and four dimensions. We analyse the properties
of the gauge groups, including the occurrence of specific individual gauge
factors and the total rank. With respect to the chiral matter content, we
establish the notion of a mean chirality and discuss their frequency
distribution.

In a second part we perform a search for models with the properties of a
supersymmetric standard model. Besides the frequency distributions in the
gauge sector we analyse the values of the gauge couplings and compare our
results to those of a recent statistical analysis of Gepner
models~\cite{dhs041,dhs042}. In addition to standard model gauge groups we
look also for models with a Pati-Salam, $SU(5)$ and flipped $SU(5)$ structure.

In the last part we consider different aspects of the question of correlations
of observables in the gauge sector and give an estimate how likely
it is to find a three generation standard model in our setup.

\subsection{Statistics of six-dimensional models}\label{ls_6d}
Before considering the statistics of realistic four-dimensional models,
let us start with a simpler construction to test the methods of analysis
developed in section~\ref{ch_landscape}.
We will use a compactification to six dimensions on
a $T^4/\bZ_2$ orientifold, defined in appendix~\ref{app_t4}.
The important question about the finiteness of solutions has been settled
in section~\ref{ls_nosol}, so we can be confident that the results we
obtain will be meaningful.
To use the saddle point approximation in this context, we have to generalise
from the eight-dimensional example in~\ref{ls_saddle2} to an
approximation in several variables, as described by
equations~(\ref{eq_partextint}) and~(\ref{eq_partextf}). In our case we will
have to deal with two variables $\vec{q}=(q_1,q_2)$, corresponding to the
two wrapping numbers $\vec{X}=(X^1,X^2)$.

\fig{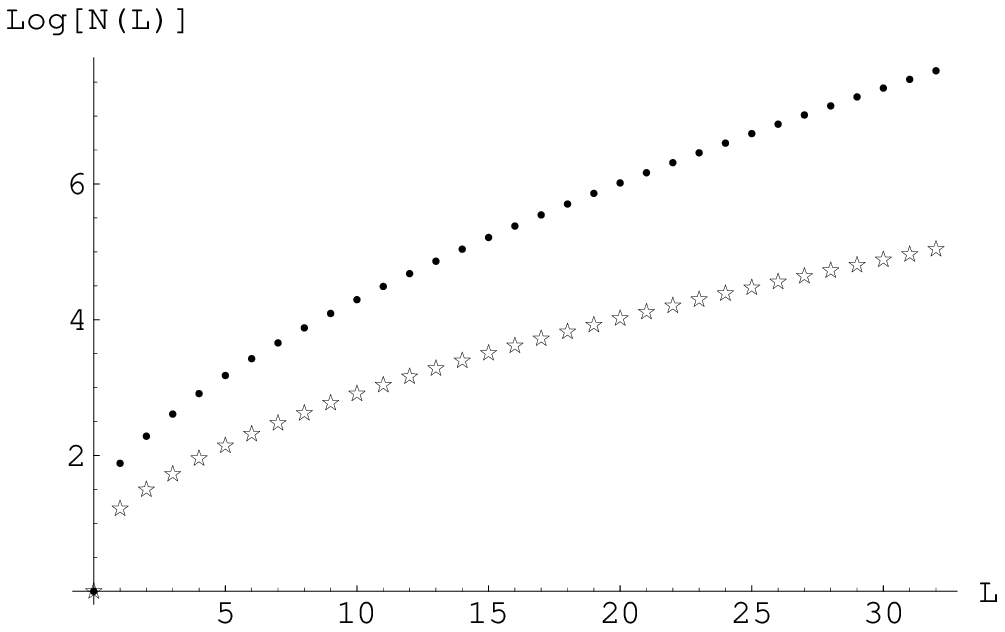}{fig_t4_numsol}{%
Logarithmic plot of the number of solutions for compactifications on
$T^4/\bZ_2$ for $L^2=8$ and
different values of $L\equiv L^1$. The complex structures are fixed to
$u_1=u_2=1$. The dotted line shows the result with multiple wrapping, the
stared line gives the result with coprime wrapping numbers.%
}

Let us briefly consider the question of multiple wrapping.
As shown in appendix~\ref{app_t4_mw}, we can derive a constraint on the
wrapping numbers $\vec{X}$ and $\vec{Y}$, such that multiply wrapping branes
are excluded. To figure out what impact this additional constraint has
on the distributions, let us compare the number of solutions for different
values of $L^1$ and $L^2$, with and without multiple wrapping. The result is
shown in figure~\ref{fig_t4_numsol}. As could have been expected, the number
of solutions with coprime wrapping numbers grows less fast then the one where
multiple wrapping is allowed.

\subsubsection{Distributions of gauge group observables}\label{ls_4d_rkun}
Using the saddle point method, introduced in section~\ref{ls_saddle},
we can evaluate the distributions for individual
gauge group factors and total rank of the gauge group in analogy to the
simple eight-dimensional example we pursued in section~\ref{ls_saddle2}.
Therefore we will fix the orientifold charges to their physical values,
$\vec{L}=(L_1,L_2)=(8,8)$.
The probability to find one $U(M)$ gauge factor can be written similar
to~(\ref{eq_sad_um}) as
\bea{eq_t4_um}\nonumber
P(M,\vec{L}) &\simeq& \frac{1}{\cN(\vec{L})(2\pi i)^2}
  \oint d\vec{q}\,\exp\Biggl[\sum_{\vec{X}\in S_U}\frac{q_1^{X_1}\, q_2^{X_2}}
  {1- q_1^{X_1}\, q_2^{X_2}}
  + \log\left( \sum_{\vec{X}\in S_U}
  q_1^{M X_1} q_2^{M X_2} \right)\\
&&    - (L_1\!+\!1)\log q_1 - (L_2\!+\!1)\log q_2\Biggl],
\eea
where we denoted with $S_U$ the set of all values for $\vec{X}$
that are
compatible with the supersymmetry conditions and the constraints on multiple
wrapping.
The number of solution $\cN(\vec{L})$ is given by
\be{eq_t4_ns}
\cN(\vec{L}) \simeq \frac{q}{(2\pi i)^2}\oint d\vec{q}\,
     \exp\Biggl[\sum_{\vec{X}\in S_U} \frac{q_1^{X_1}\,q_2^{X_2}}
     {1- q_1^{X_1}\, q_2^{X_2}}
  -(L_1\!+\!1)\log q_1 - (L_2\!+\!1)\log q_2\Biggl].
\ee

The resulting distribution for the probability of an $U(M)$ factor,
compared to the results of an exact computer
search, is shown in figure~\ref{fig_t4_rkum_a}.

\twofig{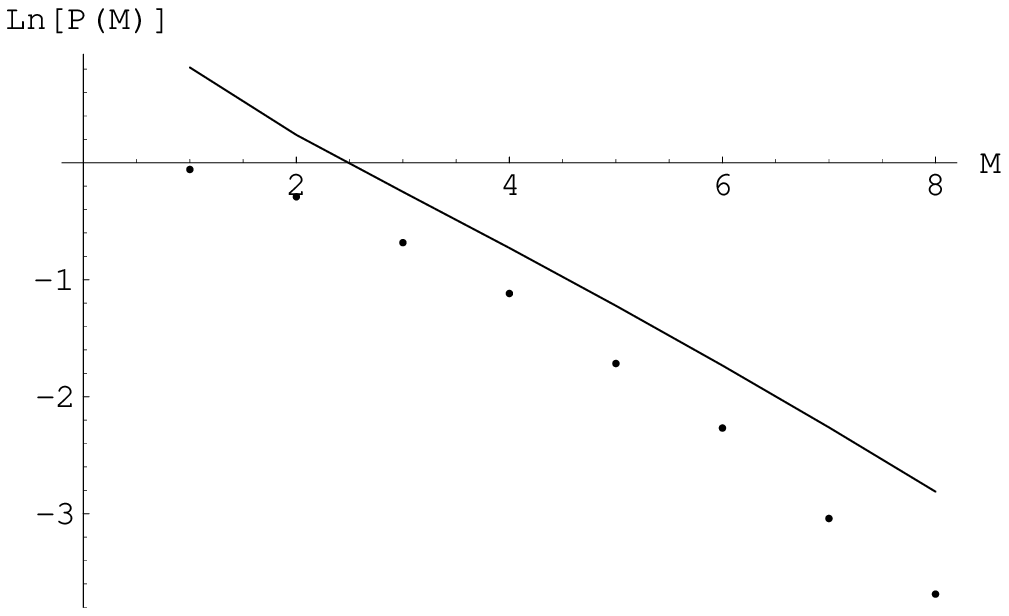}{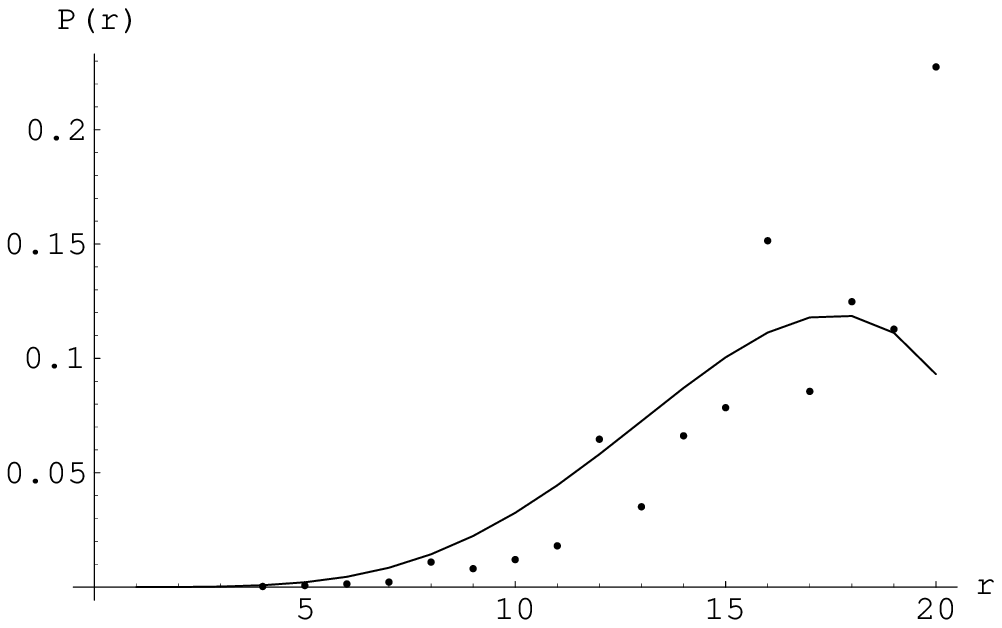}{fig_t4_rkum}{%
Distributions in the gauge sector of a compactification on $T^4/\bZ_2$. The
complex structures are fixed to $u_1=u_2=1$. The dotted line is the result of
an exact computation, the solid line shows the saddle point approximation to
second order. (a) Probability to find an $(U(M)$ gauge factor,
(b) Distribution of the total rank of the gauge group.%
}

As in the eight-dimensional example we can
evaluate the distribution of the total rank~(\ref{eq_sad_rkcon}).
As a generalisation of~(\ref{eq_sad_rk}) we obtain the following formula
\bea{eq_t4_rk}\nonumber
  P(r,\vec{L}) &\simeq& \frac{1}{\cN(\vec{L})(2\pi i)^3}\oint d\vec{q}dz\,
    \exp\Biggl[\sum_{\vec{X}\in S_U}\frac{zq_1^{X_1}q_2^{X_2}}
    {1-zq_1^{X_1}q_2^{X_2}}\\
  &&  -(L_1\!+\!1)\log q_1-(L_2\!+\!1)\log q_2
    -(r\!+\!1)\log z\Biggl].
\eea
Figure~\ref{fig_t4_rkum_b} shows the resulting distribution of the total
rank, compared to the exact result. As one can see, the results of the saddle
point analysis are much smoother then the exact results, which show a more
jumping behaviour, resulting from number theoretical effects. These are
strong at low $L$, which is also the reason that our saddle point approximation
is not very accurate. In the present six-dimensional case the deviations are
not too strong, but in the four-dimensional case their impact
is so big that the result cannot be trusted anymore. These problems can be
traced back to the small values of $L$ we are working with, but since these
are the physical values for the orientifold charge, we cannot do much about it.

\subsubsection{Chirality}\label{ls_6dchi}
Since we are ultimately interested in calculating distributions for models
with gauge groups and matter content close to the standard model, it would be
interesting to have a measure for the mean chirality of the matter content
in our models.

A good quantity to consider for this purpose would be the distribution of
intersection numbers $I_{ab}$ between different stacks of branes. This is
precisely the quantity we choose later in the four-dimensional
compactifications. In the present case we use
a simpler definition for chirality, given by
\be{eq_t4_chidef}
  \chi := X^1X^2.
\ee
This quantity counts the net number of chiral fermions in the antisymmetric
and symmetric representations.

Using the saddle point method, we can compute the distribution of values for
$\chi$, using
\bea{eq_t4_chi}\nonumber
  P(\chi,\vec{L}) &\simeq& \frac{1}{\cN(\vec{L})(2\pi i)^2}\oint d\vec{q}\,
    \exp\Biggl[\sum_{\vec{X}\in S_U}\frac{q_1^{X_1}q_2^{X_2}}
    {1-q_1^{X_1}q_2^{X_2}}
    -\log\left(\sum_{\vec{X}\in S_U}
    \frac{q_1^{X_1}q_2^{X_2}}{1-q_1^{X_1}q_2^{X_2}}\right)\\
  && +\log\left(\sum_{\vec{X}\in S_{U,\chi}}
    \frac{q_1^{X_1}q_2^{X_2}}{1-q_1^{X_1}q_2^{X_2}}\right)
    -(L_1\!+\!1)\log q_1-(L_2\!+\!1)\log q_2\Biggl],
\eea
where $S_{U,\chi}\subset S_U$ is the set of wrapping numbers that
fulfills~(\ref{eq_t4_chidef}).

\fig{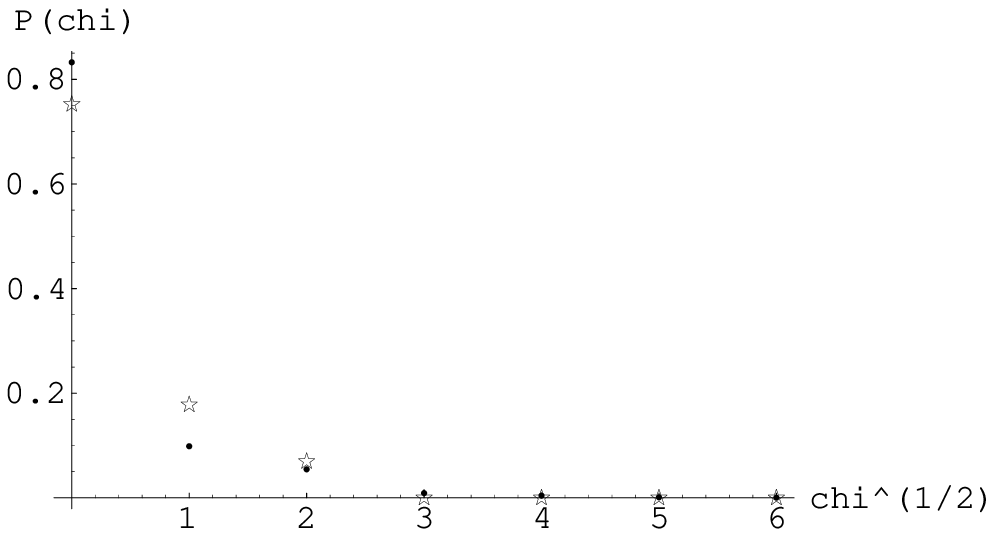}{fig_t4_chi}{%
Distribution of the mean chirality for $T^4/\bZ_2$, $L^1=L^2=8$, $u_1=u_2=1$.%
}

The resulting distribution is shown in figure~\ref{fig_t4_chi}. For the used
values of $u_1,u_2=1$, $\chi$ has to be a square, which can be directly
deduced from the supersymmetry conditions~(\ref{eq_t4_susy}). The scaling
turns out to be roughly $P(\chi)\simeq\exp(-c\sqrt{\chi})$.
From this result we can conclude that non-chiral models are exponentially
more frequent than chiral ones. This turns out to be a general property of
the orientifold models that also holds in the four-dimensional case.

\subsubsection{Correlations}\label{ls_6dcorr}
In this section we would like to address the question of correlations between
observables for the first time. We come back to this issue in
section~\ref{ls_corr}.
The existence of such correlations can be seen in figure~\ref{fig_t4_corr},
where we plotted the distributions of models with specific total rank and
chirality. The connection between both variables is given by the tadpole
cancellation conditions, which involve the $N_a$ used for the definition of
the total rank in~(\ref{eq_sad_rkcon}) and the wrapping numbers $\vec{X}_a$,
which appear in the definition of the mean chirality $\chi$
in~(\ref{eq_t4_chidef}).
The distribution can be obtained from
\bea{t4_chivsrk}\nonumber
  P(\chi,r,\vec{L})&\simeq& \frac{1}{\cN(\vec{L})(2\pi i)^3}\oint d\vec{q}dz\,
    \exp\Biggl[\sum_{\vec{X}\in S_U}\frac{zq_1^{X_1}q_2^{X_2}}
    {1-zq_1^{X_1}q_2^{X_2}}
    -\log\left(\sum_{\vec{X}\in S_U}
    \frac{zq_1^{X_1}q_2^{X_2}}{1-zq_1^{X_1}q_2^{X_2}}\right)\\\nonumber
  &&+\log\left(\sum_{\vec{X}\in S_{U,\chi}}
    \frac{zq_1^{X_1}q_2^{X_2}}{1-zq_1^{X_1}q_2^{X_2}}\right)
    -(L_1\!+\!1)\log q_1-(L_2\!+\!1)\log q_2\\
  &&-(r\!+\!1)\log z\Biggl],
\eea
which is a straightforward combination of~(\ref{eq_t4_rk})
and~(\ref{eq_t4_chi}).

\twofig{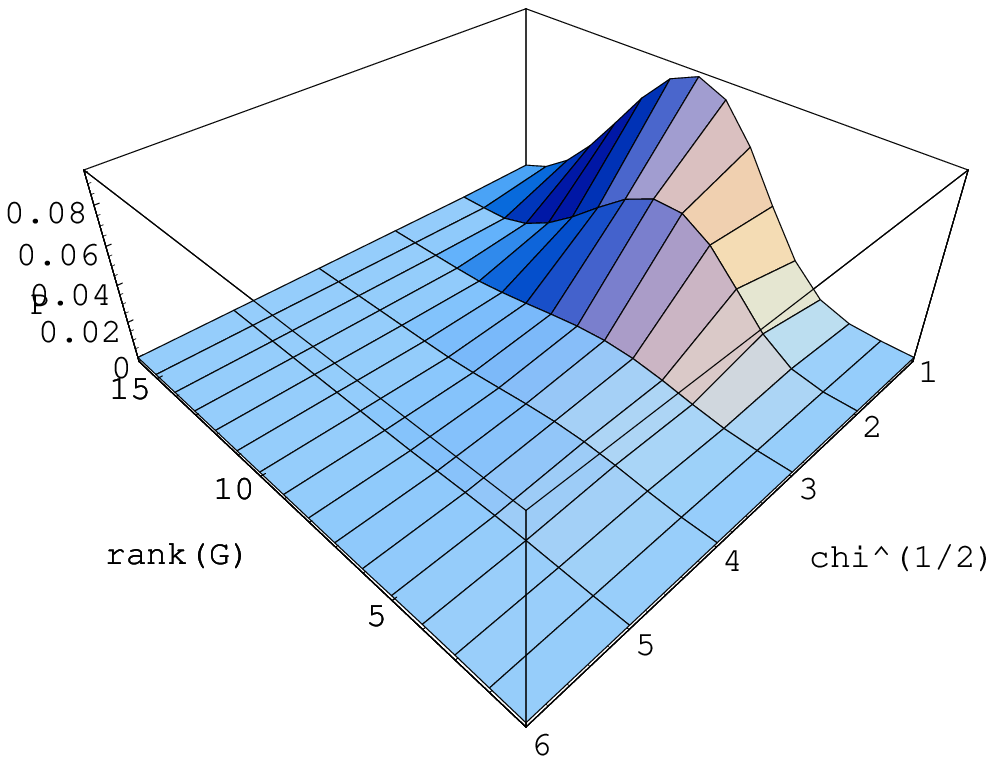}{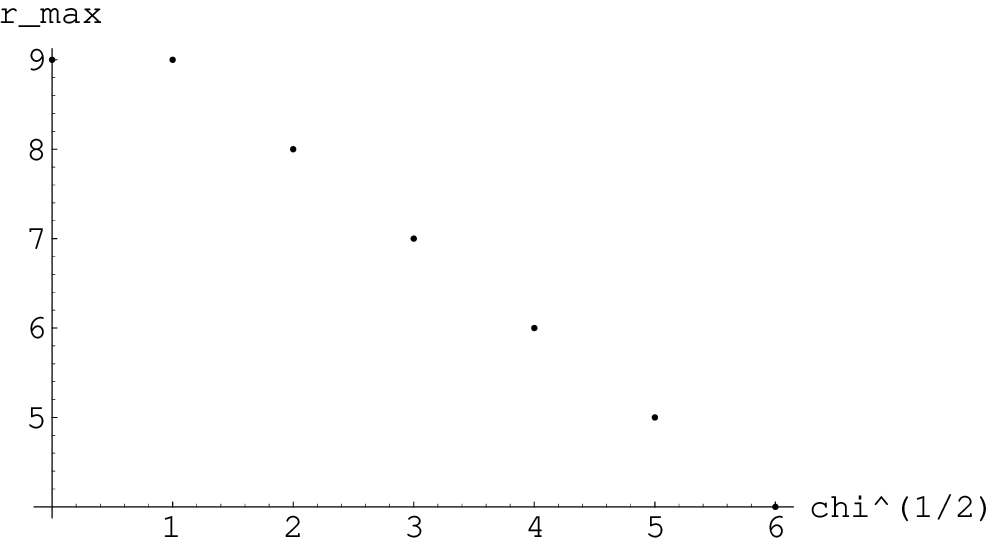}{fig_t4_corr}{%
Correlation between total rank and chirality for $L_1=L_2=8$ and $u_1=u_2=1$
for a compactification on $T^4/\bZ_2$. (b) shows the maximum of the total
rank distribution depending on $\chi$.%
}

In figure~\ref{fig_t4_corr_a} one can see that the maximum of the rank
distribution is shifted to smaller values for larger values of $\chi$.
This could have been expected, since larger
values of $\chi$ imply larger values for the wrapping numbers $\vec{X}$,
which in turn require lower values for the number of branes per
stack $N_a$, in order to fulfill the tadpole conditions.
The shift of the maximum depending of $\chi$, can be seen more directly in
figure~\ref{fig_t4_corr_b}.

\subsection{Statistics of four-dimensional models}\label{ls_4d}
Having tried our methods in compactifications down to six dimensions,
let us now switch to the phenomenologically more interesting case of
four-dimensional models.
Unfortunately we can no longer use the saddle point approximation,
since it turns out that in this more complicated case the
approximation is no longer reliable. The results deviate significantly
from what we see in exact computations. Furthermore the computer power
needed to obtain the integrals numerically in the approximation
becomes comparable to the effort needed to compute the solutions
explicitly.

\subsubsection{Properties of the gauge sector}\label{ls_dist}
Using several computer
clusters and the specifically adapted algorithm described in
section~\ref{ls_comp} for a period of several months,
we produced explicit constructions of~$\approx~1.6\times10^8$ consistent
compactifications on $T^6/\bZZ$.
The results presented in the following have been published
in~\cite{gbhlw05,gm05},
see also the analysis in~\cite{kuwe05} and more recent results using brane
recombination methods in~\cite{kuwe06}.

Using this data we can proceed to analyse the observables of these models.
The distribution of the total rank $r$ of the gauge group is shown
in figure~\ref{fig_rkundist_a}. An interesting phenomenon is the suppression
of odd values for the total rank.
This can be explained by the K-theory constraints
and the observation that the generic value for $Y^I$ is 0 or 1. Branes with
these values belong to the first class of branes in the classification of
section~\ref{ls_fin_4d} and are those which lie on top of the orientifold
planes.
Therefore equation~(\ref{eq_t6_ktheory}) suppresses solutions with an
odd value for
$r$. This suppression from the K-theory constraints is quite strong,
the total number of solutions is reduced by a factor of six compared to the
situation where these constraints are not enforced.

\twofig{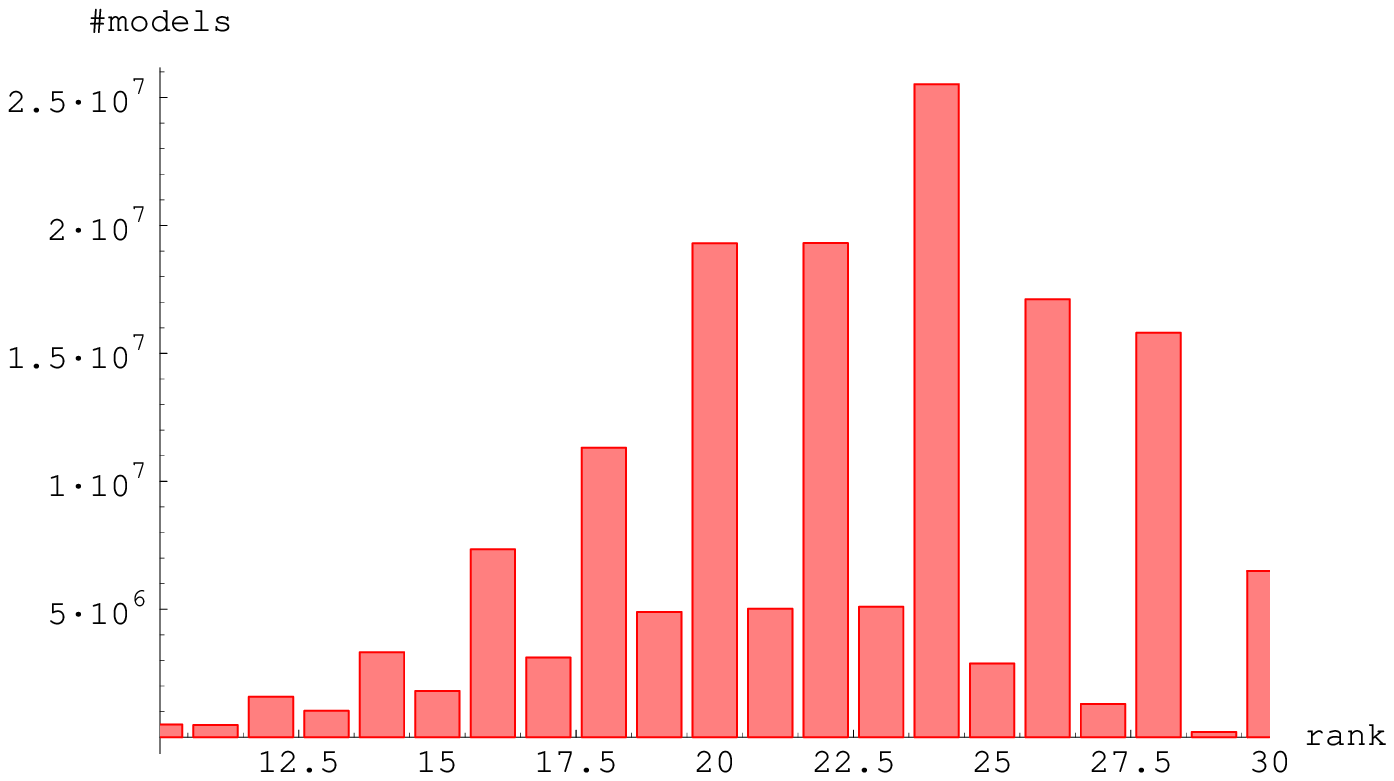}{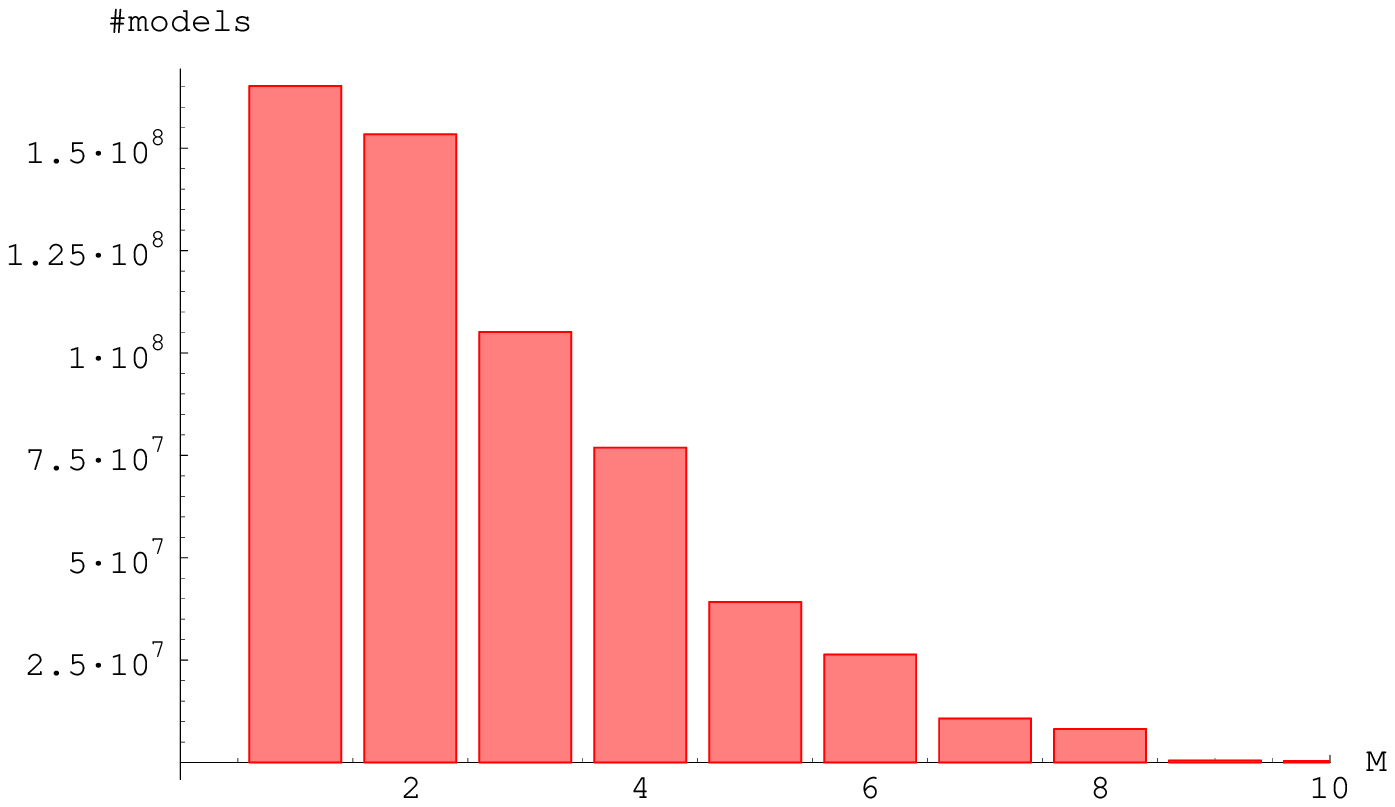}{fig_rkundist}{%
Frequency distributions of total rank and $U(M)$ gauge groups of all
models.%
}

Another quantity of interest is the
distribution of $U(M)$ gauge groups, shown in figure~\ref{fig_rkundist_b}.
We find that most models carry at least one $U(1)$ gauge group, corresponding
to a single brane, and stacks with a higher number of branes become more and
more unlikely. This could have been expected because small numbers occur
with a much higher frequency in the partition and factorisation of natural
numbers.

\subsubsection{Chirality}\label{ls_4dchi}
As in the six-dimensional case we want to define a quantity that counts
chiral matter in the models under consideration. In contrast to the very
rough estimate we used in section~\ref{ls_6dchi}, this time we are going to
count all chiral matter states, such that our definition of mean chirality
is now
\be{eq_4dchi}
  \chi := \frac{2}{k(k+1)}\sum_{a,b=0, a<b}^k I_{a'b}-I_{ab}
       = \frac{4}{k(k+1)}\sum_{a,b=0, a<b}^k \vec{Y}_a\vec{X}_b.
\ee
In this formula the states from the intersection of two branes $a$ and $b$
are counted with a positive sign, while the states from the intersection of
the orientifold image of brane $a$, denoted by $a'$, and brane $b$ are counted
negatively. As we explained in section~\ref{sb_chmatter} and summarised in
table~\ref{tab_reps}, $I_{ab}$ gives the number of bifundamental
representations $(\N_a,\overline{\N}_b)$, while $I_{a'b}$ counts
$(\overline{\N_a},\overline{\N_b})$. Therefore we compute the net number of
chiral representations with this definition of $\chi$. By summing over all
possible intersections and normalising the result we obtain a quantity that
is independent of the number of stacks and can be used for a statistical
analysis.

\fig{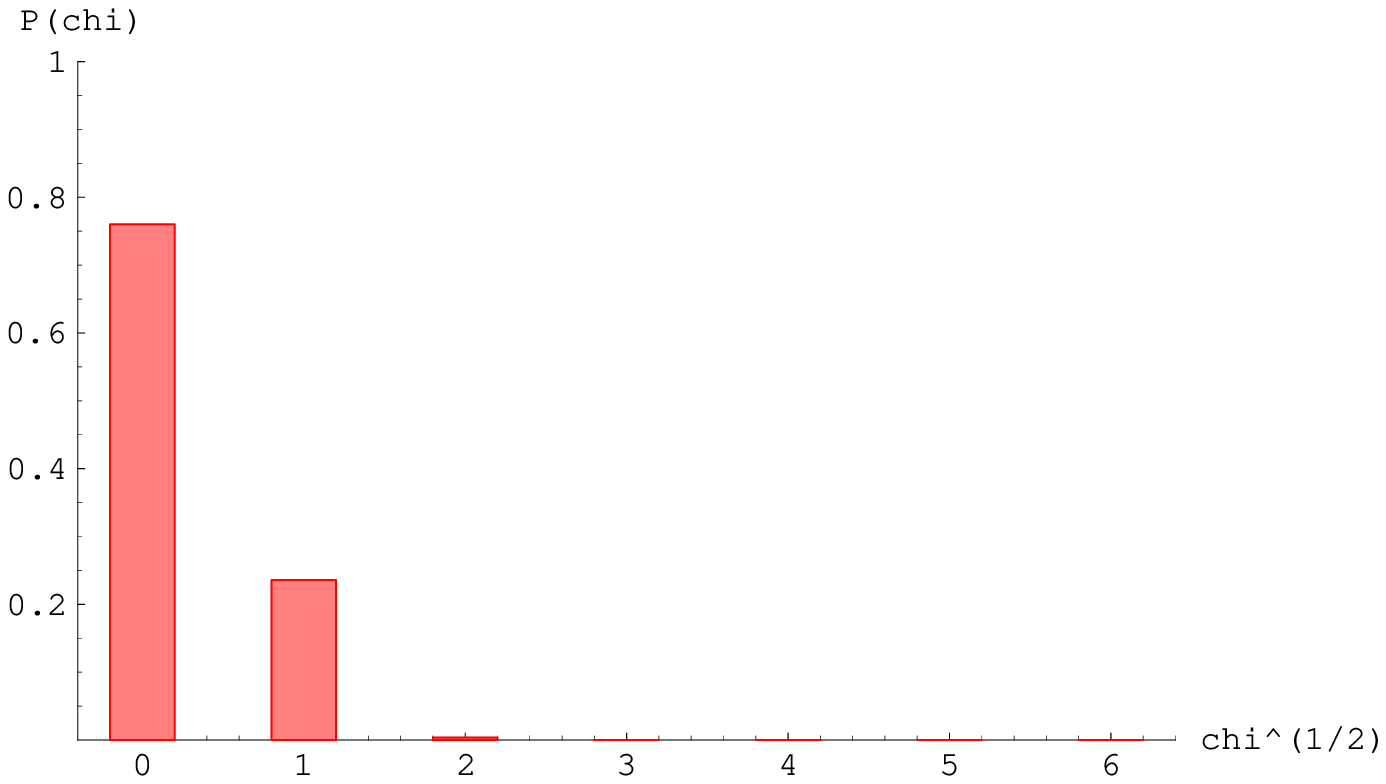}{fig_4dchi}{%
Distribution of the mean chirality $\chi$ in compactification to four
dimensions.
}

A computation of the value of $\chi$ according to~(\ref{eq_4dchi}) for all
models leads to a frequency distribution of the mean chirality as shown in
figure~\ref{fig_4dchi}. This distribution is basically identical to the one
we obtained in section~\ref{ls_6dchi}, shown in figure~\ref{fig_t4_chi}.
In particular we also find that models with a mean chirality of $0$ dominate
the spectrum and are exponentially more frequent then chiral ones.

From the similarity with the distribution of models on $T^4/\bZ_2$ we
can also conjecture that there is a correlation between the mean chirality
and the total rank,
as we found it to be the case for the six-dimensional models
in section~\ref{ls_6dcorr}.
Let us postpone this question to section~\ref{ls_corr}, where
we give a more detailed account of several questions concerning
the correlation of observables.

\subsection{Standard model constructions}\label{ls_sm}
An important subset of the models considered in the previous section are
of course those which could provide a standard model gauge group at low
energies. More precisely, since we are dealing with supersymmetric models
only, we are looking for models which might resemble the particle spectrum
of the MSSM.

To realise the gauge group of the standard model we need generically four
stacks of branes (denoted by a,b,c,d) with two possible choices for the
gauge groups:
\be{eq_smgaugegroups}
 U(3)_a \times U(2)_b \times U(1)_c \times U(1)_d,\quad\mbox{or}\quad
 U(3)_a \times Sp(2)_b \times U(1)_c \times U(1)_d.
\ee
To exclude exotic chiral matter from the first two factors we have to impose
the constraint that $\#\Sym_{a/b}=0$, i.e. the number of symmetric
representations of stacks $a$ and $b$ has to be zero.
Models with only three stacks of branes can also be
realised, but they suffer generically from having non-standard Yukawa
couplings.
Since we are not treating our models in so much detail and are more interested
in their generic distributions, we include these three-stack constructions
in our analysis.

Another important ingredient for standard model-like configurations is the
existence of a massless $U(1)_Y$ hypercharge. This is in general a
combination
\be{eq_u1y}
  U(1)_Y = \sum_{a=1}^k x_a U(1)_a,
\ee
including contributions of several $U(1)$s. Since we would like to construct
the matter content of the standard model, we are very constrained about the
combination of U(1) factors.
In order to obtain the right hypercharges for the standard model particles,
there are three different combinations of the $U(1)$s used to construct
the quarks and leptons possible, 
\bea{eq_smhyper}\nonumber
  U(1)_Y^{(1)} &=& \frac{1}{6}U(1)_a+\frac{1}{2}U(1)_c
    +\frac{1}{2}U(1)_d,\\\nonumber
  U(1)_Y^{(2)} &=& -\frac{1}{3}U(1)_a-\frac{1}{2}U(1)_b,\\
  U(1)_Y^{(3)} &=& -\frac{1}{3}U(1)_a-\frac{1}{2}U(1)_b+U(1)_d,
\eea
where choices 2 and 3 are only available for the first choice of gauge groups.
As explained in section~\ref{sb_anomalies}, we can construct a massless
combination of $U(1)$ factors, if (\ref{eq_u1rel}) is satisfied.
This gives an additional constraint on the wrapping numbers $\vec{Y}$.

\fig{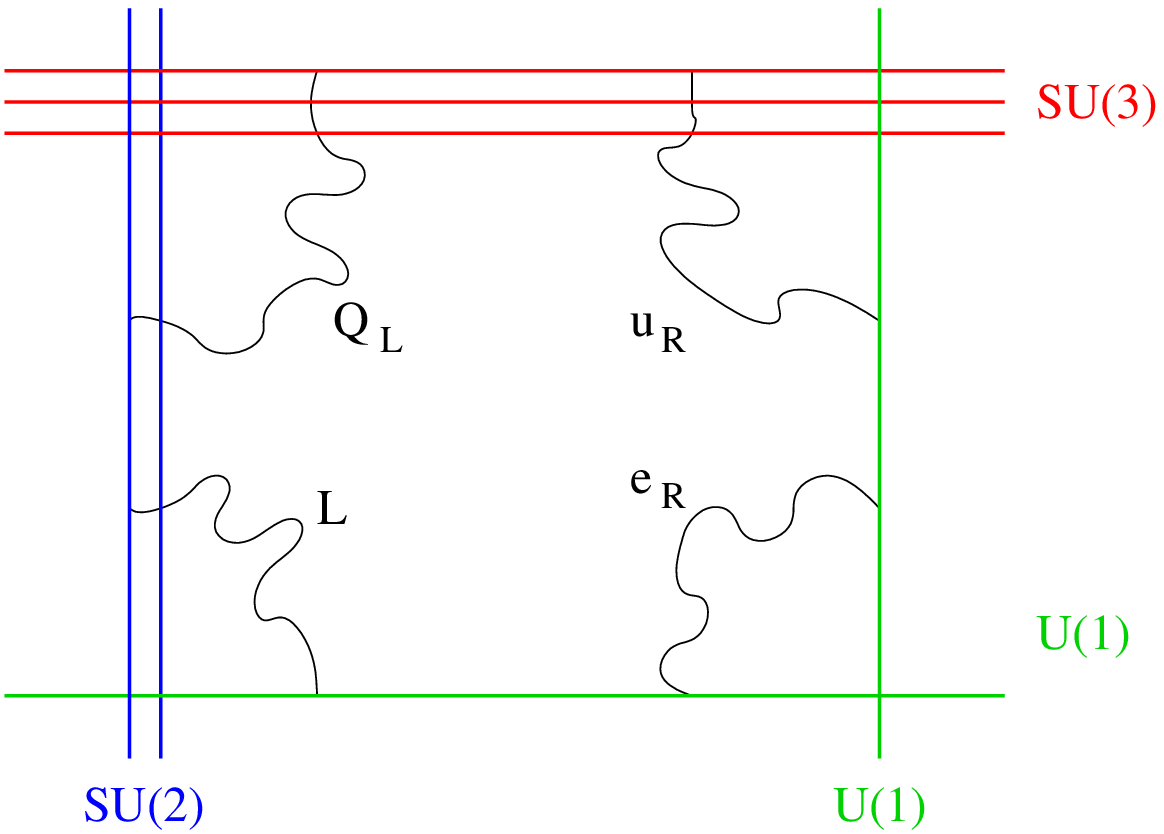}{fig_smconstr}{%
Assignment of brane intersections and chiral matter content for the first
of the possible realisations of the standard model using intersecting branes.%
}

For the different possibilities to construct the hypercharge this constraint
looks different.
In the case of $U(1)_Y^{(1)}$ the condition can be formulated as
\be{eq_q1massless}
  \vec{Y}_a + \vec{Y}_c + \vec{Y}_d =0.
\ee
For $Q_Y^{(2)}$, where the right-handed up-type quarks are realised as
antisymmetric representations of $U(3)$~\cite{akt00,bklo01}, we obtain
\be{eq_q2massless}
  \vec{Y}_a +\vec{Y}_b=0,
\ee
and for $Q_Y^{(3)}$, where we also need antisymmetric representations of $U(3)$
to realise the right-handed up-quarks, we get
\be{eq_q3massless}
  \vec{Y}_a +\vec{Y}_b -\vec{Y}_d=0.
\ee

In total we have found four ways to realise the standard model with massless
hypercharge, summarised with the explicit realisation of the fundamental
particles in tables~\ref{tab_sm1} and~\ref{tab_sm2}. The chiral matter content
arises at the intersection of the four stacks of branes. This is shown
schematically for one of the four possibilities in figure~\ref{fig_smconstr}.

\begin{table}[htb!]
\begin{center}
\begin{tabular}{|c|c|c|c|}\hline
particle & representation & mult. \\\hline\hline
\multicolumn{3}{|c|}{$U(3)_a\times Sp(2)_b \times U(1)_c \times U(1)_d$ with $Q_Y^{(1)}$}\\\hline
$Q_L$ & $(\3,\2)_{0,0}$ & $I_{ab}$ \\\hline
$u_R$ & $(\ov{\3},\1)_{-1,0}+(\ov{\3},\1)_{0,-1}$ & $I_{a'c}+I_{a'd}$ \\\hline
$d_R$ & $(\ov{\3},\1)_{1,0}+(\ov{\3},\1)_{0,1}$ & $I_{a'c'}+I_{a'd'}$ \\
$d_R$ & $(\ov{\3}_A,\1)_{0,0}$ & $\frac{1}{2}(I_{aa'} +I_{a{\rm O}6})$\\\hline
$L$ & $(\1,\2)_{-1,0}+(\1,\2)_{0,-1}$ & $I_{bc}+I_{bd}$ \\\hline
$e_R$ & $(\1,\1)_{2,0}$ & $\frac{1}{2}(I_{cc'} -I_{c{\rm O}6})$\\
$e_R$ & $(\1,\1)_{0,2}$ & $\frac{1}{2}(I_{dd'} -I_{d{\rm O}6})$\\
$e_R$ & $(\1,\1)_{1,1}$ & $I_{cd'}$ \\\hline\hline
\multicolumn{3}{|c|}{$U(3)_a\times U(2)_b \times U(1)_c \times U(1)_d$ with $Q_Y^{(1)}$}\\\hline
$Q_L$ & $(\3,\ov{\2})_{0,0}$   & $I_{ab}$\\
$Q_L$ & $(\3,\2)_{0,0}$   & $I_{ab'}$\\\hline
$u_R$ &  $(\ov{\3},\1)_{-1,0}+(\ov{\3},\1)_{0,-1}$  & $I_{a'c}+I_{a'd}$\\\hline
$d_R$ &  $(\ov{\3},\1)_{1,0}+(\ov{\3},\1)_{0,1}$  & $I_{a'c'}+I_{a'd'}$\\  
$d_R$ & $(\ov{\3}_A,\1)_{0,0}$  &  $\frac{1}{2}(I_{aa'} +I_{a{\rm O}6})$\\\hline
$L$ & $(\1,\2)_{-1,0}+(\1,\2)_{0,-1}$ & $I_{bc}+I_{bd}$  \\  
$L$ & $(\1,\ov{\2})_{-1,0}+(1,\ov{\2})_{0,-1}$ & $I_{b'c}+I_{b'd}$\\\hline 
$e_R$ & $(\1,\1)_{2,0}$ & $\frac{1}{2}(I_{cc'} -I_{c{\rm O}6})$\\
$e_R$ & $(\1,\1)_{0,2}$ & $\frac{1}{2}(I_{dd'} -I_{d{\rm O}6})$\\
$e_R$ & $(\1,\1)_{1,1}$ & $I_{cd'}$ \\
\hline
\end{tabular}
\caption{Realisation of quarks and leptons for the two different choices of gauge groups (\ref{eq_smgaugegroups}) and hypercharge (1) in (\ref{eq_smhyper}).}
\label{tab_sm1}
\end{center}
\end{table}
\begin{table}[htb!]
\begin{center}
\begin{tabular}{|c|c|c|c|}\hline
particle & representation & mult. \\\hline\hline
\multicolumn{3}{|c|}{$U(3)_a\times U(2)_b \times U(1)_c \times U(1)_d$ with $Q_Y^{(2)}$}\\\hline
$Q_L$ & $(\3,\ov{\2})_{0,0}$   & $I_{ab}$  \\\hline
$u_R$ & $(\ov{\3}_A,\1)_{0,0}$ & $\frac{1}{2}(I_{aa'} +I_{a{\rm O}6})$ \\\hline
$d_R$ &  $(\ov{\3},\1)_{-1,0}+(\ov{\3},1)_{0,-1}$   & $I_{a'c}+I_{a'd}$ \\ 
$d_R$ &  $(\ov{\3},\1)_{1,0}+(\ov{\3},1)_{0,1}$ & $I_{a'c'}+I_{a'd'}$ \\\hline
$L$ & $(\1,\2)_{-1,0}+(\1,\2)_{0,-1}$ & $I_{bc}+I_{bd}$ \\
$L$ & $(\1,\2)_{1,0}+(\1,\2)_{0,1}$ & $I_{bc'}+I_{bd'}$ \\ \hline
$e_R$ & $(\1,\ov{\1}_A)_{0,0}$  &  $-\frac{1}{2}(I_{bb'} +I_{b{\rm O}6})$ \\
\hline\hline
\multicolumn{3}{|c|}{$U(3)_a\times U(2)_b \times U(1)_c \times U(1)_d$ with $Q_Y^{(3)}$}\\\hline
$Q_L$ & $(\3,\ov{\2})_{0,0}$   & $I_{ab}$  \\\hline
$u_R$ & $(\ov{\3}_A,\1)_{0,0}$ & $\frac{1}{2}(I_{aa'} +I_{a{\rm O}6})$ \\\hline
$d_R$ &  $(\ov{\3},\1)_{-1,0}$  & $I_{a'c}$ \\  
$d_R$ &  $(\ov{\3},\1)_{1,0}$ & $I_{a'c'}$ \\\hline
$L$ & $(\1,\ov{\2})_{0,-1}$  & $I_{b'd}$ \\ \hline
$e_R$ & $(\1,\ov{\1}_A)_{0,0}$  &  $-\frac{1}{2}(I_{bb'} +I_{b{\rm O}6})$   \\ 
$e_R$ & $(\1,\1)_{1,1}$   & $I_{cd'}$  \\
$e_R$ & $(\1,\1)_{-1,1}$   &  $I_{c'd'}$ \\\hline
\end{tabular}
\caption{Realisation of quarks and leptons for hypercharges (2) and (3) of (\ref{eq_smhyper}), which can only be realised for the first choice of gauge groups in (\ref{eq_smgaugegroups}).}
\label{tab_sm2}
\end{center}
\end{table}

\fig{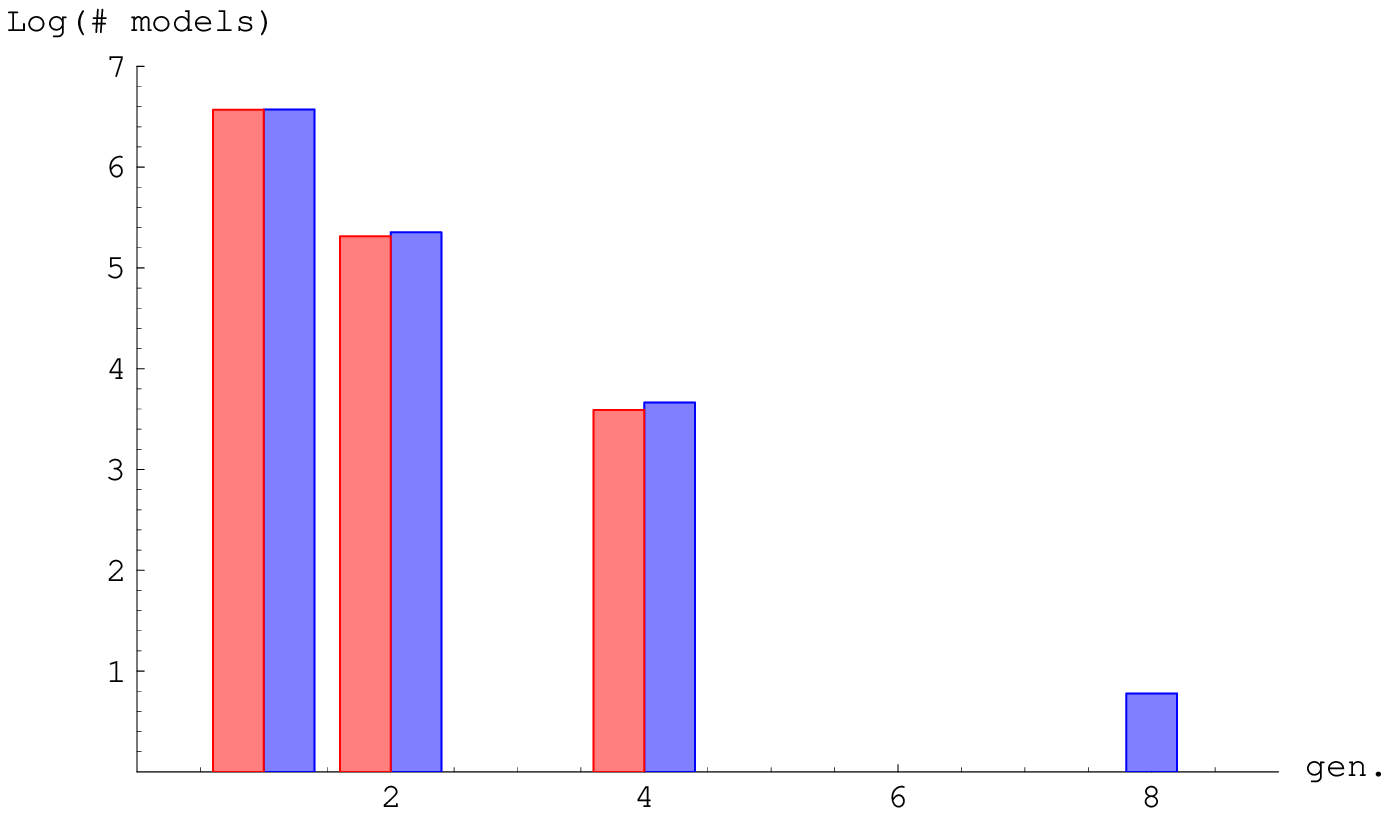}{fig_gensm}{%
Number of quark and lepton generations with (red bars on the left) and
without (blue bars on the right) enforcing a massless $U(1)$.%
}

\subsubsection{Number of generations}\label{ls_sm_nogen}
The first question one would like to ask, after having defined what a
''standard model'' is in our setup, concerns the frequency
of such configurations in the space of all solutions.
Put differently: \emph{How many standard models
with three generations of quarks and leptons do we find?} The answer to
this question is \emph{zero}, even if we relax our constraints and allow for
a massive hypercharge (which is rather fishy from a phenomenological point of
view). The result of the analysis can be seen in figure~\ref{fig_gensm}.

To analyse this result more closely, we relaxed our constraints further and
allowed for different numbers of generations for the quark and lepton sector.
This is of course phenomenologically no longer relevant, but it helps to
understand the structure of the solutions. The three-dimensional plot of this
analysis is shown in figure~\ref{fig_genmass}. Actually there exist solutions
with three generations of either quarks or leptons, where models with only
one generation of quarks clearly dominate. The suppression of three generation
models can therefore be pinned down to the construction of models with three
generations of quarks, which arise at the intersection of the $U(3)$ with the
$SU(2)/Sp(2)$ branes and the $U(1)$ branes respectively. models with three
generations of either quarks or leptons are shown in table~\ref{tab_genmass}.
\begin{table}[htb]
\begin{center}
\begin{tabular}{|r|r|r|}\hline
\# of quark gen. & \# of lepton gen. & \# of models\\\hline
1               &                3 &      183081\\
2               &                3 &           8\\
3               &                4 &         136\\
4               &                3 &          48\\\hline
\end{tabular}
\caption{Number of models found with either three quark or three lepton
generations.}
\label{tab_genmass}
\end{center}
\end{table}

This result is rather strange, since we know that models with three families
of quarks and leptons have been constructed in our setup
(e.g. in~\cite{blt03,caur03,mash04,cll05}).
A detailed analysis of the models in the literature shows that all models
which are known use (in our conventions) large values for the complex structure
variables $U_I$ and therefore did not appear in our analysis (see section
\ref{ls_comp}). On the other hand we know that the number of models decreases
exponentially with higher values for the complex structures. Therefore we
conclude that standard models with three generations are highly suppressed in
this specific setup.

This brings up a natural question, namely: How big is this suppression factor?
We postpone this question
to section~\ref{ls_smestimate}, where we analyse this issue more
closely and finally give an estimate for the probability to find a
three generation standard model in our setup.
For now let us just notice that this probability has to be
smaller than the inverse of the total number of models we analysed,
i.e.~$< 10^{-8}$.

\figclip{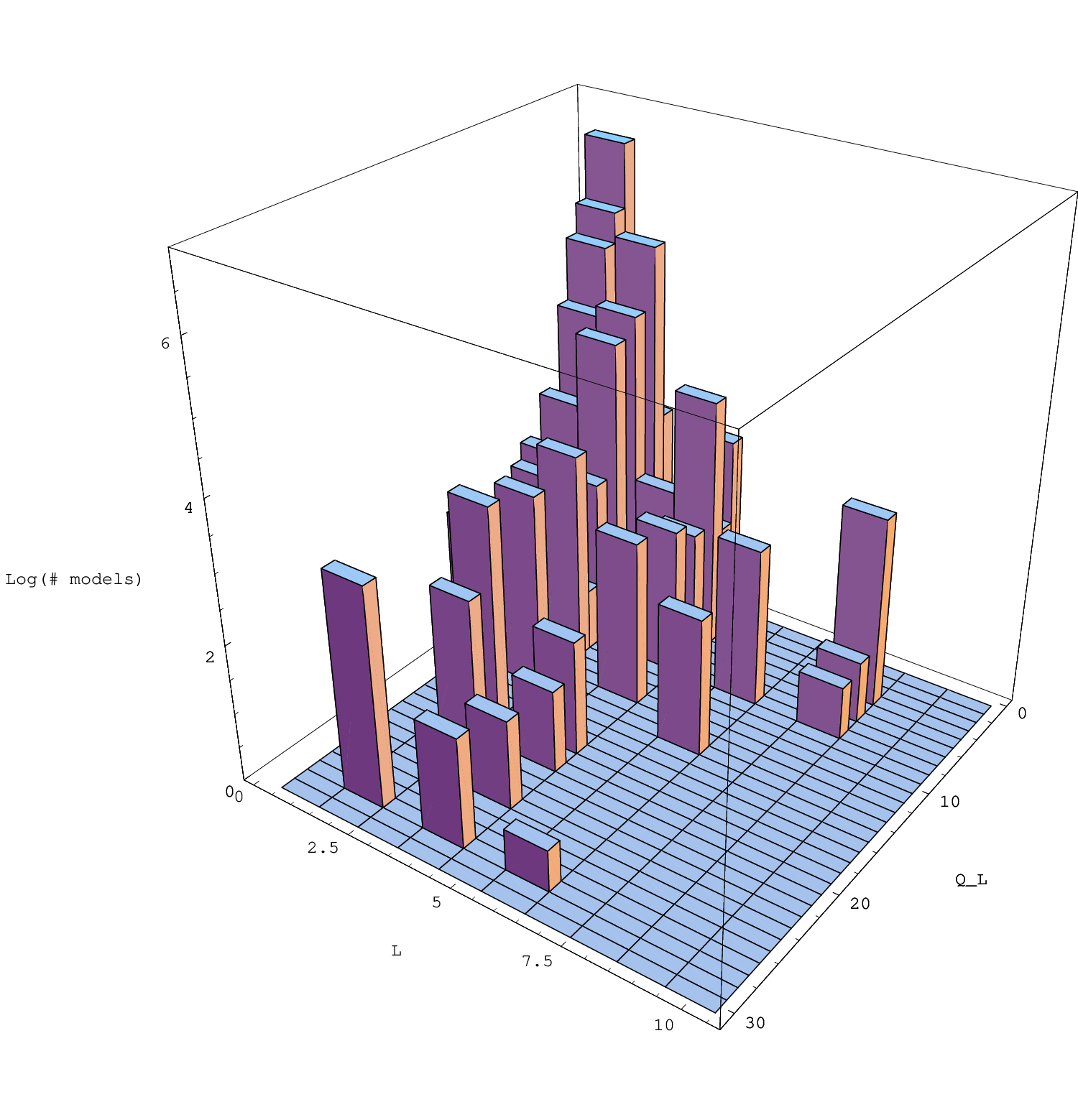}{fig_genmass}{%
Logarithmic plot of the number of models with different numbers of
generations of quarks and leptons. $Q_L$ denotes the number of quark
families, $L$ is the number of lepton generations.%
}

\subsubsection{Hidden sector}\label{ls_smhidden}
Besides the so called ``visible sector'' of the model, containing the
standard model gauge group and particles, we have generically additional
chiral matter, transforming under different gauge groups. This sector is
usually called the ``hidden sector'' of the theory, assuming that the masses
of the additional particles are lifted and therefore unobservable at low
energies.

\twofig{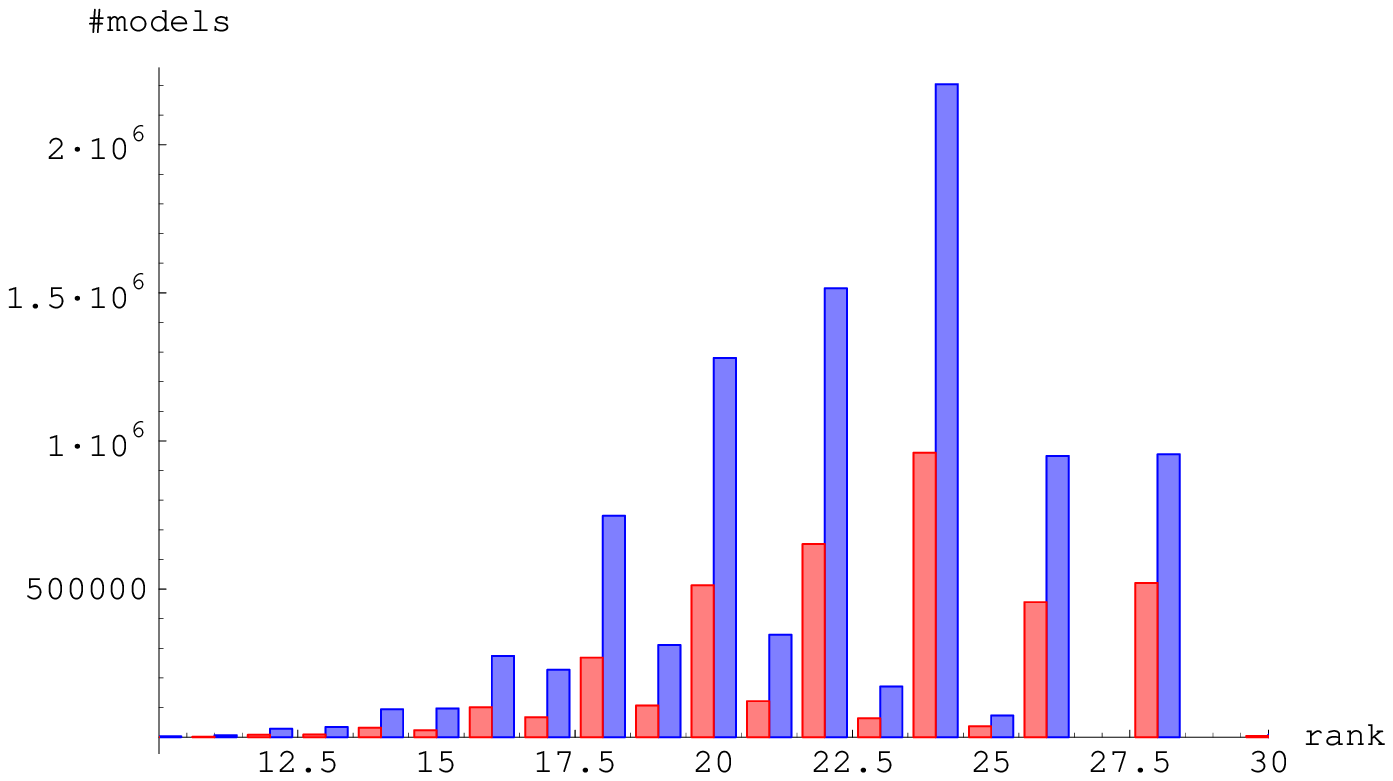}{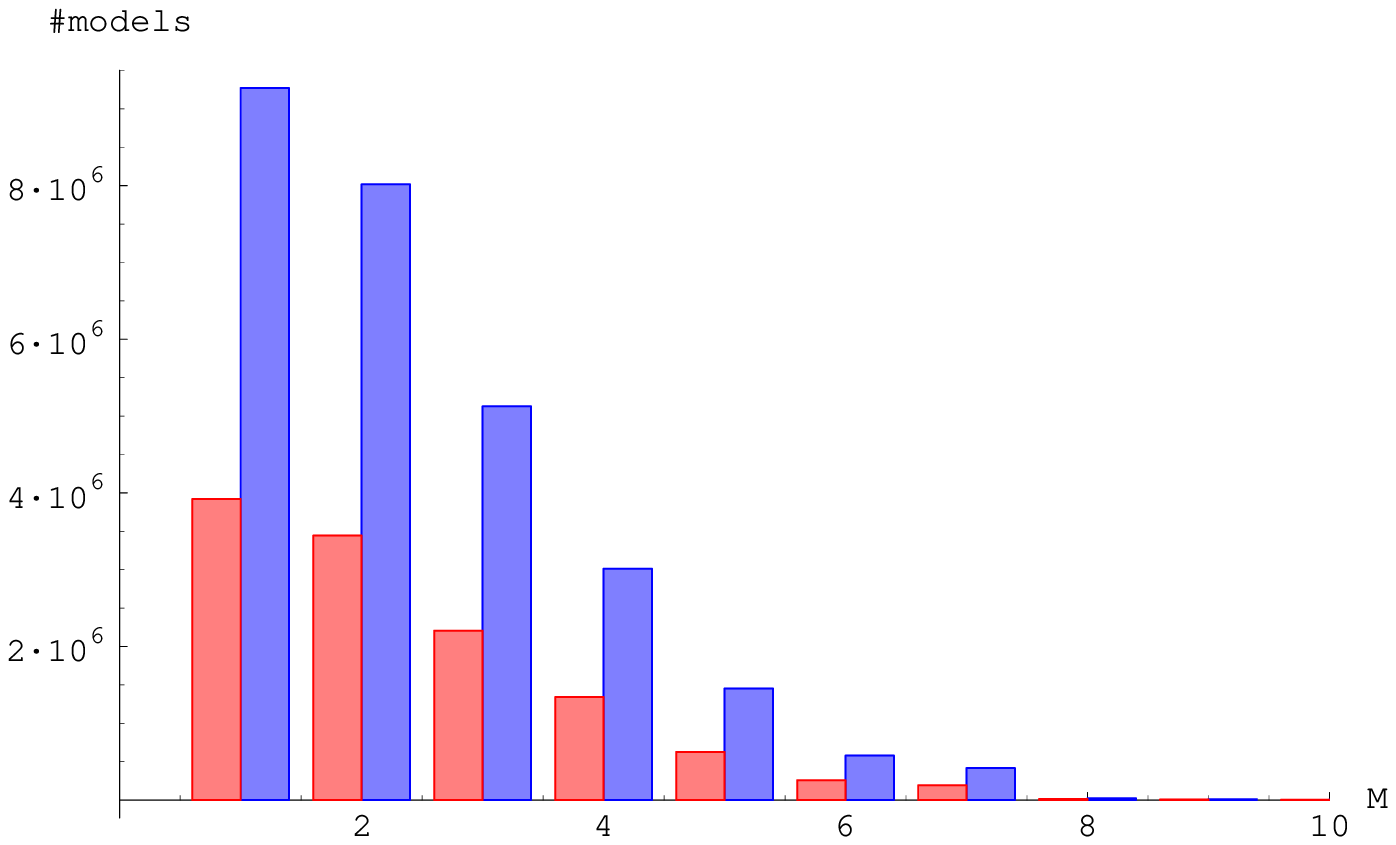}{fig_smhidden}{%
Frequency distributions of (a) total rank and (b) single gauge group factors
in the hidden sector of MSSM-models (red bars on the left) and MSSM models with
massive $U(1)$ (blue bars on the right).%
}

In figure~\ref{fig_smhidden_a} we show the frequency distributions of the total
rank of gauge groups in the hidden sector. In~\ref{fig_smhidden_b} we show the
frequency distribution of individual gauge group factors.
Comparing these results with the distributions of the full set of
models in figure~\ref{fig_rkundist}, we observe that at a qualitative
level the restriction to
the standard model gauge group in the visible sector did not change the
distribution of gauge group observables.
The number of constructions in the standard model case is of course much
lower, but the frequency distributions of the hidden sector properties behave
pretty much like those we obtained for the complete set of models.

As we argue in section~\ref{ls_corr}, this is not a coincidence,
but a generic feature of the class of models we analysed. Many
of the properties of our models can be regarded to be independent of each
other, which means that the statistical analysis of the hidden sector of 
any model with specific visible gauge group leads to very similar
results.

\subsubsection{Gauge couplings}\label{ls_gcoup}
The gauge sector considered so far belongs
to the topological sector of the theory, in the sense that its observables
are defined
by the wrapping numbers of the branes and independent of the geometric
moduli. This does not apply to the gauge couplings, which explicitly
do depend on the complex structures, following the derivation
in~\cite{bls03}, which in our conventions reads
\be{eq_gcoup}
  \frac{1}{\alpha_a}=\frac{M_{Planck}}{2\sqrt{2}M_s\kappa_a}\,\frac{1}{c\sqrt{\prod_{i=1}^3R_1^{(i)}R_2^{(i)}}}\sum_{I=0}^{3}X^I U_I,
\ee
where $\kappa_a=1$ or $2$ for an $U(N)$ or $Sp(2N)$ stack respectively.

If one wants to perform an honest analysis of the coupling constants,
one would have to compute their values at low energies using the
renormalization group equations. We are not going to do this, but look instead
at the distribution of $\ga_s/\ga_w$ at the string scale.
A value of one at the string scale does of course not necessarily mean
unification at lower energies, but it could be taken as a hint in
this direction.

To calculate the coupling $\ga_Y$ we have to include contributions from all
branes used for the definition of $U(1)_Y$. Therefore we need to distinguish
the different possible constructions defined in~(\ref{eq_smhyper}). In general
we have
\be{eq_aYgen}
  \frac{1}{\alpha_Y}=\sum_{a=1}^k 2N_a x_a^2\frac{1}{\alpha_a},
\ee
which for the three different possibilities reads explicitly
\bea{eq_aYdef}\nonumber
  \frac{1}{\alpha_Y^{(1)}} &=&  \frac{1}{6}\frac{1}{\alpha_a}
                               +\frac{1}{2}\frac{1}{\alpha_c}
                               +\frac{1}{2}\frac{1}{\alpha_d},\\\nonumber
  \frac{1}{\alpha_Y^{(2)}} &=&  \frac{2}{3}\frac{1}{\alpha_a}
                               +\frac{1}{\alpha_b},\\
  \frac{1}{\alpha_Y^{(3)}} &=&  \frac{2}{3}\frac{1}{\alpha_a}
                               +\frac{1}{\alpha_b}
                               +2\frac{1}{\alpha_d}.
\eea
The result is shown in figure \ref{fig_gc_a} and it turns out that only
2.75\% of all models actually do show gauge unification at the string
scale.

\twofig{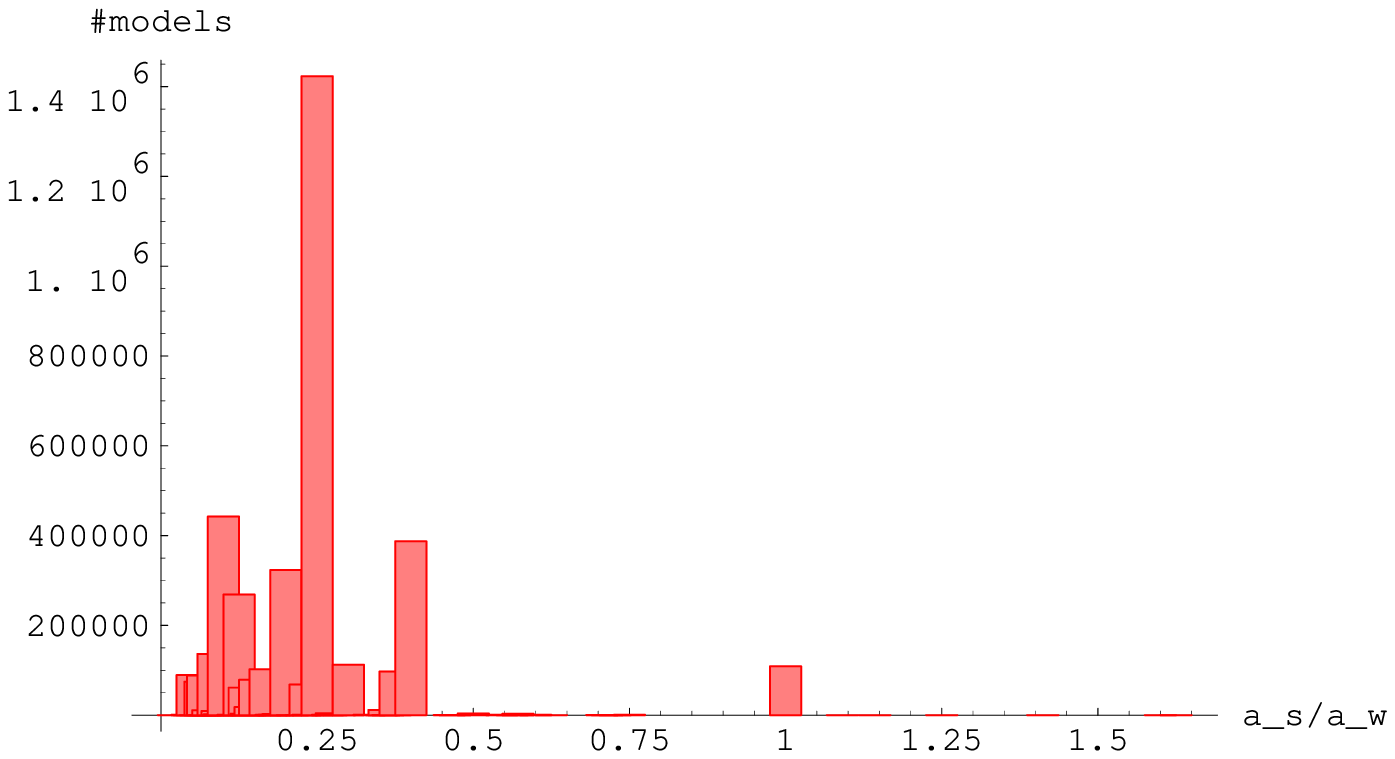}{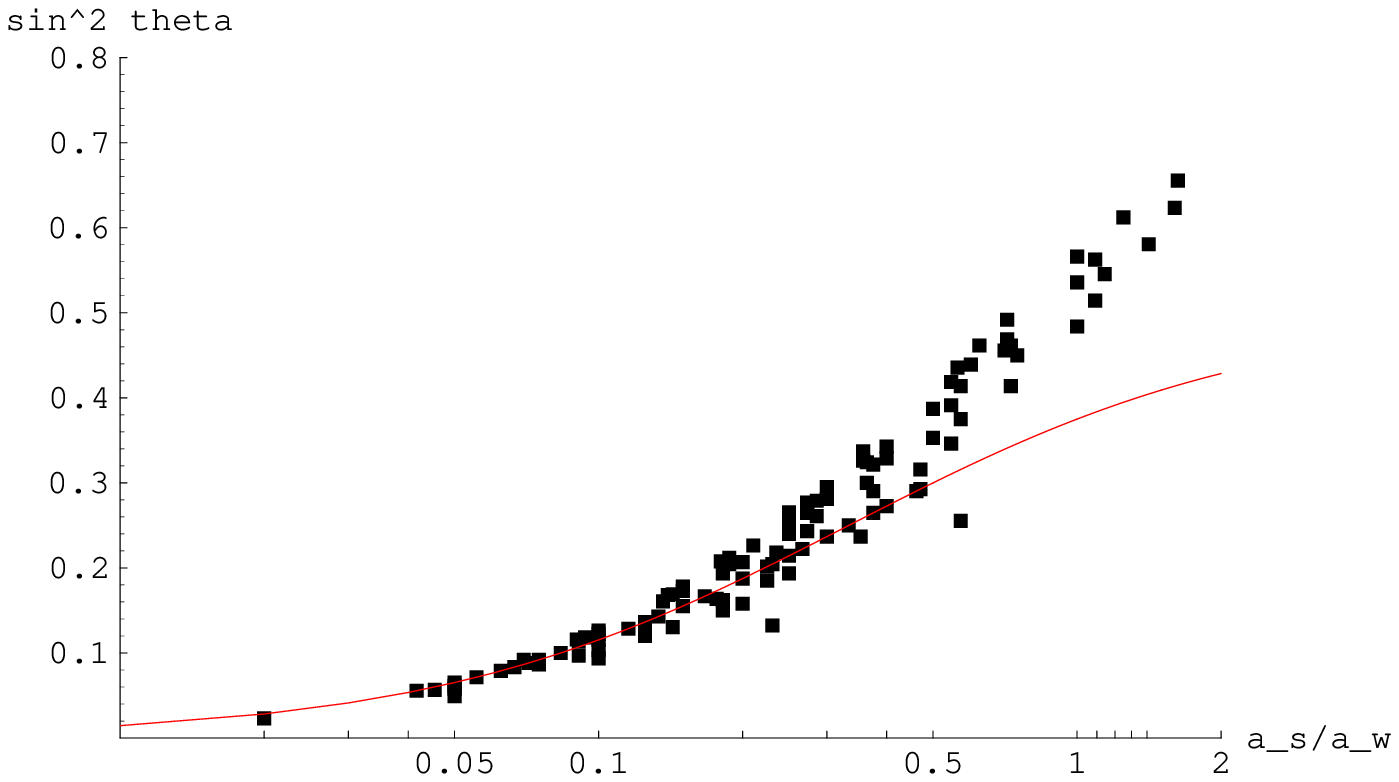}{fig_gc}{%
(a) Frequency distribution of $\ga_s/\ga_w$ in standard model-like configurations.
(b) Values of $sin^2\theta$ depending on $\ga_s/\ga_w$. Each dot represents a class of models with these values.%
}

Furthermore we analyse the distribution of values for the
Weinberg angle
\be{eq_weinberg}
  \sin^2\theta=\frac{\ga_Y}{\ga_Y+\ga_w},
\ee
which depends on the ratio
$\ga_s/\ga_w$. We want to check the following 
relation between the three couplings, which was proposed
in~\cite{bls03} and is supposed to hold for a large class
of intersecting brane models
\be{eq_crel}
  \frac{1}{\ga_Y}=\frac{2}{3}\frac{1}{\ga_s}+\frac{1}{\ga_w}.
\ee
From this equation we can derive a relation for the weak mixing angle
\be{eq_sinthetarel}
  \sin^2\theta=\frac{3}{2}\,\frac{1}{\ga_w/\ga_s+3}.
\ee
The result is shown in figure~\ref{fig_gc_b}, where we included a red line
that represents the relation (\ref{eq_crel}). The fact that actually 88\% of
all models obey this relation is a bit obscured by the plot, because each dot
represents a class of models and small values for $\ga_s/\ga_w$ are highly
preferred, as can be seen from figure~\ref{fig_gc_a}.

\subsubsection{Comparison with the statistics of Gepner models}\label{ls_gepner}
In this paragraph we would like to compare our results with the analysis
of~\cite{dhs041,dhs042}, where a search for standard model-like features in
Gepner model constructions~\cite{ge187,ge287,bhhw04,blwe04} has been performed.

To do so, we have to take only a subset of the data analysed in the previous
sections, since the authors of~\cite{dhs041,dhs042} restricted their analysis
to a special subset of constructions. Due to the complexity of the problem
they restricted their analysis to models with a maximum of three branes in the
hidden sector and focussed on three-generation models only.
Since the number of generations does not modify the frequency distributions
and we obtained no explicit results for three generation models, we
include models of an arbitrary number of generations in the analysis.
To match the first constraint we filter our results and include only
those models with a maximum of three hidden branes. But, as we will see, this
does also not change the qualitative behaviour of the frequency distributions.

\twofig{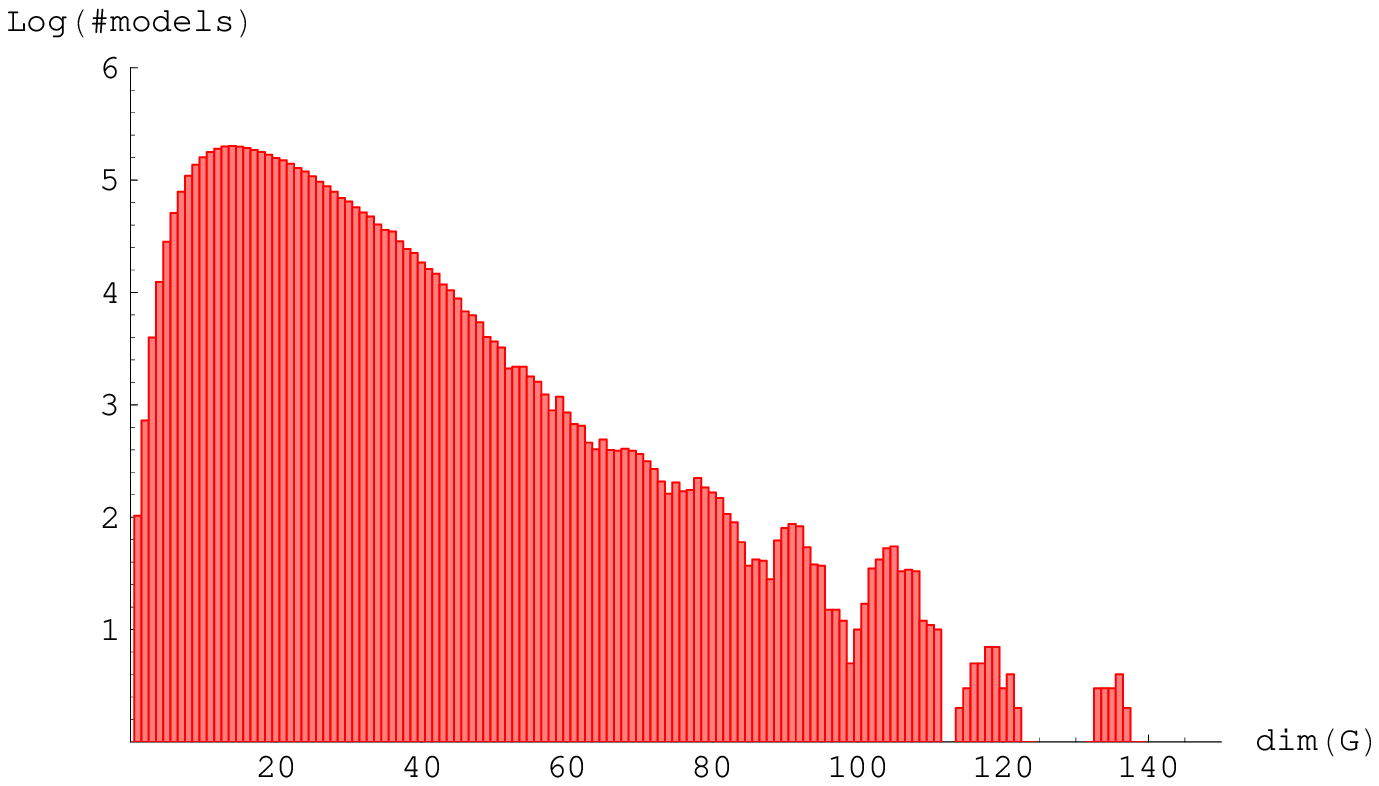}{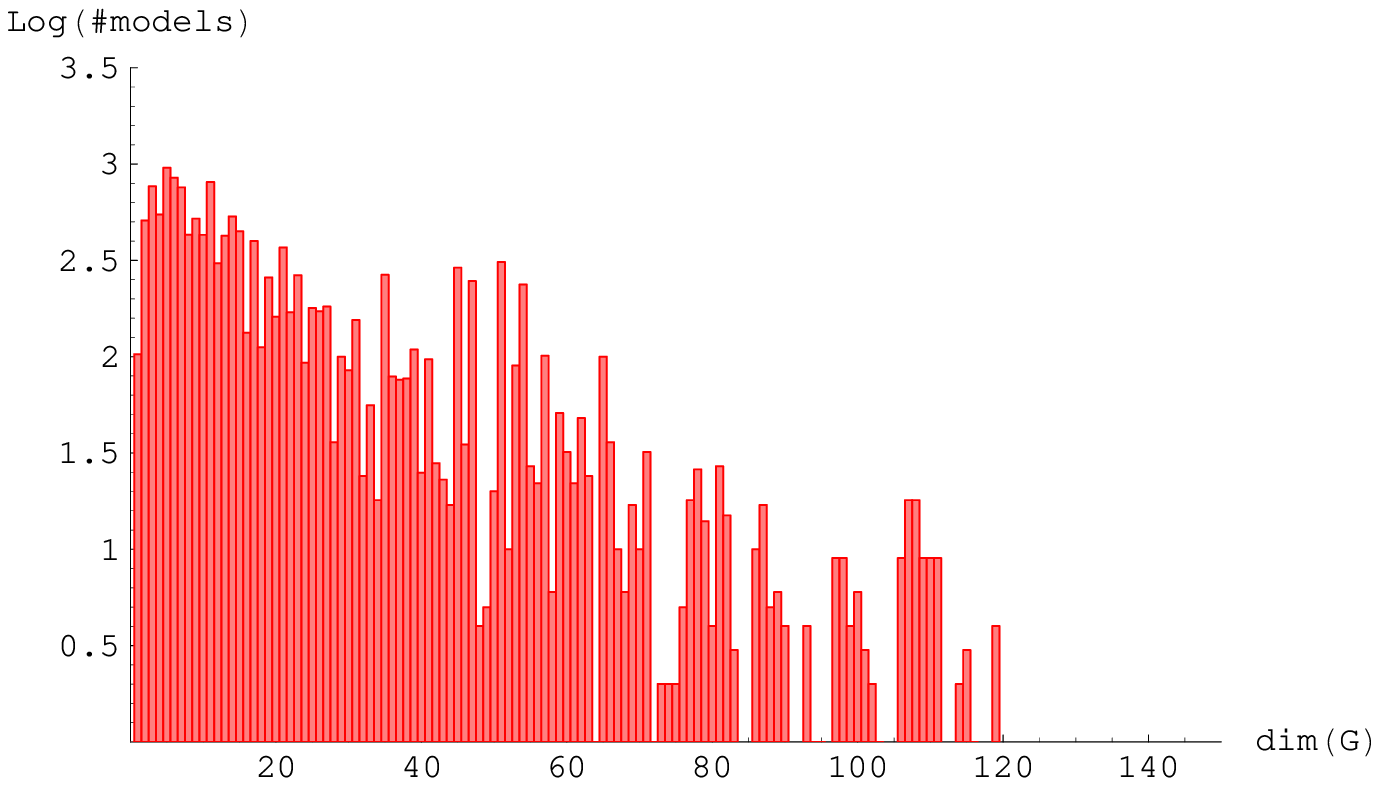}{fig_ggdim}{%
Frequency distribution of the dimension of the hidden sector gauge group. Figure (a) is the full set of models, figure (b) shows the subset of solutions with a maximum of three branes in the hidden sector.}

In figure~\ref{fig_ggdim} we show the frequency distribution of the dimension
of the hidden sector gauge group before (a) and after (b) the truncation to
a maximum of three hidden branes. Obviously the number of models drops
significantly, but the qualitative shape of the distribution remains the same.
Figure~\ref{fig_ggdim_b} can be compared directly with figure~5
of~\cite{dhs042}.
From a qualitative point of view both distributions are very
similar, which could have been expected since the Gepner model construction
is from a pure topological point of view quite similar to intersecting
D-branes. A major difference can be observed in the absolute values of models
analysed. In the Gepner case the authors of~\cite{dhs042} found a significantly
larger amount of candidates for a standard model.

Besides the frequency distribution of gauge groups we can also compare the
analysis of the distribution of gauge couplings.
In particular, the distribution of values for
for $\sin^2\theta$ depending on the ratio $\ga_w/\ga_s$, figure~\ref{fig_gc_b},
can be compared with figure~6 of~\cite{dhs042}.
We find, in contrast to the case of hidden sector gauge groups,
very different distributions. While almost all of our models are distributed
along one curve, in the Gepner case a much larger variety of values is
possible. The fraction of models obeying
(\ref{eq_crel}) was found to be only about 10\% in the Gepner model case,
which can be identified as a very thin line in figure~6 of~\cite{dhs042}.
This discrepancy might be traced back to the observation that in contrast to
the topological
data of gauge groups we are dealing with geometrical aspects here.

As explained in the last paragraph, the gauge couplings do depend explicitly
on the geometric moduli. A major difference between the Gepner construction
and our intersecting D-brane models lies in the different regimes of internal
radius that can be assumed.
In our approach we rely on the fact that we are in a perturbative
regime, i.e. the compactification radius is much larger than the string
length and the string coupling is small.

%
%
\subsection{Pati-Salam models}\label{ls_ps}
As in the case of a $SU(3)\times SU(2)\times U(1)$ gauge group, we can try
to construct models with a gauge group of Pati-Salam type
\be{eq_patisalam}
  SU(4) \times SU(2)_L \times SU(2)_R.
\ee
Analogous to the case of a standard model-like gauge group, we analysed the
statistical data for Pati-Salam constructions, realised via the intersection
of three stacks of branes. One brane with $N_a=4$ and two stacks with
$N_{b/c}=2$, such that the chiral matter of the model can be realised as
\be{eq_psmatter}
  Q_L=(\rep{4},\rep{2},\rep{1}),\quad Q_R=(\ov{\rep{4}},\rep{1},\rep{2}).
\ee

One possibility to obtain the standard model gauge group in this setup is
given by breaking the $SU(4)$ into $SU(3)\times U(1)$ and one of the $SU(2)$
groups into $U(1)\times U(1)$. This can be achieved by separating the four
branes of stack $a$ into two stacks consisting of three and one branes,
respectively, and the two branes of stack $b$ or $c$ into two stacks consisting
of one brane each. The separation corresponds to giving a vacuum expectation
value to the fields in the adjoint representation of the gauge groups
$U(N_a)$ and $U(N)_{b/c}$, respectively.

Models of this type have been constructed explicitly in the literature,
see e.g~\cite{csu011,csu012,cp03,cll04,cll05,cln061}.
However, one has to be careful comparing these models with our results,
since our constraints are stronger compared to those usually imposed.
In particular, we do not allow for symmetric or antisymmetric representations
of $SU(4)$, a constraint that is not always fulfilled for the models that can
be found in the references above. 

\fig{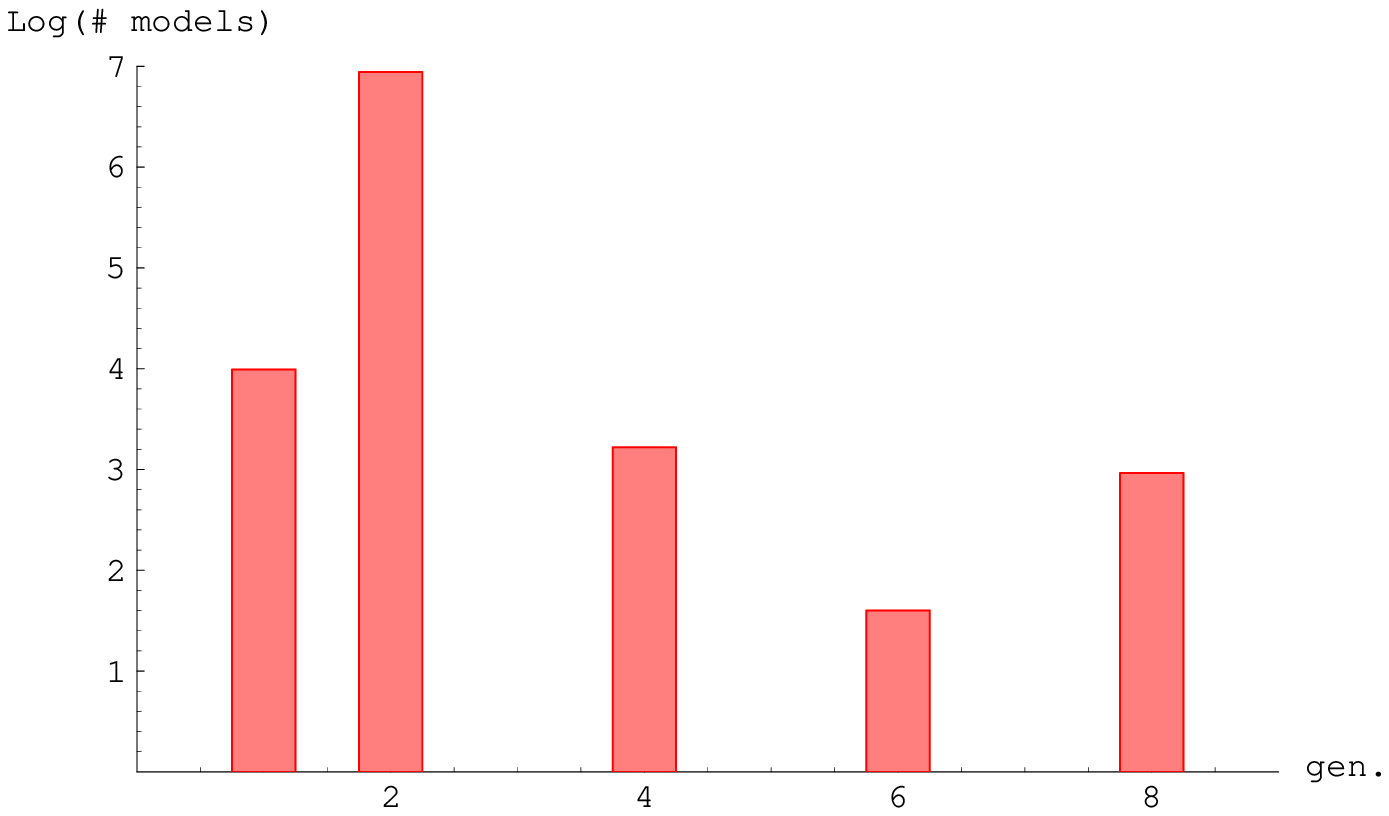}{fig_genps}{%
Logarithmic plot of the number of Pati-Salam models found, depending on the
number of generations. The solutions have been restricted to an equal number
of left- and right-handed fermions, i.e. $gen.=Q_L\hypeq Q_R$%
}

A restriction on the possible models, similar to the standard model case,
is provided by the constraint that there should be no additional antisymmetric
matter and the number of chiral fermions transforming under
$SU(2)_L$ and $SU(2)_R$ should be equal.

As can be seen in figure~\ref{fig_genps}, we found models with up to eight
generations, but no three-generation models. The conclusion is the
same as in section~\ref{ls_sm} -- the suppression of three generation models
is extremely large and explicit models show up only at very large values of
the complex structure parameters.
The distribution differs from the standard model case in the domination
of two-generation models. This is an interesting phenomenon, which can be
traced back to the specific construction of the models using two $N=2$ stacks
of branes. This example shows that the number of generations, in contrast
to the distribution of gauge groups in the hidden sector (see also
section~\ref{ls_corr}), does depend on the
specific visible sector gauge group we chose.

%
%
\subsection{SU(5) models}\label{ls_su5}
From a phenomenological point of view a very interesting class of low-energy
models consist of those with a grand unified gauge group\footnote{For an
introduction see e.g.~\cite{ro85} or the corresponding chapters
in~\cite{chli84,pdg04}.}, providing
a framework for the unification of the strong and electro-weak forces.

The minimal simple Lie group that could be used to achieve this is
$SU(5)$~\cite{gegl74} or also the so-called flipped $SU(5)$~\cite{ba81,dkn83},
consisting
of the gauge group $SU(5)\times U(1)_X$. They
represent the two possibilities how to embed an $SU(5)$ gauge group into
$SO(10)$.
The flipped construction is more interesting phenomenologically,
because models based on this gauge group might survive the experimental limits
on proton decay.
Several explicit constructions of supersymmetric $SU(5)$ models in the context
of intersecting D-brane models are present
in the literature~\cite{cps02,afk03,ckmnw051,ckmnw052,cmn05,cln062}, as well as
some non-supersymmetric ones~\cite{bklo01,ekn02}.

In the remainder of this section we present some results on the
distribution
of the gauge group properties of $SU(5)$ and flipped $SU(5)$ models, using the
same $T^6/\bZZ$ orientifold setting as
in the previous sections. This part is based on~\cite{gmst06}.

\subsubsection{Construction}
In the original $SU(5)$ construction, the standard model particles are embedded in a $\fivebar$ and
a $\ten$ representation of the unified gauge group as follows
\bea{eq_su5embed}\nonumber
SU(5) &\to& SU(3)\times SU(2)\times U(1)_Y,\\\nonumber
\fivebar &\to& (\rep{\bar{3}},\rep{1})_{2/3} + (\rep{1},\rep{2})_{-1},\\
\ten &\to& (\rep{\bar{3}},\rep{1})_{-4/3} + (\rep{3},\rep{2})_{1/3} + (\rep{1},\rep{1})_{2},
\eea
where the hypercharge is generated by the $SU(3)\times SU(2)$-invariant generator
\be{eq_z}
  Z=\mathrm{diag}(-1/3,-1/3,-1/3,1/2,1/2).
\ee
In the flipped $SU(5)$ construction, the embedding is given by
\bea{eq_su5xembed}\nonumber
SU(5) \times U(1)_X &\to& SU(3)\times SU(2) \times U(1)_Y,\\\nonumber
\fivebar_{-3} &\to& (\rep{\bar{3}},\rep{1})_{-4/3} + (\rep{1},\rep{2})_{-1},\\\nonumber
\ten_1 &\to& (\rep{\bar{3}},\rep{1})_{2/3} + (\rep{3},\rep{2})_{1/3} + (\rep{1},\rep{1})_{0},\\
\rep{1}_5 &\to& (\rep{1},\rep{1})_2,
\eea
including a right-handed neutrino~$(\rep{1},\rep{1})_{0}$. The hypercharge is in this case given by
the combination
\be{eq_su5hyper}
  Y=-\frac{2}{5} Z+\frac{2}{5}X.
\ee

We would like to realise models of both type within our orientifold setup.
The $SU(5)$ case
is simpler, since in principle it requires only two branes, a $U(5)$ brane $a$
and a $U(1)$ brane $b$, which
intersect such that we get the $\fivebar$ representation at the intersection.
The $\ten$ is
realised as the antisymmetric representation of the $U(5)$ brane.
To get reasonable models,
we have to require that the number of antisymmetric representations is equal
to the number of $\fivebar$ representations,
\be{eq_su5constr}
  I_{ab} = -\#\Anti_a.
\ee

In a pure $SU(5)$ model one should also include a restriction to configurations
with $\#\Sym_a=0$ to exclude
$\fifteen$ representations from the beginning.
Since it has been proven in~\cite{cps02} that in this
case no three generation models can be constructed and symmetric
representations might also
be interesting from a phenomenological point of view,
we include these in our discussion.

The flipped $SU(5)$ case is a bit more involved since in addition to the
constraints of the $SU(5)$
case one has to make sure that the $U(1)_X$ stays massless and the $\fivebar$
and $\ten$ have the
right charges, summarised in (\ref{eq_su5xembed}).
To achieve this, at least one additional brane $c$ is needed. Generically, the
$U(1)_X$ can be constructed as a combination of all $U(1)$s present in the model
\be{eq_u1constrgen}
  U(1)_X=\sum_{a=1}^k x_a U(1)_a.
\ee
The simplest way to construct a combination which gives the right charges would be
\be{eq_u1constr}
  U(1)_X=\frac{1}{2}U(1)_a - \frac{5}{2}U(1)_b + \frac{5}{2}U(1)_c,
\ee
but a deeper analysis shows~\cite{st06}, that this is in almost all cases
not enough to ensure that the hypercharge remains massless.
The condition for this can be formulated as
\be{eq_u1massless}
  \sum_{a=1}^k x_aN_a\vec{Y}_a = 0,
\ee
with the coefficients $x_a$ from (\ref{eq_u1constr}). To fulfill this
requirement we need generically one or more additional $U(1)$ factors.

\fig{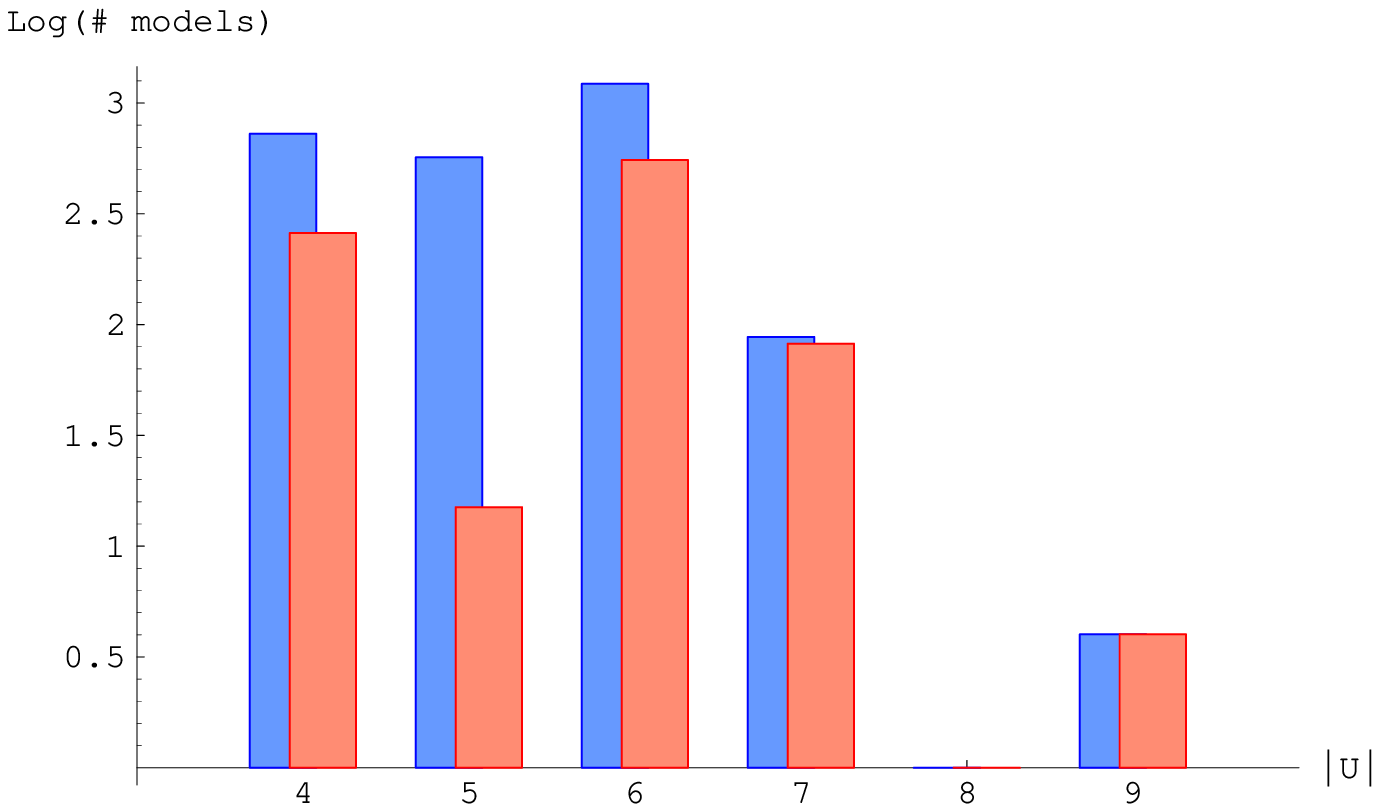}{fig_numsol_su5}{%
Logarithmic plot of the number of solutions with an $SU(5)$ factor depending
on the absolute value of the parameters $U$. We give the results with
(blue bars to the left) and without (red bars to the right)
symmetric representations of $SU(5)$.
}

\subsubsection{General results}
Having specified the additional constraints, we use the techniques described
in section~\ref{ls_comp} to
generate as many solutions to the tadpole, supersymmetry and K-theory
conditions as possible.
The requirement of a specific set of branes to generate the $SU(5)$ or
flipped $SU(5)$ simplifies the
computation and gives us the possibility to explore a larger part of the
moduli space as compared to the more general analysis we described above.

Before doing an analysis of the gauge sector properties of the models under
consideration, we would like to check if the number of solutions decreases
exponentially for large values
of the $U_I$, as we observed in section~\ref{ls_dist} for the
general solutions.
In figure~\ref{fig_numsol_su5}
the number of solutions with and without
symmetric representations are shown.
The scaling holds in our present case as well, although
the result is a bit obscured by the much smaller statistics.
In total we found 2590 solutions without restrictions on the
number of generations and the presence of symmetric representations.
Excluding these representations reduces the number of solutions to 914.
Looking at the flipped $SU(5)$ models, we found 2600 with and 448 without
symmetric representations.
Demanding the absence of symmetric representations is obviously a much severer
constraint in the flipped case.

\twofig{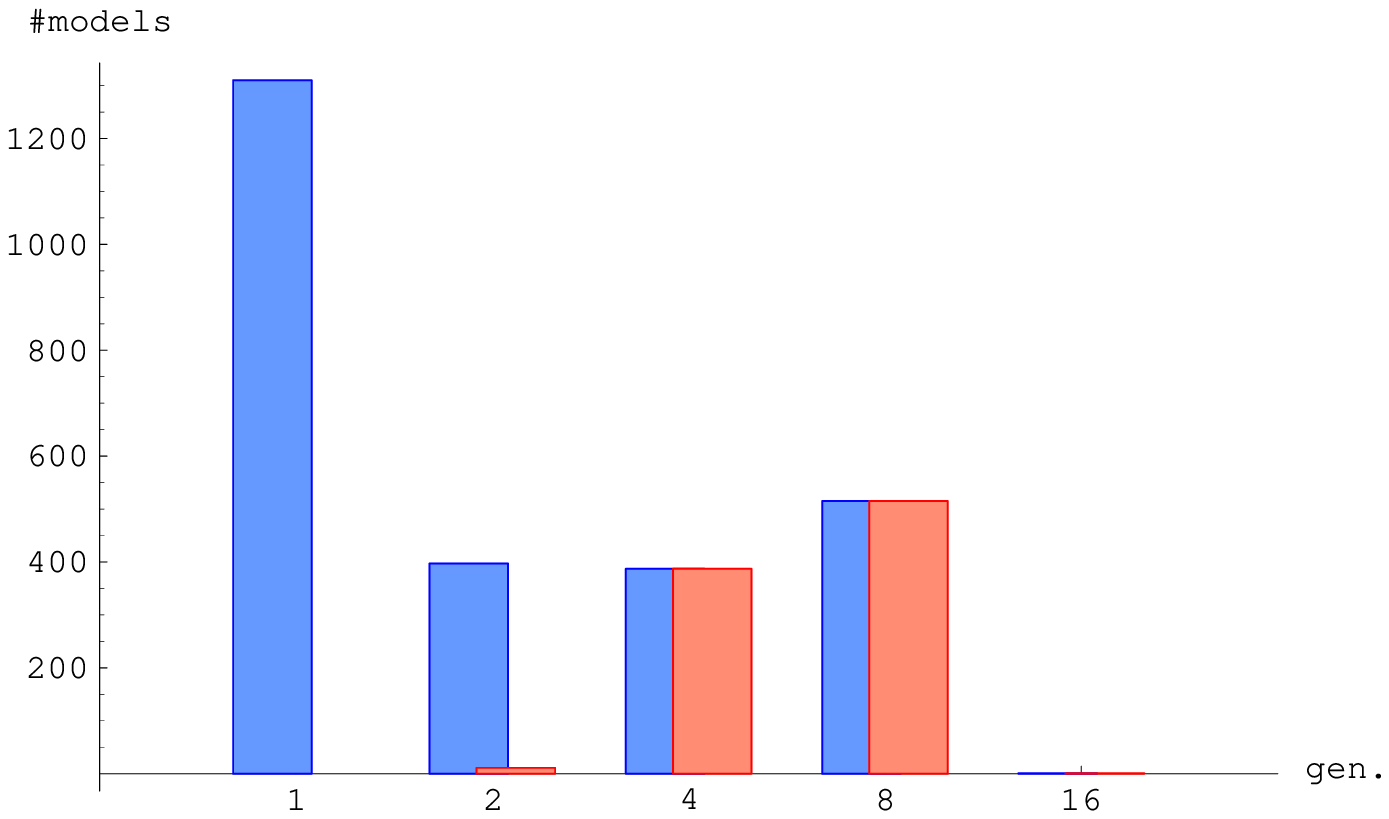}{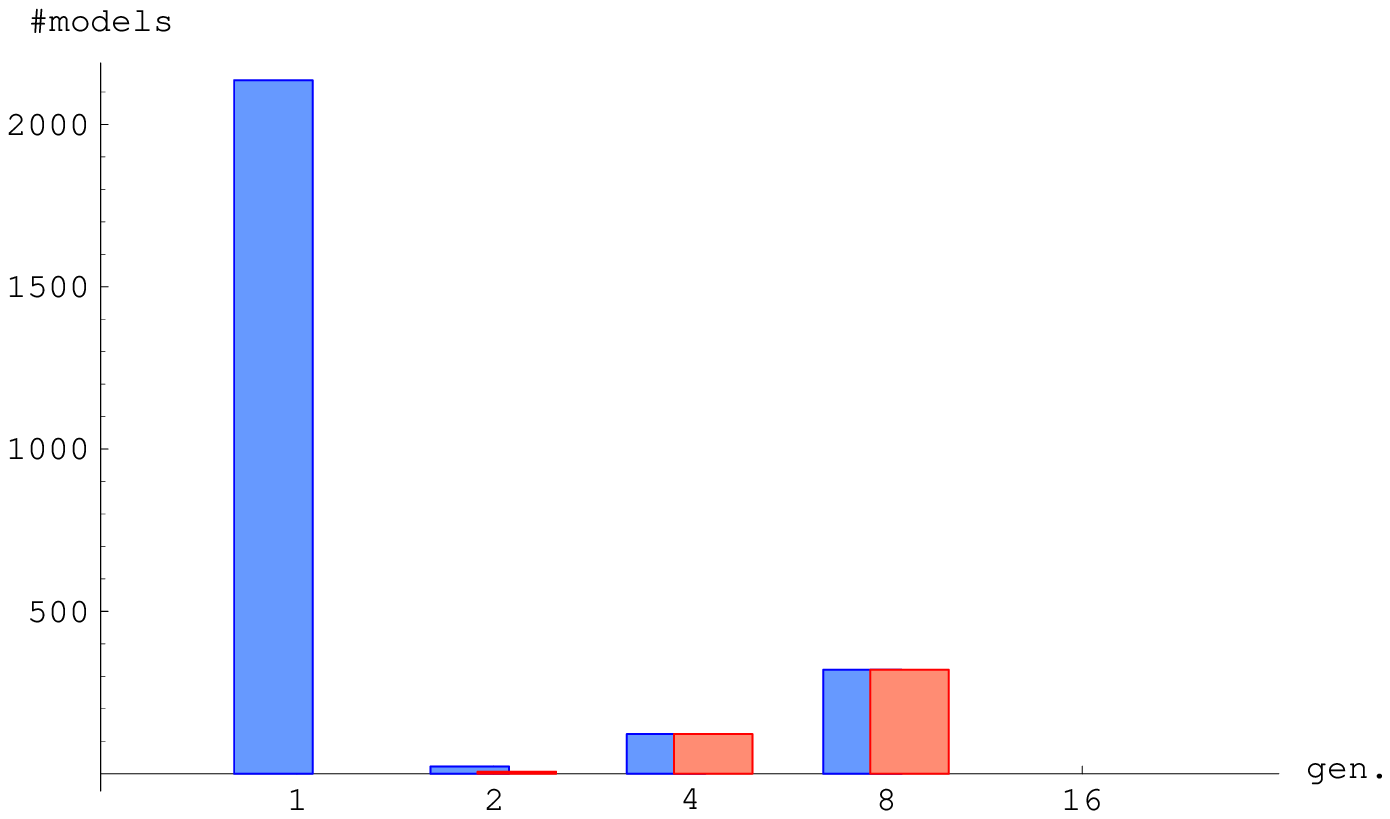}{fig_gen_su5}{%
Plots of the number of solutions for different numbers of generations for
(a) $SU(5)$ and (b) flipped $SU(5)$ models with (blue bars to the left)
and without (red bars to the right) symmetric representations of $SU(5)$.%
}

The correct number of generations turned out to be the strongest constraint
on the statistics in our previous work on standard model constructions.
The $SU(5)$ case is not different in this aspect.
In figure~\ref{fig_gen_su5}
we show the number of solutions for different numbers
of generations.
We did not find any solutions with three $\fivebar$ and $\ten$
representations. This
situation is very similar to the one we encountered in our previous analysis
of models with a standard model gauge group in section \ref{ls_sm}.
An analysis of the
models which have been explicitly constructed showed that they exist only for
very large values of the complex structure parameters. The same is true in
the present case. Because the number of models decreases rapidly for higher
values of the parameters, we can draw the conclusion that these models are
statistically heavily suppressed.

Comparing the standard and the flipped $SU(5)$ construction
the result for models with one generation might be surprising, since there are
more one generation models in the flipped than in the standard case.
This is due to the fact that there are generically different possibilities to
realise the additional $U(1)_X$ factor for one geometrical setup, which we
counted as distinct models.

\twofig{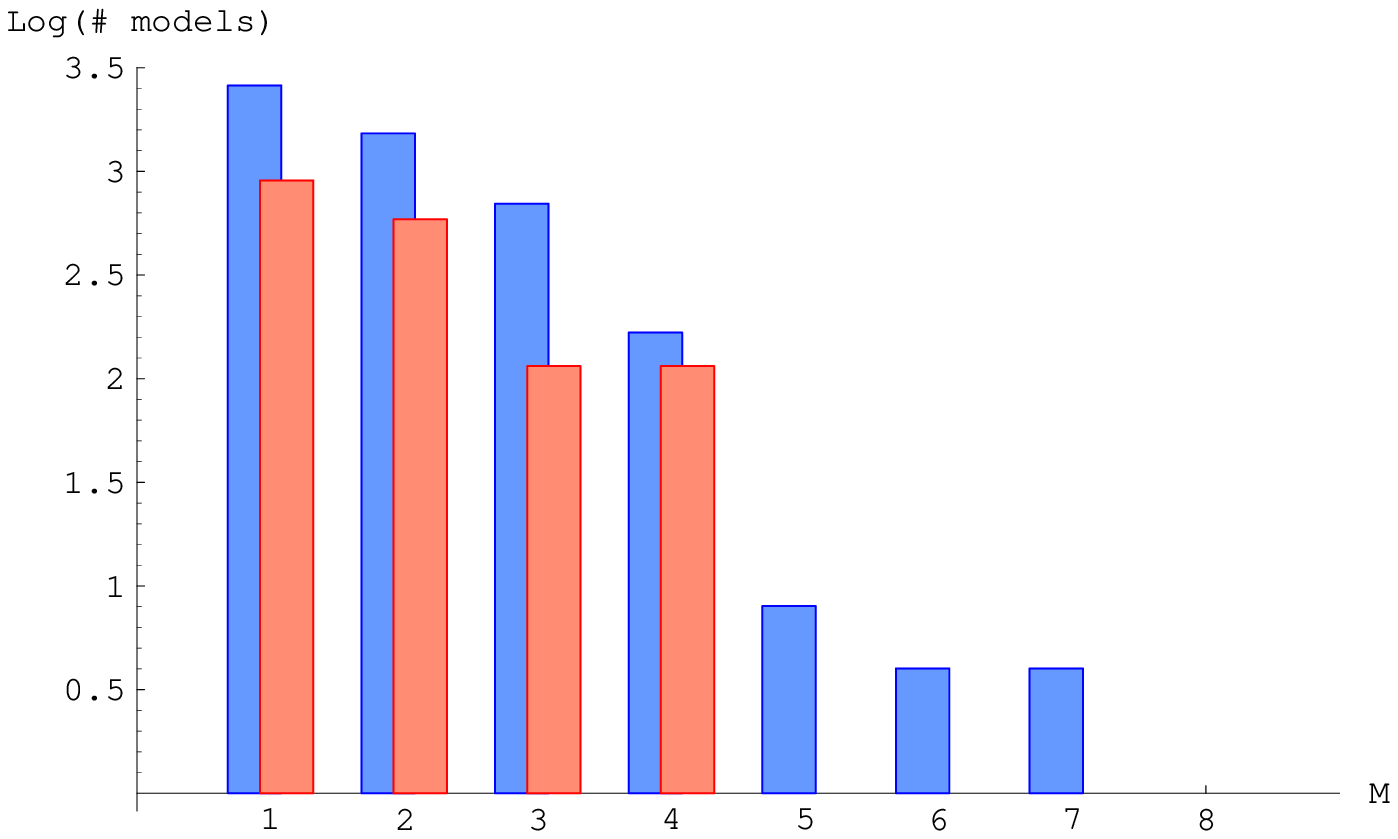}{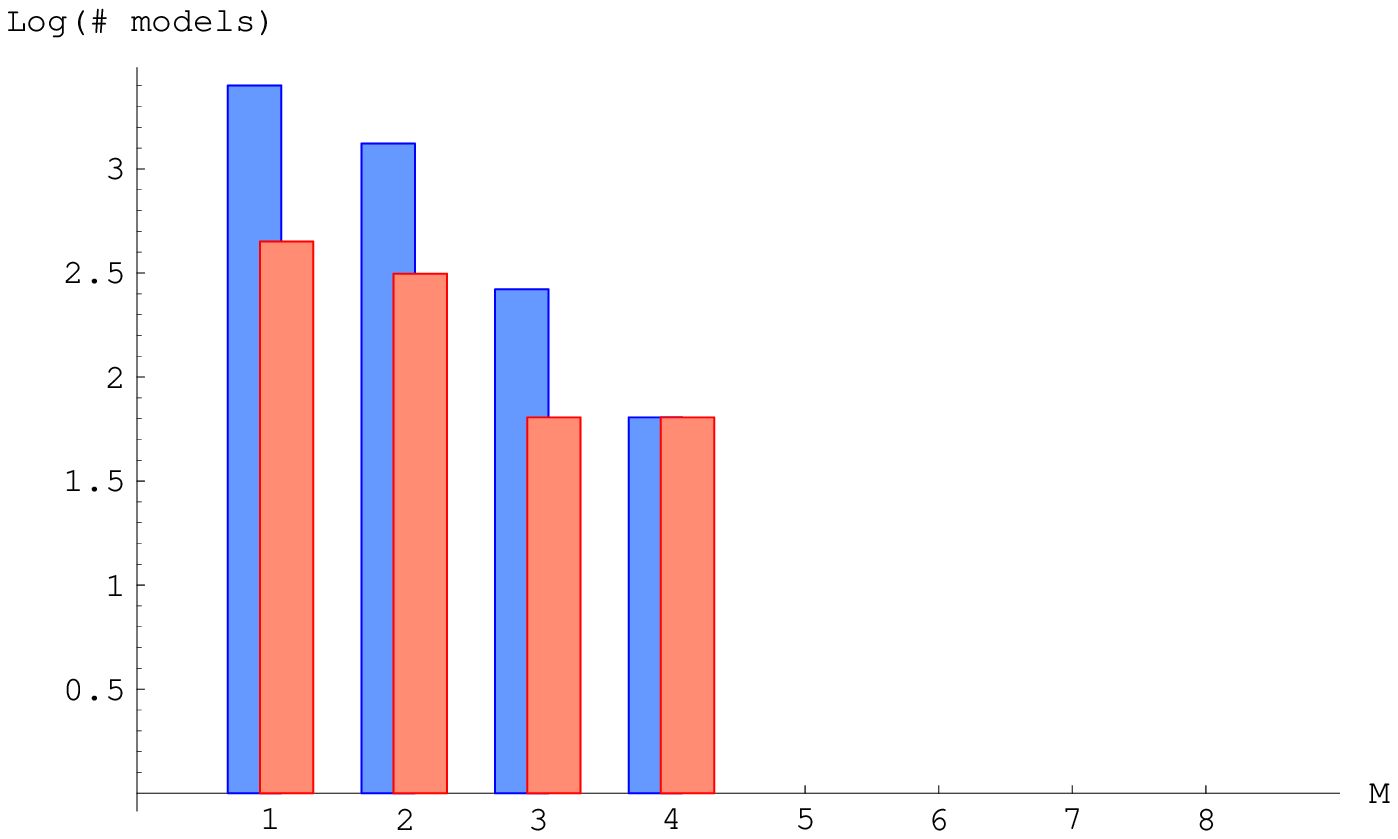}{fig_hum_su5}{%
Logarithmic plots of the number of solutions with a specific rank $M$ gauge
factor in the hidden sector in (a) $SU(5)$ and (b)
flipped $SU(5)$ models with (blue bars to the left) and without (red bars to
the right) symmetric representations of $SU(5)$.%
}
Regarding the hidden sector, we found in total only four $SU(5)$ models which
did not have a hidden sector at all - one with 4, two with 8 and one with 16
generations. For the flipped $SU(5)$ case such a model cannot exist, because
it is not possible to solve the condition for a massless $U(1)_X$ without
hidden sector gauge fields.

The frequency distribution of properties of the hidden sector gauge group,
the probability to find a gauge group of specific rank $M$ and the
distribution of the total rank, are shown in
figures~\ref{fig_hum_su5} and~\ref{fig_hrk_su5}.
The distribution for individual gauge factors is qualitatively very similar
to the one
obtained for all possible solutions above (see figures~\ref{fig_rkundist}).
One remarkable difference between standard and flipped $SU(5)$ models
is the lower
probability for higher rank gauge groups. This is due to the above mentioned
necessity to have a sufficient number of hidden branes for the construction
of a massless $U(1)_X$.

\twofig{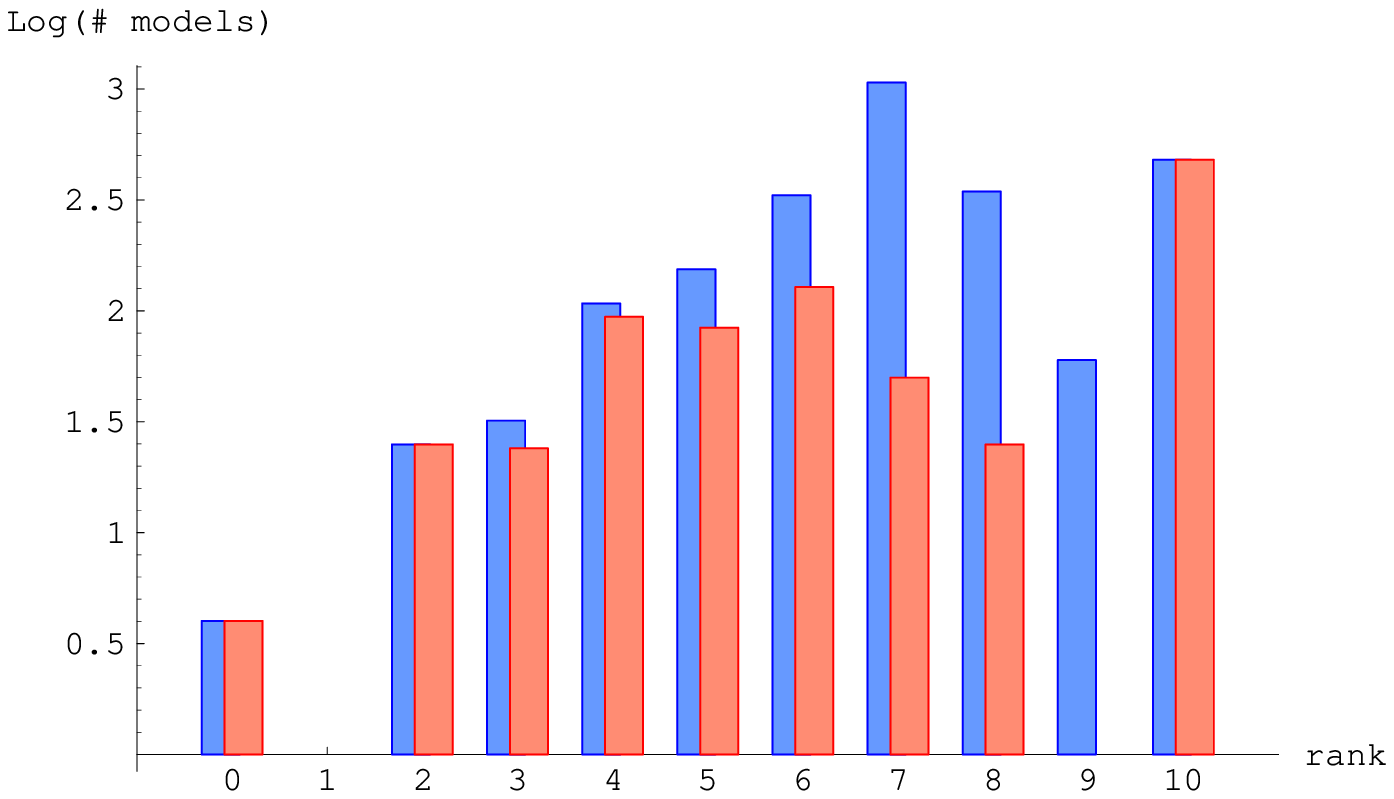}{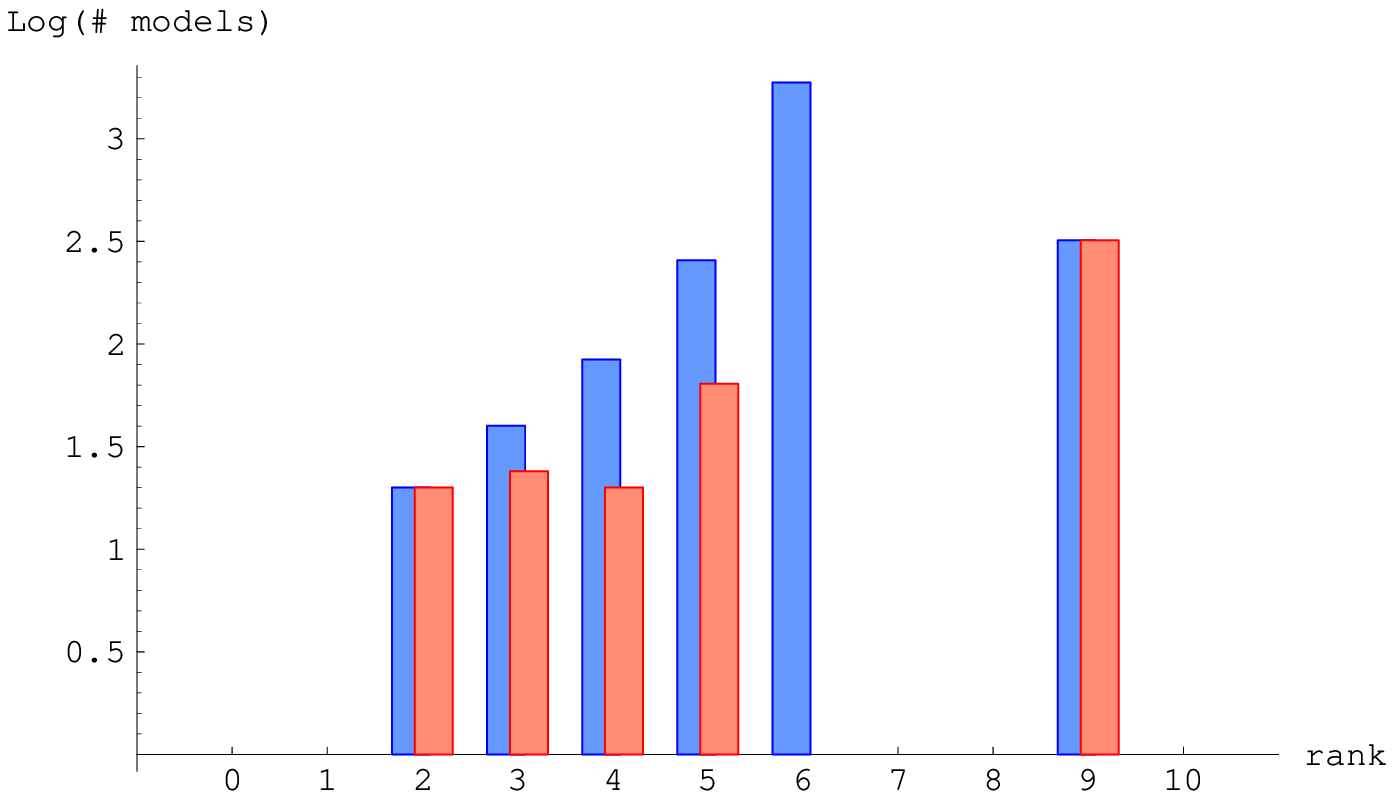}{fig_hrk_su5}{%
Plots of the number of solutions for given values of the total rank of the
hidden sector gauge group in (a) $SU(5)$ and (b) flipped
$SU(5)$ models with (blue bars to the left) and without (red bars to the right)
symmetric representations of $SU(5)$.%
}
The total rank distribution for both, the standard and the flipped version,
differs in one aspect from the one obtained in~\ref{ls_dist},
namely in the large fraction of hidden sector groups with a total
rank of 10 or 9, respectively.
This can be explained by just one specific construction which is possible for
various values of the complex structure parameters.
In this setup the hidden sector
branes are all except one on top the orientifold planes on all three tori.
If we exclude this specific feature of the $SU(5)$ construction, the remaining
distribution shows the behaviour estimated from the prior results.

Note that while comparing the distributions one has to take
into account that
the total rank of the hidden sector gauge group in the $SU(5)$ case is
lowered by the contribution from the visible sector branes to
the tadpole cancellation conditions. In the flipped case, the additional
$U(1)$-brane contributes as well.

\subsubsection{Restriction to three branes in the hidden sector}%
\label{ls_su53stack}
In order to compare our results for the statistics of constructions with a
standard model-like gauge group with Gepner models in section~\ref{ls_gepner},
we truncated the full set of models to those with only three stacks of branes
in the hidden sector.
In the following we also perform a restriction to a maximum of three
branes in the hidden sector in the $SU(5)$ case, but with a different
motivation and in a different way. We do not truncate our original results,
but instead impose the constraint to a maximum of three branes from the very
beginning in the computational process.
It turns out that such a restriction can greatly improve the
performance of the partition algorithm and allows us therefore to analyse a
much bigger range of complex structures. This is highly desirable, since it
opens up the possibility to check some claims about the growths of solutions
that we made in section~\ref{ls_nosol}. The method has also some
drawbacks. Since we do not compute the full distribution of models, but with
an artificial cutoff, we can not be sure that the frequency distributions of
properties in the gauge sector are the same as in the full set of models.
As we will see in the following, there are indeed some deviations.

\fig{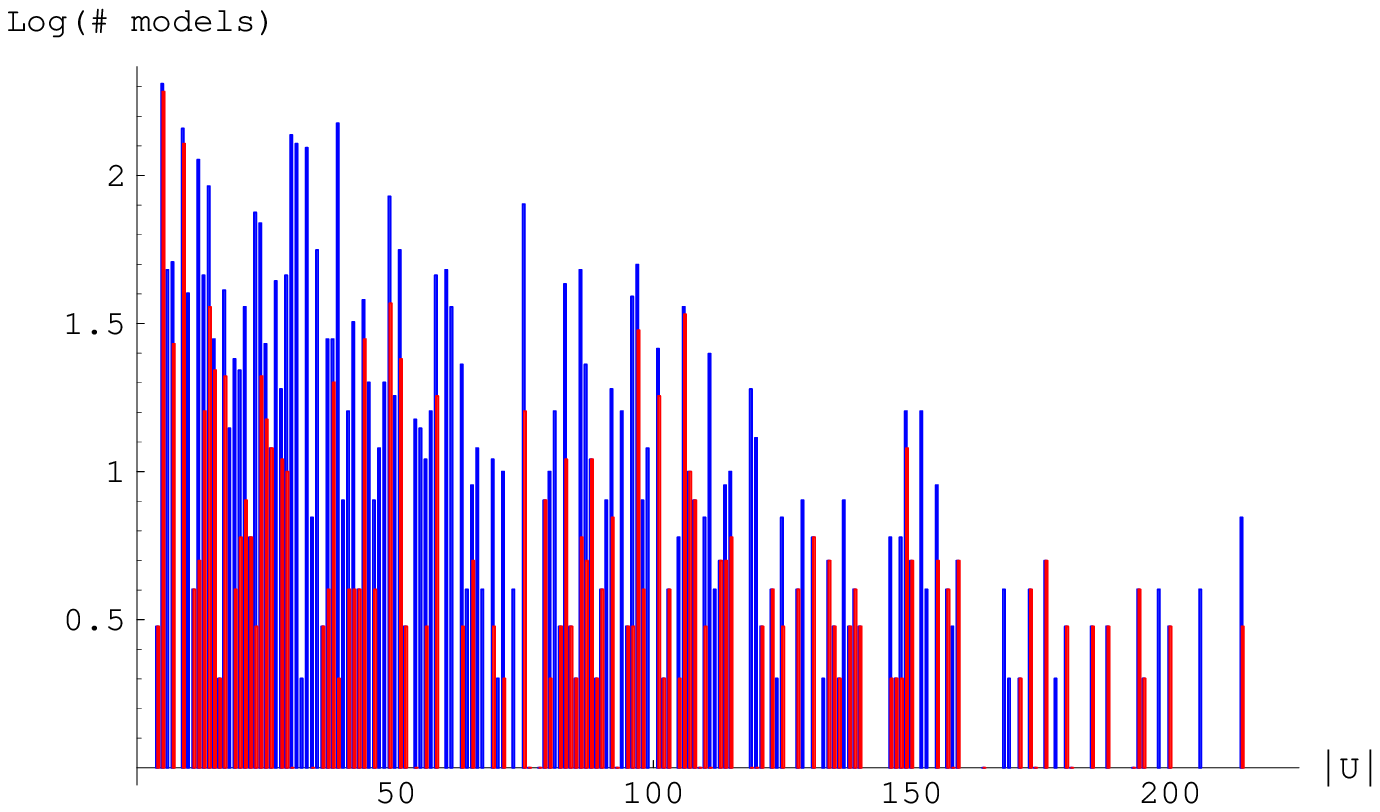}{fig_su5_mh3_nosol}{%
Logarithmic plot of the number of solutions with an $SU(5)$ factor depending
on the absolute value of the parameters $U$. The number of brane stacks in the
hidden sector is restricted to three and the results are shown for models with
(blue spikes) and without (red spikes) symmetric representations of $SU(5)$.
}

In figure~\ref{fig_su5_mh3_nosol} we plotted the total number of models with
a maximum of three stacks of branes in the hidden sector.
As in our analysis above we
show the models without symmetric representations separately.
This plot should be compared with figure~\ref{fig_numsol_su5}, the number
of solutions for $SU(5)$ models without restrictions.
In the restricted case we were able to compute up to much higher values of the
complex structures and confirm the assertion of~\ref{ls_nosol}, that the
number of solution drops exponentially with $|U|$. This provides another hint
that the total number of solutions is indeed finite.
In total we found 3275 solutions, which is more then in the case without
restrictions, but in contrast to a range of complex structures which is 25
times bigger, the amount of additional solutions is comparably small.

\twofig{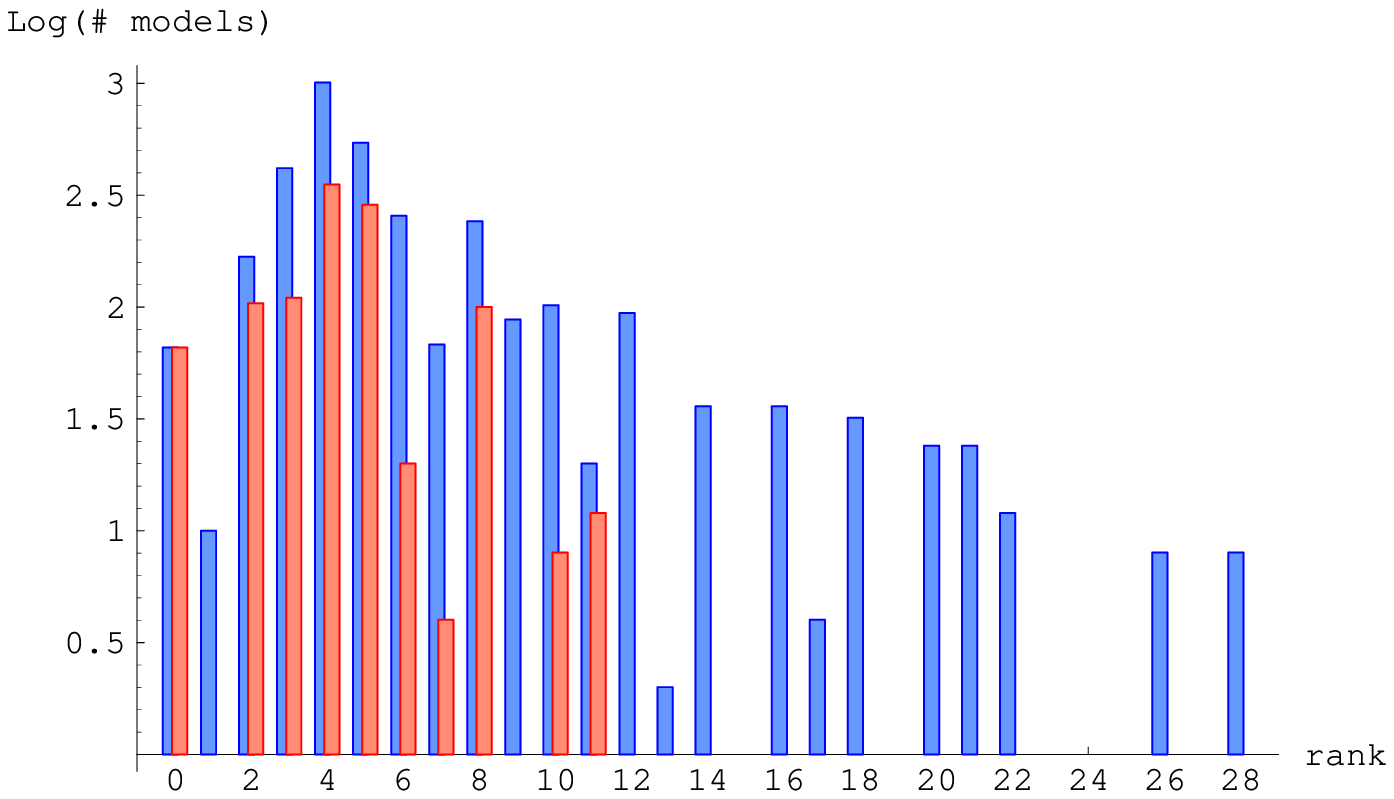}{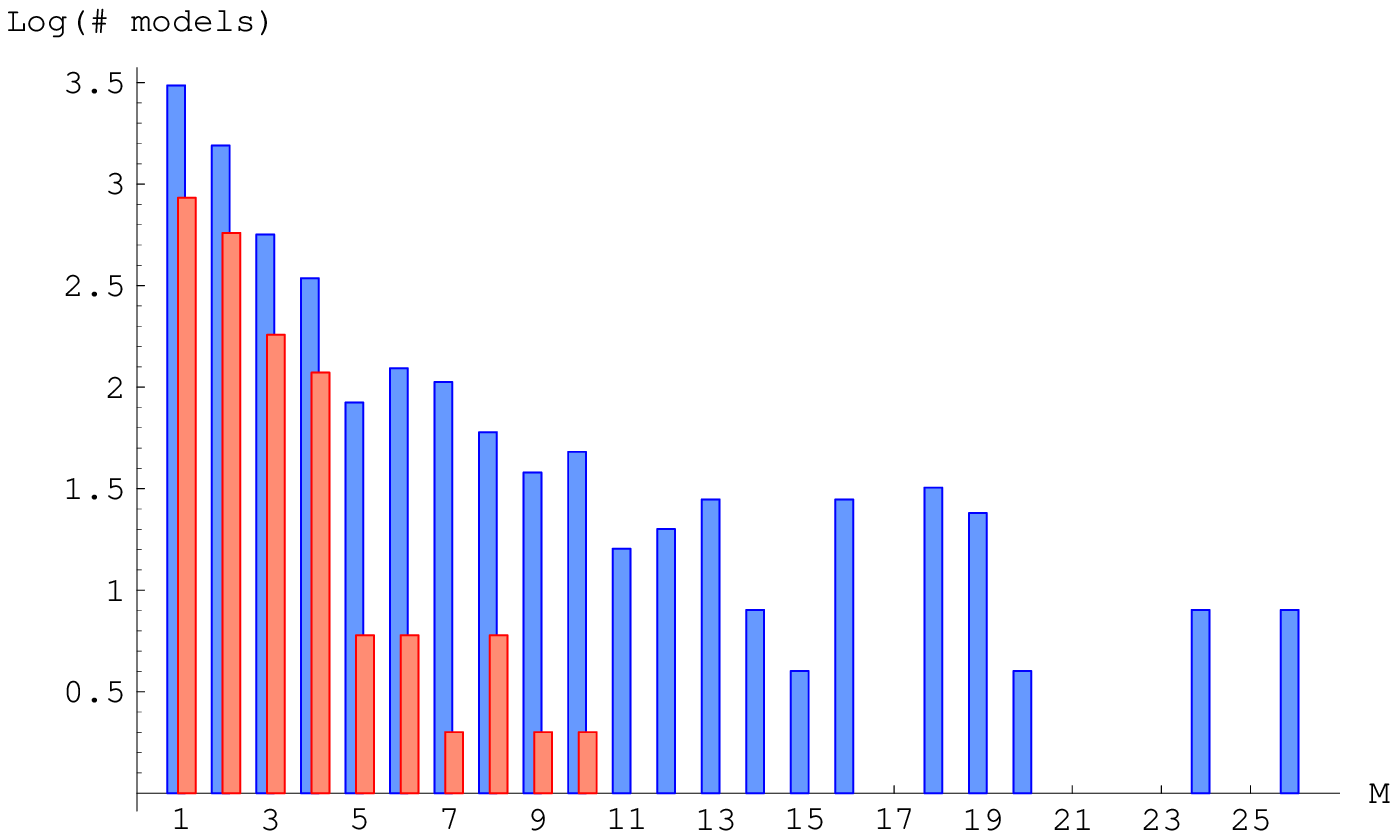}{fig_su5_mh3_hrkum}{%
Logarithmic plots of the frequency distributions in the hidden sector of
$SU(5)$ models with a maximum of three hidden branes.
(a) Specific rank $M$ gauge factors, (b) Total rank of the hidden sector
gauge group.%
}

Comparing the distributions for individual gauge factors
(figure~\ref{fig_su5_mh3_hrkum_a}) and the total rank in the hidden sector
(figure~\ref{fig_su5_mh3_hrkum_b}), we see some interesting differences to
figures~\ref{fig_hum_su5_a} and~\ref{fig_hrk_su5_a}. The distribution
of individual gauge factors is just extended to higher factors in the
restricted case. This was to be expected, since larger values for the complex
structure parameters allow for larger gauge factors to occur, since they
provide us with very long branes with negative wrapping numbers $X$ that can
compensate these large numbers in the tadpole cancellation conditions. The
general shape of the distribution remains unchanged. In the case of the total
rank the situation is different. The distribution also shows larger values for
the total rank, which is directly correlated to the larger individual ranks
of the factors, but moreover the maximum of the distribution is shifted from
around seven in the unrestricted case to about four. This can be explained
by the fact that the restriction to a maximum of three branes in the hidden
sector also restricts the possible contributions from models with many gauge
factors of small rank, especially the contribution of $U(1)$ gauge factors.

What about models with a flipped $SU(5)$ gauge group? Repeating the analysis
for these models in the case of a restriction in the hidden sector can of
course be done, be the results might not be very predictive. For a consistent
flipped $SU(5)$ model, we need a massless $U(1)_X$, which also depends on a
combination of $U(1)$ factors from the hidden sector. After choosing an
additional $U(1)$ brane for the visible sector of flipped $SU(5)$ there
remain only two hidden sector branes. This restriction is too drastic to
give meaningful results, since it turned out in the analysis of flipped
$SU(5)$ models that we need more than two hidden sector branes to solve the
equations for the $U(1)_X$ to be massless.

The analysis in this section showed that three generation models with a
minimal grand unified gauge group
are heavily suppressed in this specific orientifold setup.
This result was expected, since we know that the
explicit construction of three generation $SU(5)$ models using the $\bZZ$
orbifold has turned out to be difficult.

The analysis of the hidden sector showed that the frequency distributions of
the total rank of the
gauge group and of single gauge group factors are quite similar to the results
for generic models in section~\ref{ls_dist}.
Differences in the qualitative picture result from specific
effects in the $SU(5)$ construction.

Comparing the results for the standard and flipped $SU(5)$ models, we find no
significant differences.
If we allow for symmetric representations, there is basically no additional
suppression factor. If we restrict ourselves to models without these
representations, flipped constructions are three times less likely then the
standard ones.

\subsection{Correlations}\label{ls_corr}
An interesting question that we raised in the introduction concerns the
correlation of observables.
If different properties of our models were correlated,
independently of the specific visible gauge group, this would provide us
with some information about the generic behaviour of this class of models.
In the following discussion we would like to clarify this point, emphasizing
a crucial difference between correlations
of phenomenologically interesting observables in the gauge sector of our
models on the one hand,
and correlations between basic properties used as constraints to characterize
a specific visible sector on the other hand. Finally we use the
observations on the second class of correlations to estimate the number of
models with a standard model gauge group and three generations of quarks
and leptons for the $T^6/\bZZ$ orientifold.

\fourfigclip{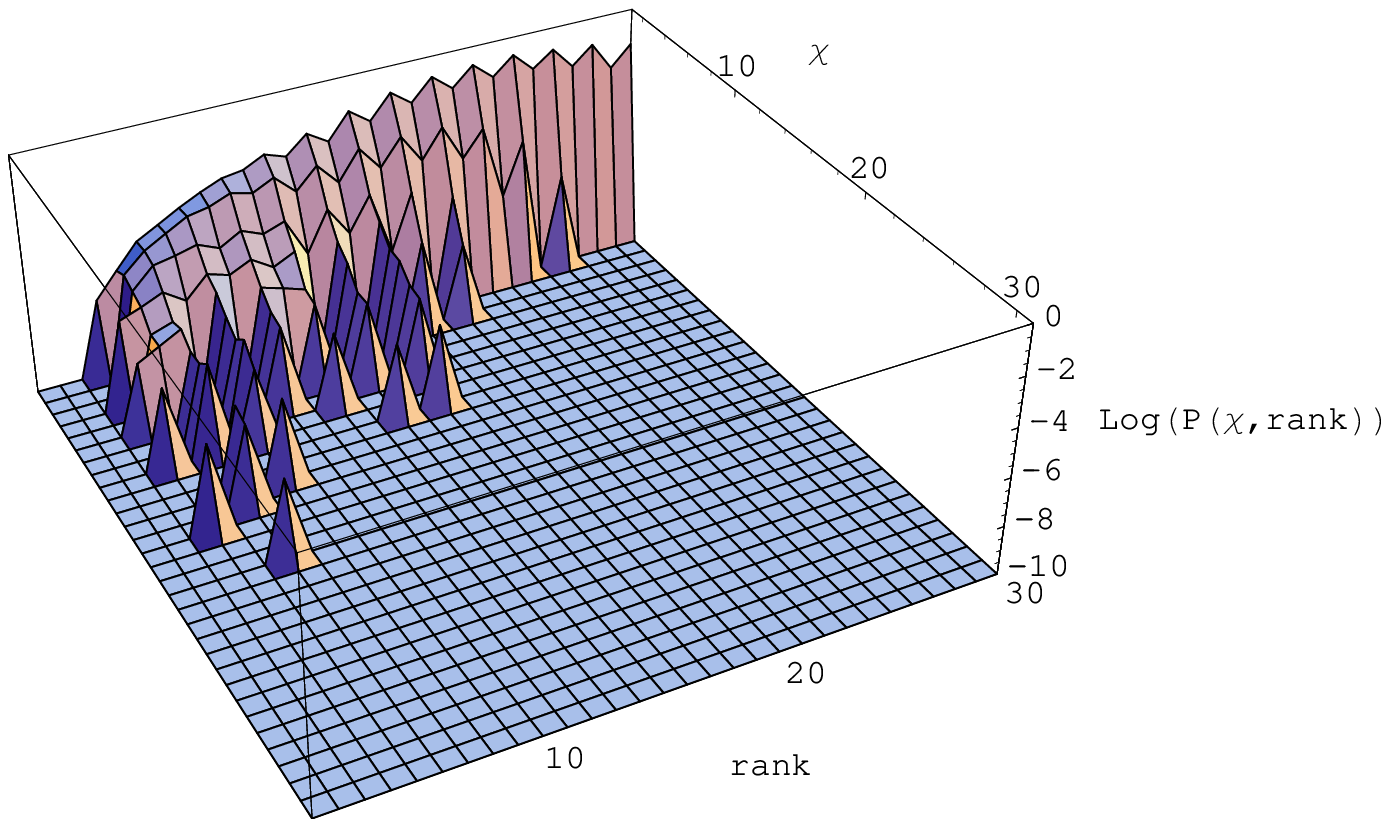}{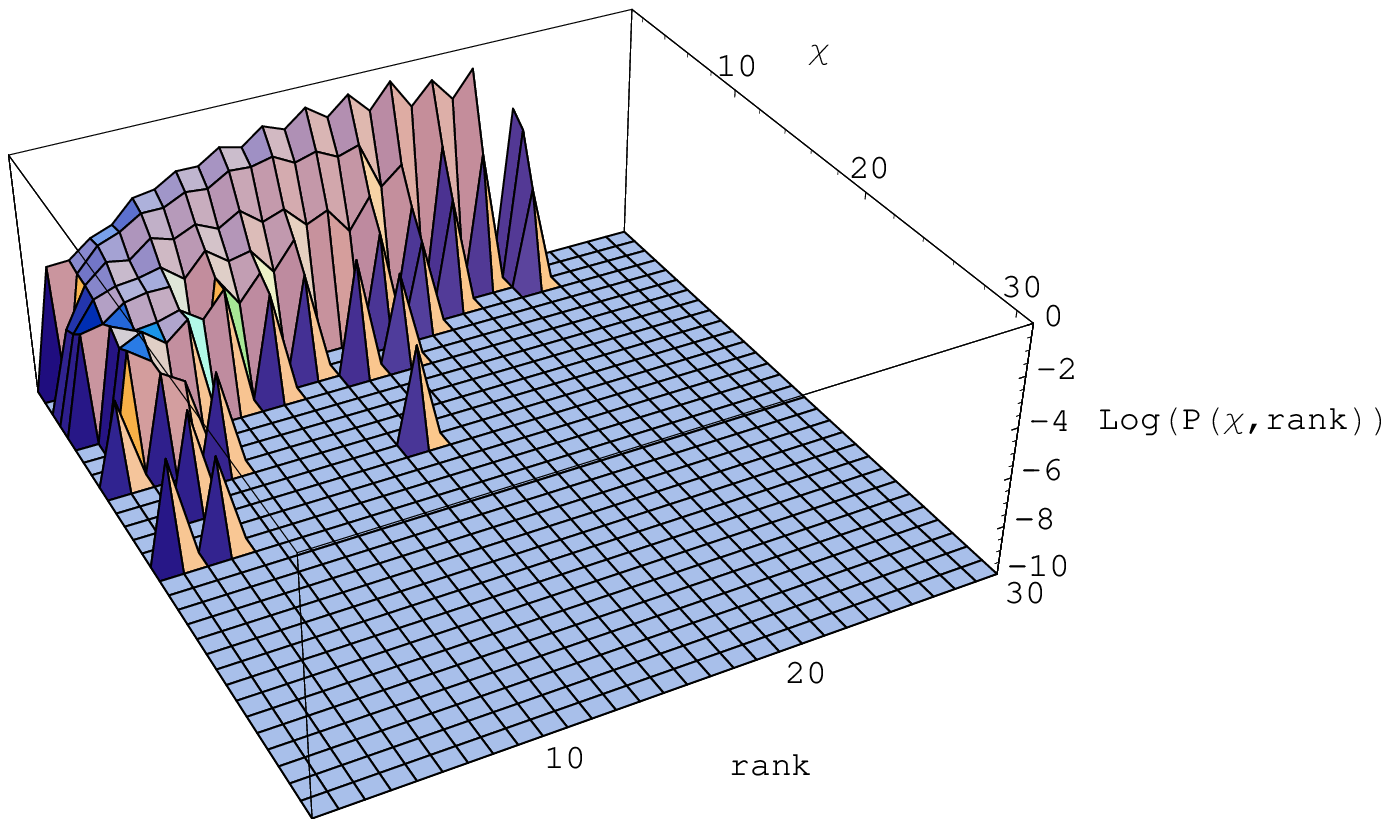}{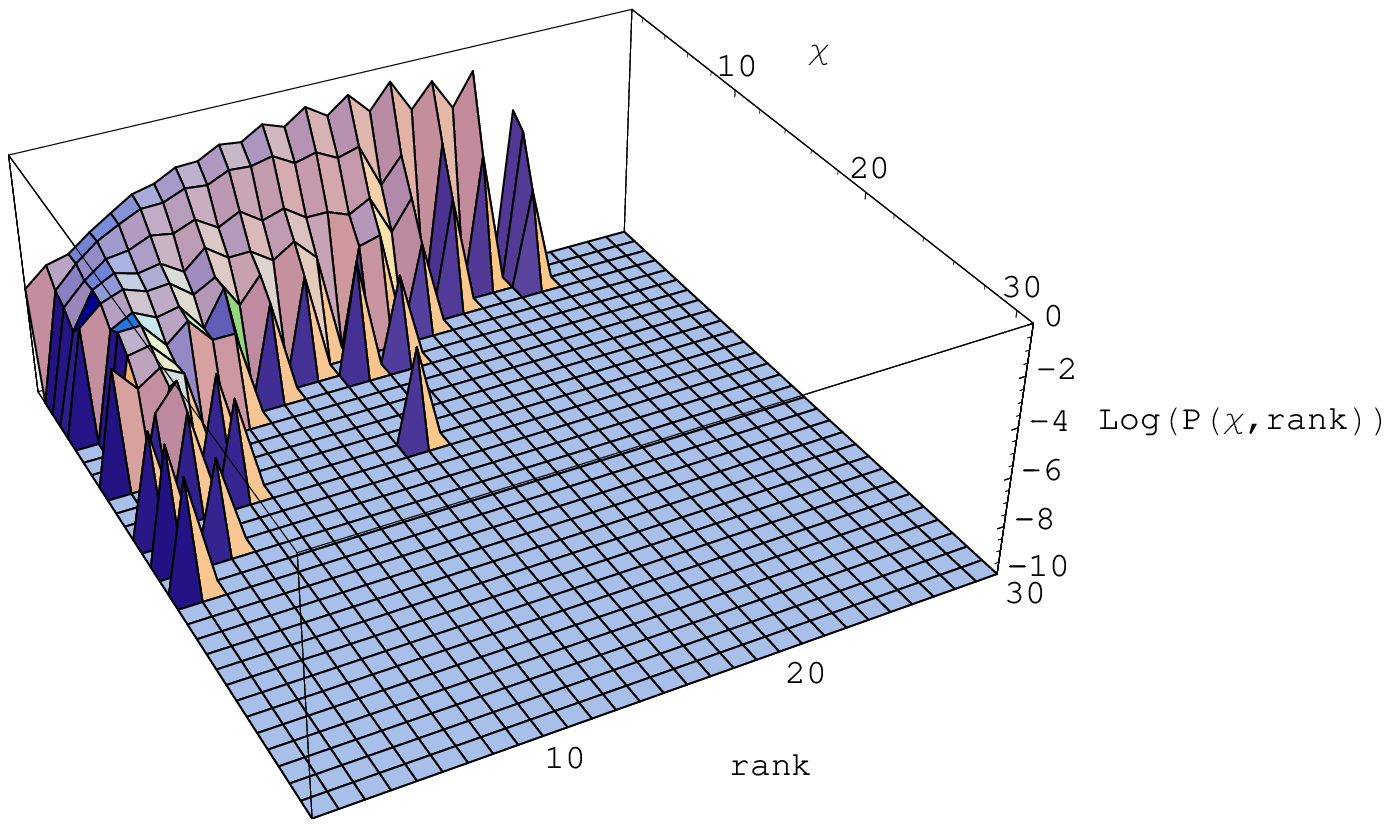}{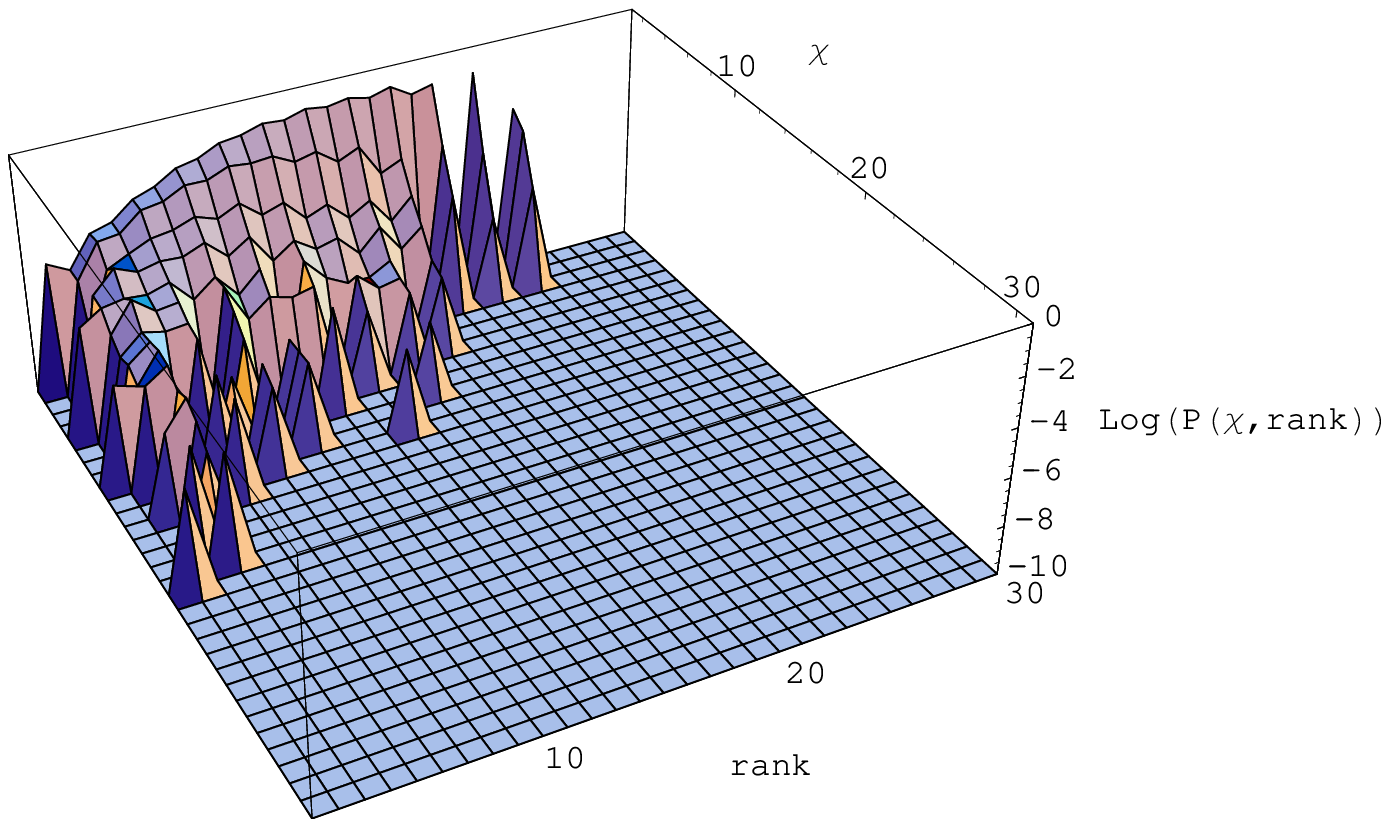}{fig_chivsrk}{%
Logarithmic plots of the relative frequency distributions of models with
specific total rank of the gauge group and mean chirality. Plot (a) shows the
analysis for the full gauge group of all models, figures (b), (c) and (d) give
the results for the hidden sector gauge groups of standard model-like
constructions with and without a massive hypercharge and Pati-Salam models,
respectively.%
}

\subsubsection{Rank and chirality}\label{ls_chivsrk}
To give an example of correlations between gauge group observables let us
consider the mean chirality $\chi$, defined by~(\ref{eq_4dchi}),
and the total rank of the gauge group.
As we already saw using the saddle point
approximation on $T^4$ in section~\ref{ls_6dcorr}, these two quantities
should be correlated. To confirm this in the four-dimensional case, we
use our explicit results and compute the frequency distributions for
the different visible sectors considered above, standard model-like
constructions with and
without a massless hypercharge and Pati-Salam models.
The result is shown in figure~\ref{fig_chivsrk}.
Please note that we have normalised the distributions in order to make
the results better comparable.

We find two striking results here, which illustrate the two points we made
in the introduction to this section. Firstly the two observables are
clearly correlated, a large value for the mean chirality is much more
likely to find if the total rank is small. Secondly the results for the
full set of models, figure~\ref{fig_chivsrk_a}, and the different visible
sectors, figures~\ref{fig_chivsrk_b}, (c) and~(d),
show qualitatively very similar results.
This last observation is intriguing, since we might use this to conjecture
that the specific properties used to define an individual visible sector
do not influence the distributions. Put differently, we might speculate that
these properties could be regarded independent of each other.
If this would be indeed the case, it could simplify some specific analysis
dramatically. Instead of constructing solutions for one specific setup with
some set of properties it would be enough to know the probabilities for
each property. Since they would be independent of each other we could just
multiply the results and get an answer to our more difficult question.

\subsubsection{Estimates}\label{ls_smestimate}
We would like to test this conjecture using the properties of a standard model
construction. These include several constraints on the models, in particular
the existence of specific $U(N)$ gauge factors, the vanishing of antisymmetric
representations, a massless hypercharge and
three generations of chiral matter.
How can we check whether two of these properties $A$ and $B$,
are independent? A good measure for this would be
to calculate the correlation between the probabilities $P(A)$ and $P(B)$
to find these properties. This can be expressed as
\be{eq_pabcorr}
  P_{AB}=\frac{P(A) P(B)-P(A\wedge B)}{P(A) P(B) + P(A\wedge B)},
\ee
where $P(A\wedge B)$ is the probability to find both properties realised at
the same time.

\twofig{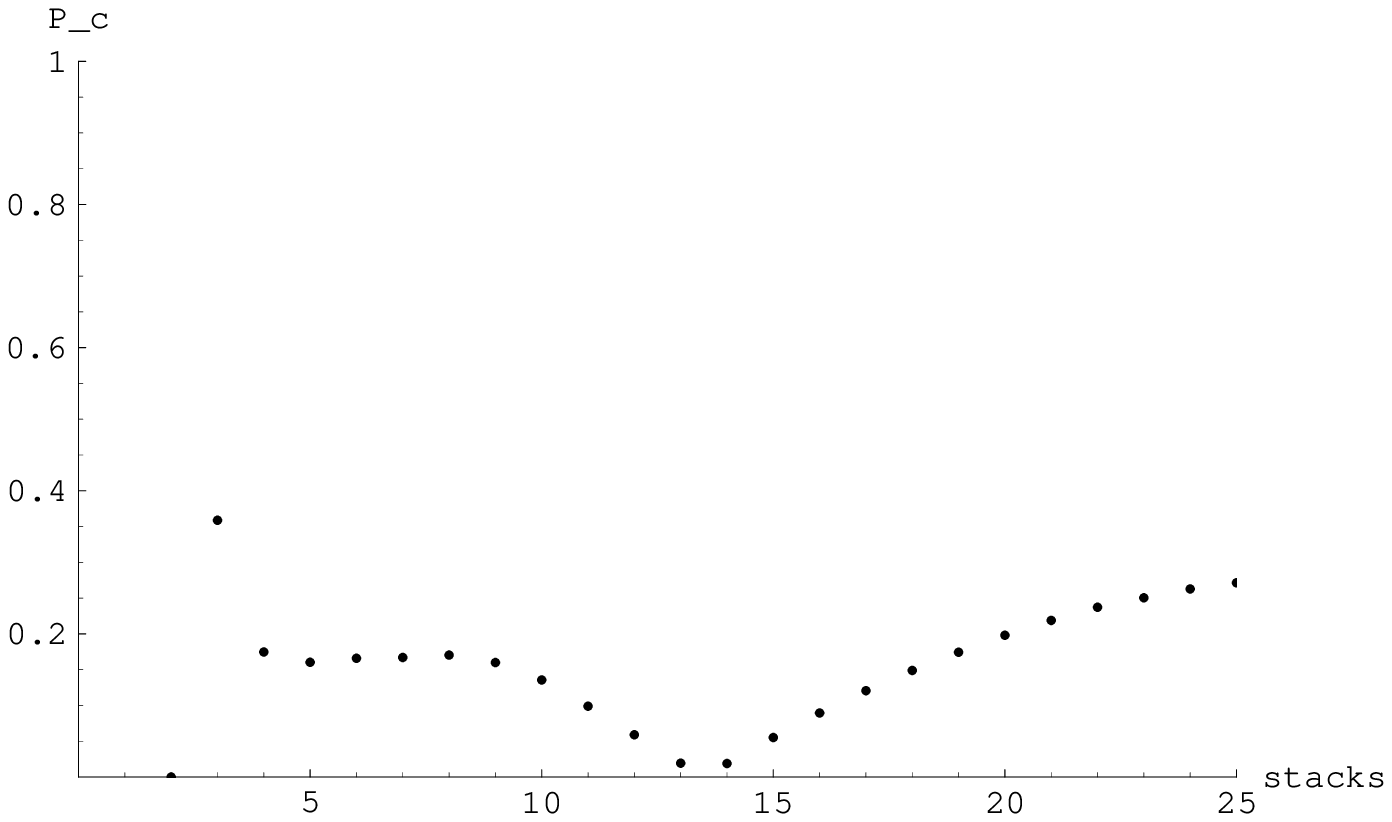}{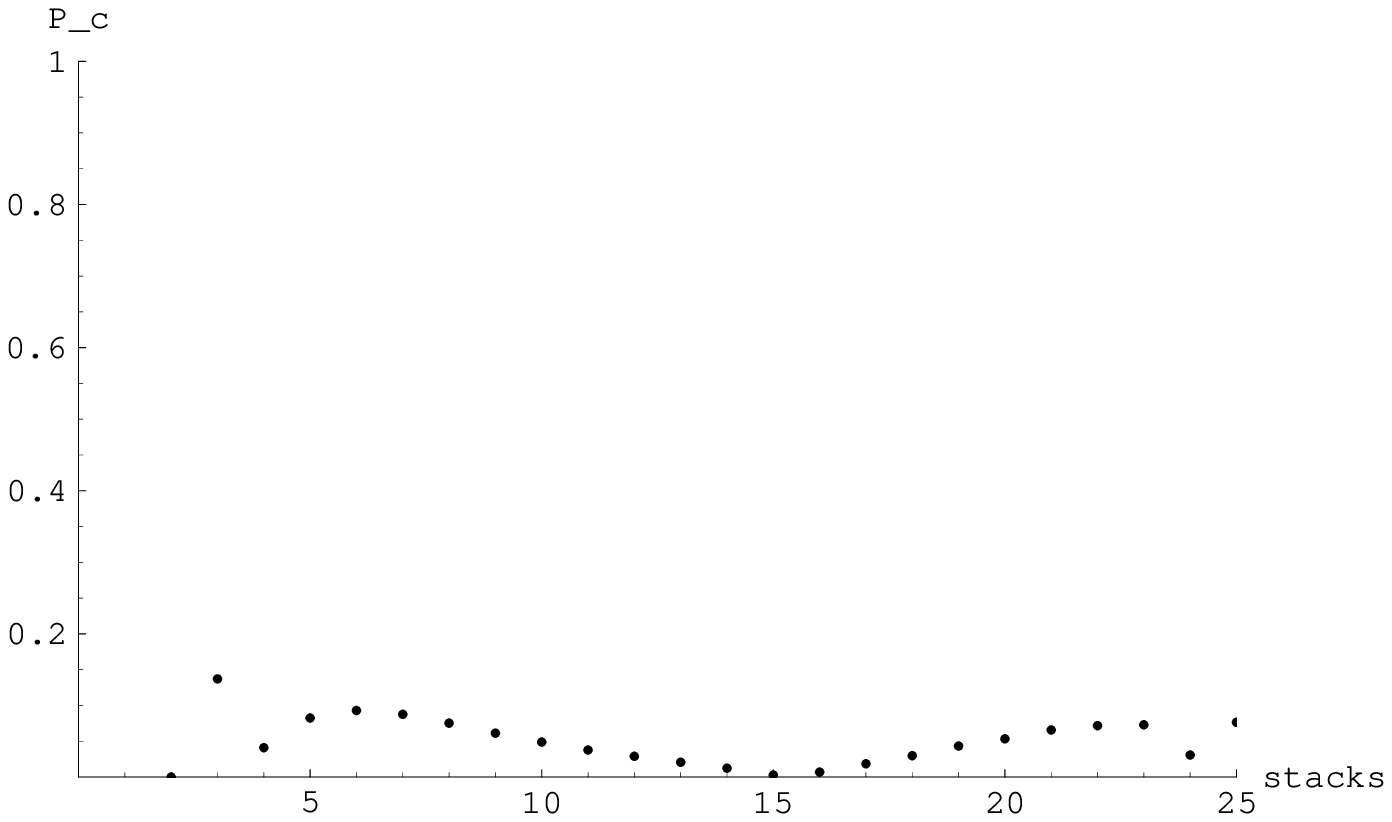}{fig_corr}{%
Correlations between properties of standard model-like configurations.
(a) Correlation between the existence of an $SU(3)$ and an $SU(2)$ or $Sp(2)$
gauge group. (b) Correlation between the existence of an $SU(3)$ gauge group
and the absence of symmetric representations.%
}

For concreteness let us take the following properties as examples:
The existence of a $U(3)$
gauge group, existence of a $U(2)$ or $Sp(2)$ gauge group and the vanishing
of antisymmetric representations. In figure~\ref{fig_corr} we plotted the
value of $P_{AB}$ in the set of all models for different values of the
number of stacks. As can be derived from these plots the two properties are
not really independent, but values of about $0.1$ and $0.2$, respectively,
which are also the order of magnitude for other possible correlations,
suggest that one could give it a try and treat these properties as
independent in an estimate\footnote{Note that the independence of different
properties have been an assumption that was used in the original work on
vacuum statistics~\cite{do03}.}.

\begin{table}[htb]
\begin{center}
\begin{tabular}{|l|r|}\hline
Restriction                    & Factor\\\hline
gauge factor $U(3)$            & $0.0816$\\
gauge factor $U(2)/Sp(2)$      & $0.992$\\
No symmetric representations   & $0.839$\\
Massless $U(1)_Y$	       & $0.423$\\
Three generations of quarks    & $2.92\times10^{-5}$\\
Three generations of leptons   & $1.62\times10^{-3}$\\\hline
\emph{Total}                   & $1.3\times10^{-9}$\\\hline
\end{tabular}
\caption{Suppression factors for various constraints of standard model
properties.}
\label{tab_corr}
\end{center}
\end{table}

In table~\ref{tab_corr} we summarised the properties of a three-generation
standard model, including the suppression factor calculated using the
probability to find this property in the set of all models and their total
number, $1.66\times10^{8}$.
The two $U(1)$ gauge groups required for a standard model setup
are not included in this, since the probability to find a $U(1)$ in
one of the constructions is essentially one.
Multiplying all these factors, we get a probability of
$\approx\,1.3\times10^{-9}$, i.e. one in a billion,
to find a three-generation standard model in the
$T^6/\bZZ$ setup.

\emph{How reliable is this estimate?}
This is of course an important question, since we concluded from the analysis
above that the basic properties are only approximately independent and we can
not really make a quantitative statement about the possible error in our
estimate. So let us compare the result we obtain with this method for models
with standard model gauge group and two or four generations of quarks and
leptons
with the actual numerical results we have obtained in these cases.

The result is shown in table~\ref{tab_smtest}. As can be read of this table,
the estimate for the two- and four-generation case deviates by around 20\%
from the correct value. Keeping this in mind and further noting that we are
making an estimate only at an order-of-magnitude level, a suppression
factor of $\approx 10^{-9}$ seems to be a reliable value.

\begin{table}[th]
\begin{center}
\begin{tabular}{|r|r|r|r|}\hline
\# generations & \# of models found & estimated \# & suppression factor\\\hline
$2$ & $162921$ & $188908$ & $\approx 10^{-3}$\\
$3$ & $0$ & $0.2$ & $\approx 10^{-9}$\\
$4$ & $3898$ & $3310$ & $\approx 2\times10^{-5}$\\\hline
\end{tabular}
\caption{Comparison between the estimated number of solutions and the actual
number of solutions found for models with two, three and four generations.}
\label{tab_smtest}
\end{center}
\end{table}

%
%
\section{Conclusions and outlook}\label{ch_conclusions}
In this work we have reviewed the results on the statistics of the gauge
sector of a specific class of orientifold models. We have presented two
different methods to derive these results. The saddle point
approximation~\cite{bghlw04},
working well in the eight- and six-dimensional case, is not powerful enough
to deal with four-dimensional compactifications, forcing us to perform a more
direct computer aided analysis~\cite{gbhlw05}.
Using this approach, we discussed various
aspects of frequency distributions in the gauge sector. After exploring
the most general case, we focused on models containing phenomenologically
interesting gauge groups.
In the particular case of a standard model gauge
group~\cite{gm05}, an estimate of the number of three generation models in
this setup was given, using the fact that the basic properties of such a
model are sufficiently uncorrelated. For this class of models we also
analysed the values of the gauge couplings at the string scale.
In the case of models containing a Pati-Salam or $SU(5)$ gauge
group~\cite{gmst06}, it was shown that the frequency distributions of
gauge factors do not change, indicating that the specific choice of a
visible sector does not alter the statistics.

Concerning the universality of these results, it should be stressed that
it is very likely that some of them depend strongly on the specific
geometry that has been chosen. The comparison of some of our results with
a study of Gepner models~\cite{dhs041} confirms this conjecture, showing
that only those results show similarities, which are fairly independent
of the geometry. Especially the amount of three generation standard
models might be very different if one chooses other orbifold groups.
To illuminate this point, a study of models on $T^6/\bZ_6$ is currently
under way~\cite{gls06}.
It would be of course very desirable to do similar studies in other corners
of the landscape, in order to see whether the distribution of models
resembling our four dimensional world is uniform or rather sharply peaked
at special points in moduli space.

Interesting directions of future studies would be on the one hand to include
fluxes in our discussion, which probably give rise to a much larger set of
models. On the other hand one could try to obtain frequency
distributions of other important aspects of the (supersymmetric) standard
model, such as Yukawa couplings or soft supersymmetry breaking terms.
Including these in our discussion will certainly reduce the number of
acceptable models and might also give additional hints which parts of the
landscape are worth to be investigated in greater detail.

%
%
\begin{appendix}
\settocdepth{section}
\section{Orientifold models}\label{app_models}
In this appendix we summarise the concrete examples of orientifold models
that are used in this paper.
We fix the notation and translate the conditions
explained in general in section~\ref{ls_orientifolds} into variables that
suit the specific cases and simplify the computations.

\subsection[$T^2$]{$\mathbf{T^2}$}\label{app_t2}
For compactification on $T^2$, a special Lagrangian submanifold is specified
by two wrapping numbers $(n_a,m_a)$ around the fundamental one-cycles. In this
case these numbers are precisely identical to the numbers $(X_a,Y_a)$ used
in section~\ref{ls_orientifolds}.

The tadpole cancellation condition (\ref{eq_tad}) reads
\be{eq_t2_tad}
  \sum_a N_a\, X_a = L,
\ee
where the physical value is $L=16$.

The first supersymmetry condition of~(\ref{eq_susy}) reads just
\be{eq_t2_susy1}
  Y_a=0,
\ee
and is independent of the complex structure $U=R_2/R_1$
on the rectangular torus. This implies that all supersymmetric
branes must lie along the x-axis, i.e. on top of the orientifold
plane.
The second supersymmetry  condition in~(\ref{eq_susy}) becomes
\be{eq_t2_susy2}
  X_a > 0.
\ee
From these conditions we can immediately deduce that
if one does not allow for multiple wrapping, as it is usually done
in this framework, there would only exist one supersymmetric brane,
namely the one with $(X,Y)=(1,0)$.

\subsection[$T^4/\bZ_2$]{$\mathbf{T^4/\bZ_2}$}\label{app_t4}
In this case a class of special Lagrangian branes is given by so-called
factorisable branes, which can be defined by two pairs of
wrapping numbers $(n_i,m_i)$ on two $T^2$s.
The wrapping numbers $(X^i,Y^i)$ with $i=1,2$ for the $\bZ_2$ invariant
two-dimensional cycles are then given by
\be{eq_t4_XYdef}
  X^1=n_1\, n_2, \quad\quad X^2=m_1\, m_2, \quad\quad
  Y^1=n_1\, m_2, \quad\quad Y^2=m_1\, n_2.
\ee
To simplify matters we sometimes use a vector notation
$\vec{X}=(X^1,X^2)^T$ and $\vec{Y}=(X^1,X^2)^T$.

Note that these branes do not wrap the most general homological class,
for the 2-cycle wrapping numbers satisfy the relation
\be{eq_t4_XYrel}
  X^1\, X^2 = Y^1\, Y^2.
\ee
However, for a more general class we do not  know how the
special Lagrangians look like. Via brane recombination it is known
that there exist flat directions in the D-brane moduli space,
corresponding to branes wrapping non-flat special
Lagrangians. Avoiding these complications, we use
the well understood branes introduced above only.

The untwisted tadpole cancellation conditions read
\be{eq_t4_tad_old}
  \sum_a N_a\, X^1_a = L^1, \qquad
  \sum_a N_a\, X^2_a = -L^2,
\ee
with the physical values $L_1=L_2=8$.
In order to put these equations on the same footing, we change the sign
of $X_2$ to get
\be{eq_t4_tad}
  \sum_a N_a\, X^1_a = L^1, \qquad
  \sum_a N_a\, X^2_a = L^2.
\ee

Note that in contrast to models discussed for example
in~\cite{gipo96}, we are only considering bulk branes without
any twisted sector contribution for simplicity%
\footnote{For a treatment of fractional branes in this framework see
e.g.~\cite{bbkl021,bbkl022}.}.
Defining the two form $\Omega_2=(dx_1+i U_1 dy_1)(dx_2+i U_2 dy_2)$, the
supersymmetry conditions become
\be{eq_t4_susy}
  U_1\, Y^1 + U_2\, Y^2 =0, \qquad
  X^1 + U_1\, U_2\, X^2 >0.
\ee
The intersection number between two bulk branes has an extra
factor of two
\be{eq_t4_Iab}
  I_{ab}=-2\left( X^1_a\,X^2_b+X^a_2\,X^1_b
                 +Y^1_a\,Y^2_b+Y^2_a\,Y^1_b\right).
\ee

\subsubsection{Multiple wrapping}\label{app_t4_mw}
In the case of $T^2$ it made no sense to restrict the analysis of
supersymmetric branes to those which are not multiply wrapped around the
torus, because there would have been just one possible construction.
In the case of $T^4/\bZ_2$ the situation is different and we would like
to derive the constraints on the wrapping numbers $\vec{X}$ and $\vec{Y}$.

For the original wrapping numbers $n_i, m_i$ the constraint to forbid
multiple wrapping is $\gcd(n_i,m_i)=1\,\,\forall\,i=1,2$. Without losing
information we can multiply these two to get
\be{eq_t4_mw_org}
  \gcd(n_1,m_1)\gcd(n_2,m_2)=1.
\ee
Using the definitions (\ref{eq_t4_XYdef}) of $\vec{X}$ and $\vec{Y}$,
we can rewrite this as
\be{eq_t4_mw}
  \gcd(X^1,Y^2)\gcd(X^2,Y^2)=Y_2,
\ee
which is invariant under an exchange of $X$ and $Y$.

\subsection[$T^6/\bZZ$]{$\mathbf{T^6/\bZZ}$}\label{app_t6}
In the case of compactifications on this six-dimensional
orientifold, which has been studied by many authors
(see e.g.~\cite{fhs00,csu011,csu012,lapr03,duti05,bcms05}) the
situation is very similar to the four-dimensional case above.
We can describe factorisable branes by their wrapping numbers $(n_i,m_i)$
along the basic one-cycles $\pi_{2i-1}, \pi_{2i}$ of the three
two-tori $T^6=\Pi_{i=1}^3T_i^2$.
To preserve the
symmetry generated by the orientifold projection
$\Omega\bar{\sigma}$, only two different shapes
of tori are possible, which can be parametrised by $b_i\in\{0,1/2\}$
and transform as
\be{eq_cyctr}
  \Omega\bar{\sigma}: \left\{\begin{array}{rcl}
    \pi_{2i-1} &\to& \pi_{2i-1}-2b_i\pi_{2i}\\
    \pi_{2i} &\to& -\pi_{2i}
  \end{array}\right..
\ee
For convenience we work with the combination
$\tilde{\pi}_{2i-1}=\pi_{2i-1}-b_i\pi_{2i}$ and modified
wrapping numbers $\tilde{m}_i=m_i+b_in_i$.
Furthermore we introduce a rescaling factor
\be{eq_t6_cfactor}
c:=\left(\prod_{i=1}^3(1-b_i)\right)^{-1}
\ee
to get integer-valued coefficients.
These are explicitly given by ($i,j,k\in\{1,2,3\}$ cyclic)
\bea{eq_t6_xydef}\nonumber
X^0=cn_1n_2n_3,&\quad& X^i=-cn_i\tilde{m}_j\tilde{m}_k,\\
Y^0=c\tilde{m}_1\tilde{m}_2\tilde{m}_3,&\quad& Y^i=-c\tilde{m}_in_jn_k.
\eea
The wrapping numbers $\vec{X}$ and $\vec{Y}$ are not independent, but
satisfy the following relations:
\be{eq_t6_xyrel}\begin{array}{rclcrcl}
X_I\, Y_I &=& X_J\, Y_J,&\quad\quad&
X_L\, (Y_L)^2&=&X_I\, X_J\, X_K,\\
X_I\, X_J &=& Y_K\, Y_L,&&
Y_L\, (X_L)^2&=&Y_I\, Y_J\, Y_K,
\end{array}\ee
for all $I,J,K,L\in\{0,\ldots 3\}$ cyclic.

Using these conventions the intersection numbers can be written as
\be{eq_t6_Iab}
I_{ab}=\frac{1}{c^2}\left(\vec{X}_a\vec{Y}_b-\vec{X}_b\vec{Y}_a\right).
\ee

The tadpole cancellation conditions read
\be{eq_t6_tad}
\sum_{a=1}^kN_a\vec{X}_a=\vec{L},\quad\quad
\vec{L}=\begin{pmatrix}8c\\\{8/(1-b_i)\}\end{pmatrix},
\ee
where we used that the value of the physical orientifold charge is $8$ in
our conventions.

The supersymmetry conditions can be written as
\be{eq_t6_susy}
\sum_{I=0}^3\frac{Y^I}{U_I}=0,\qquad
\sum_{I=0}^3X^IU_I>0,
\ee
where we used that the complex structure moduli $U_I$ can be defined in terms
of the radii $(R_i^{(1)},R_I^{(2)})$ of the three tori as
\bea{eq_t6_udef}\nonumber
U_0&=&R_1^{(1)}R_1^{(2)}R_1^{(3)},\\
U_i&=&R_1^{(i)}R_2^{(j)}R_2^{(k)},\quad i,j,k\in\{1,2,3\}\,\mbox{cyclic}.
\eea

Finally the K-theory constraints can be expressed as
\be{eq_t6_ktheory}
\sum_{a=1}^kN_aY_a^0\in2\bZ,\quad\quad
\frac{1-b_i}{c}\sum_{a=1}^kN_aY_a^i\in2\bZ,\quad
  i\in\{1,2,3\}.
\ee

\subsubsection{Multiple wrapping}\label{app_t6_mw}
We can define the condition to exclude multiple wrapping in a way similar
to the $T^4$-case. A complication that arises is the possibility to have
tilted tori. In the definition of $\vec{X}$ and $\vec{Y}$
in~(\ref{eq_t6_xydef}) we used the wrapping numbers $\tilde{m}_i$, which have
been defined to include the possible tilt. To analyse coprime wrapping numbers,
however, we have to deal with the original wrapping numbers $m_i$, such that
\be{eq_t6_mw_org}
  \prod_{i=1}^3\gcd(n_i,m_i)=1.
\ee

We can express this condition in terms of the variables $\vec{\tilde{X}}$
and $\vec{\tilde{Y}}$, defined as
\bea{eq_t6_xytildedef}\nonumber
  \tilde{X}^0 = n_1n_2n_3,&\quad&\tilde{Y}^0 = m_1m_2m_3,\\
  \tilde{X}^i = n_in_jn_k,&\quad&\tilde{Y}^i = m_in_jn_k,
\eea
where $i,j,k\in\{1,2,3\}$ cyclic,
analogous to section~\ref{app_t4_mw}
\be{eq_t6_mw}
  \prod_{i=1}^3\gcd(\tilde{Y}^0,\tilde{X}^i)=(\tilde{Y}^0)^2.
\ee
The $\vec{\tilde{X}}$ and $\vec{\tilde{Y}}$ can be expressed in terms of the
$\vec{X}$ and $\vec{Y}$ of~(\ref{eq_t6_xydef}), using their
definition~(\ref{eq_t6_mw_org}) and the rescaling factor~(\ref{eq_t6_cfactor}),
as
\bea{eq_t6_xytilderel}\nonumber
\tilde{X}^0 &=& c^{-1}X^0,\\\nonumber
\tilde{X}^i &=& c^{-1}\left(-X^i+b_jY^k+b_kY^j+b_jb_kX^0\right),\\\nonumber
\tilde{Y}^0 &=& c^{-1}\left(Y^0 + \sum_{i=1}^3 b_i X^i
                                - \sum_{i=1}^3 b_j b_kY^i
                                - b_1b_2b_3 X^0\right),\\
\tilde{Y}^i &=& c^{-1}\left(-Y^i - b_i X^0\right).
\eea

\section{Partition algorithm}\label{app_pa}
In this part of the appendix we briefly outline the partition algorithm
used in the
computer analysis of vacua\footnote{The complete program used to generate
the solutions, which is written in C, can be obtained from the author upon
request.}.
It is designed to calculate the unordered partition of a
natural number $n$, restricted to a maximal number of $m$ factors,
using only a subset $F\subset\bN$ of allowed factors to appear in the
partition.

To describe the main idea, let us drop the additional constraints on the length
and factors of the partition. They can be added easily to the algorithm, for
details see the comments in listing~\ref{lst_main}.
The result is stored in a list $\{a_i\}$, which is initialized with
$a_i=n \delta_{1,i}$. An internal pointer $q$ is set to the first element at
the beginning and after each call of the main routine the list $a$ contains
the next partition. The length of this partition is stored in a variable $m$,
which is set to $m=0$, after the last partition has been generated.

The main routine contains the following steps. It checks if the element $a_q$
is equal to $1$ -- if yes, it sets $q=q-1$. This is repeated until $a_q>1$ or
$q=0$ -- in this case no new partitions exist, $m$ is set to $0$ and the
algorithm terminates. In the second step the routine sets $a_q=a_q-1$,
$a_{q+1}=a_{q+1}+1$ and $q=q+1$.
But this operation is only
performed if $a_{q+1}<a_q$ and $a_q>1$, otherwise the counter $q$ is reduced
by one and the algorithm starts over.

Let us give an example to illustrate this procedure. Consider the unordered
partitions of $5$:
\be{eq_app_pfive}
  \left\{\,
  \{5\},\{4,1\},\{3,2\},\{3,1,1\},\{2,2,1\},\{2,1,1,1\},\{1,1,1,1,1\}\,
  \right\}.
\ee
Starting with $5$ itself, the first time we call the algorithm, it decreases
$a_1$ to
$a_1=4$, increases $a_2$ to $a_2=1$, which generates the partition $\{4,1\}$.
The pointer $q$ is increased to $q=2$. The next time we call the routine,
the element $a_q=a_2$ is equal to $1$, which leads to $q=1$. Now the condition
$a_q>1$ is satisfied and the result of $a_q=a_q-1$, $a_{q+1}=a_{q+1}+1$ gives
the partition $\{3,2\}$.
Continuing in this way, four more partitions of $5$ are generated, until we
reach $\{1,1,1,1,1\}$. We have $a_i=1$ for all $i=1,\ldots,5$, which leads to
the termination of the algorithm in the first step.

\subsection{Implementation}\label{app_paimp}
\lstset{language=c,basicstyle=\footnotesize,aboveskip=2ex,belowskip=3ex,
        xleftmargin=3em, captionpos=b, float=ht, fontadjust}
The algorithm
uses a data structure \texttt{partition} to collect the necessary parameters
and internal variables:
\begin{lstlisting}
typedef struct _partition { long n,m,q,*fac,*a,min; } partition;
\end{lstlisting}
Here \texttt{n}$\,\in\bN$ is the number to be partitioned and \texttt{m} holds
the length of the partition list \texttt{a}.
The array \texttt{fac} contains the set $F$ of allowed values of
partition factors.
\texttt{min} and \texttt{q} are internal variables to be
explained below. Besides these internal variables, a global variable
\texttt{maxp} is used, which contains the maximal length of the
partition.

The algorithm itself is split into two parts. The function
\texttt{apartitions\_first} is called once at the beginning of the program loop
that runs through all partitions. It initializes the internal variables
\texttt{n} and \texttt{fac} and calculates the minimum possible
value for a partition factor from the list \texttt{fac}.
Finally it checks if \texttt{n} itself is contained in \texttt{fac} and
calls the main routine \texttt{apartitions\_next} if this is not the case.

\lstset{stepnumber=5}
\begin{lstlisting}[firstnumber=last,caption={Partition algorithm, initial routine},label=lst_first]
void apartitions_first(long n, long *f, partition *p) {
  long i;
/* check if we're supposed to do anything */
  if ((n>0)&&(maxp==0)) {
    p->m=0;
    return;
  }
/* find minimum and check consistency */
  p->min=n+1;
  i=1;
  while (i<=n) {
    if (f[i]>0) {
      p->min=i;
      i=n+1;
    } else {
      i++;
    }
  }
  if (p->min>n) {
    p->m=0;
    return;
  }
/* init data structure */
  p->n=n;
  p->fac=f;
  p->a=malloc((n+1)*sizeof(long));
  p->a[0]=p->n;
  p->m=1;
  p->a[1]=p->n;
  p->q=1;
/* generate first partition (check if n is allowed...) */
  if (f[n]<=0) {
    apartitions_next(p);
  }
}
\end{lstlisting}

The main routine can be called subsequently as long as the length
\texttt{m} of the partition list \texttt{a} is positive.
Each call will produce a new partition of \texttt{n}.
Special care has to be taken if elements of the partition are not contained
in \texttt{fac} -- see the comments in the source code for
these subtleties.

\begin{lstlisting}[firstnumber=last,caption={Partition algorithm, main routine},label=lst_main]
void apartitions_next(partition *p) {
/* set the number n what we have to distribute to 0. */
  p->n=0; 
/* go back until there is a value bigger then the minimum min to distribute
and the partition doesn't get too long. */
  while ((p->q>=maxp)||((p->q>0)&&(p->a[p->q]==p->min))) {
    p->n=p->n+p->a[p->q];
    p->q=p->q-1;
  }
/* loop through the distribution process as long as we're not back at the
beginning of the factor list. */
  while (p->q>0) {
/* lower the actual value at q we're trying to distribute by 1 and add 1 to
the distribution account. then increase the list-length m by one. */
    p->a[p->q]=p->a[p->q]-1;
    p->n=p->n+1;
    p->m=p->q+1;
/* as long as the new factor is > then the one before or it is not in
fac, subtract 1 from it (and add 1 to n). do this as long as it is >
then the minimum. */
    while (((p->a[p->q]>p->a[p->q-1])||(p->fac[p->a[p->q]]<=0))
           &&(p->a[p->q]>=p->min)) {
      p->a[p->q]=p->a[p->q]-1;
      p->n=p->n+1;
    }
/* check if the new factor is lower or equal then the one before and it's
in fac (the loop above might have terminated on the minimum condition).
if yes, add the distribution sum to the new factor at q+1. if not, add the
whole factor at q to n and go one step back in the list. */
    if ((p->a[p->q]<=p->a[p->q-1])&&(p->fac[p->a[p->q]]>0)) {
      p->q=p->q+1;
      p->a[p->q]=p->n;
/* if the new factor is < then the one before and in our list return. */
      if ((p->a[p->q]<=p->a[p->q-1])&&(p->fac[p->a[p->q]]>0)) {
        return;
      } else {
/* so the new factor is not smaller or in our list - means we have to
redistribute some of it to a new factor. but if we are already at the
maximum length of the partition we have to go one step back! */
        if (p->q < maxhidden) {
          p->n=0;
        } else {
          p->q=p->q-1;
        }
      }
    } else {
      p->n=p->n+p->a[p->q];
      p->q=p->q-1;
    }
  }
/* if the pointer is q is 0 there is nothing left to do - free memory and
return 0 for the length of the partition */
  if (p->q <= 0) {
    free(p->a); p->m=0;
  }
}
\end{lstlisting}
\end{appendix}

%
%
\begin{acknowledgement}
The author would like to thank Dieter L{\"u}st for his support during the
last three years. This work is based on collaborations with Ralph Blumenhagen,
Gabriele Honecker, Maren Stein and Timo Weigand.
\end{acknowledgement}

\end{document}